\documentclass[12pt,preprint]{aastex}

\usepackage{amssymb}
\usepackage{graphicx}

\shorttitle{Mid-IR Spectroscopy of the 12$\mu$m Seyferts}
\shortauthors{Wu et al.}

\begin{document}

\title{Spitzer/IRS 5-35$\mu$m Low-Resolution
  Spectroscopy of the 12$\mu$m Seyfert Sample}

\author{Yanling Wu\altaffilmark{1}, Vassilis Charmandaris\altaffilmark{2,3},
Jiasheng Huang\altaffilmark{4}, Luigi Spinoglio\altaffilmark{5}, Silvia Tommasin\altaffilmark{5,6}}

\altaffiltext{1}{Infrared Processing and Analysis Center, California
  Institute of Technology, 1200 E. California Blvd, MC 314-6,
  Pasadena, CA 91125}

\altaffiltext{2}{University of Crete, Department of Physics, GR-71003,
  Heraklion, Greece}

\altaffiltext{3}{IESL/Foundation for Research and Technology - Hellas,
  GR-71110, Heraklion, Greece, and Chercheur Associ\'e, Observatoire
  de Paris, F-75014, Paris, France}

\altaffiltext{4}{Harvard-Smithsonian Center for Astrophysics, 60
  Garden Street, Cambridge, MA, 02138}

\altaffiltext{5}{Istituto di Fisica dello Spazio Interplanetario,
  INAF, Via Fosso del Cavaliere 100, I-00133 Rome, Italy}

\altaffiltext{6}{Dipartimento di Fisica, Universit di Roma, La Sapienza, Roma,
Italy}

\email{yanling@ipac.caltech.edu, vassilis@physics.uoc.gr,
  jhuang@cfa.harvard.edu, luigi.spinoglio@ifsi-roma.inaf.it,
  Silvia.Tommasin@ifsi-roma.inaf.it}

\begin{abstract}

  We present low-resolution 5.5-35\,$\mu$m spectra for 103 galaxies
  from the 12\,$\mu$m Seyfert sample, a complete unbiased 12\,$\mu$m
  flux limited sample of local Seyfert galaxies selected from the {\em
    IRAS} Faint Source Catalog, obtained with the Infrared
  Spectrograph (IRS) on-board {\it Spitzer} Space Telescope.  For 70
  of the sources observed in the IRS mapping mode, uniformly extracted
  nuclear spectra are presented for the first time. We performed an
  analysis of the continuum emission, the strength of the Polycyclic
  Aromatic Hydrocarbon (PAH) and astronomical silicate features of the
  sources.  We find that on average, the 15-30\,$\mu$m slope of the
  continuum is $<\alpha_{15-30}>$=-0.85$\pm$0.61 for Seyfert 1s and
  -1.53$\pm$0.84 for Seyfert 2s, and there is substantial scatter in
  each type. Moreover, nearly 32\% of Seyfert 1s, and 9\% of Seyfert
  2s, display a peak in the mid-infrared spectrum at 20\,$\mu$m, which
  is attributed to an additional hot dust component.  The PAH
  equivalent width decreases with increasing dust temperature, as
  indicated by the global infrared color of the host
  galaxies. However, no statistical difference in PAH equivalent width
  is detected between the two Seyfert types, 1 and 2, of the same
  bolometric luminosity.  The silicate features at 9.7 and 18\,$\mu$m
  in Seyfert 1 galaxies are rather weak, while Seyfert 2s are more
  likely to display strong silicate absorption. Those Seyfert 2s with
  the highest silicate absorption also have high infrared luminosity
  and high absorption (hydrogen column density N$_H>$10$^{23}$
  cm$^{-2}$) as measured from the X-rays.  Finally, we propose a new
  method to estimate the AGN contribution to the integrated 12\,$\mu$m
  galaxy emission, by subtracting the "star formation" component in
  the Seyfert galaxies, making use of the tight correlation between
  PAH 11.2\,$\mu$m luminosity and 12\,$\mu$m luminosity for star
  forming galaxies.

\end{abstract}

\section{Introduction}

for which most of their nuclear and often
bolometric luminosities is produced in

Active galaxies are galaxies in which one detects radiation from their
nucleus which is due to accretion onto a super-massive black hole
(SMBH) located at the center. The spectrum of an Active Galactic
Nucleus (AGN) is typically flat in $\nu$f$_{\nu}$. The fraction of the
energy emitted from the AGN compared with the total bolometric
emission of the host can range from a few percent in moderated
luminosity systems (L$_{bol} <10^{11}$L$_{\odot}$) to more than 90\%
in quasars (L$_{bol} >10^{12}$L$_{\odot}$) \citep[see][and references
therein]{Ho08}. As a subclass, Seyfert galaxies are the nearest and
brightest AGNs, with 2-10keV X-ray luminosities less than
$\sim10^{44}$ergs$^{-1}$ and their observed spectral line emission
originates principally from highly ionized gas. Seyferts have been
studied at many wavelengths, from X-rays, ultraviolet,optical, to
infrared (IR) and radio. The analysis of their optical spectra has
lead to the identification of two types, Seyfert 1s (Sy 1s) and
Seyfert 2s (Sy 2s), with the type 1s displaying features of both broad
(FWHM$>$2000 km s$^{-1}$) and narrow emission lines, while the type 2s
only narrow-line emission.

The differences between the two Seyfert types have been an intense
field of study for many years. Are they due to intrinsic differences
in their physical properties, or are they simply a result of dust
obscuration that hides the broad-line region in Sy 2s? A so-called
unified model has been proposed \citep[see][]{Antonucci93, Urry95},
suggesting that Sy 1s and Sy 2s are essentially the same objects
viewed at different angles. A dust torus surrounding the central
engine blocks the optical light when viewed edge on (Sy 2s) and allows
the nucleus to be seen when viewed face on (Sy 1s). Optical spectra in
polarized light \citep{Antonucci85} have indeed demonstrated for
several Sy 2s the presence of broad lines, confirming for these
objects the validity of the unified model.  However, the exact nature
of this orientation-dependent obscuration is not clear yet. Recently,
more elaborate models, notably the ones of \citet{Elitzur08},
\citet{Nenkova08}, and \citet{Thompson09} suggest that the same
observational constraints can also be explained with discrete dense
molecular clouds, without the need of a torus geometry.

The study of Seyfert galaxies is interesting also from a cosmological
perspective, as they trace the build up of SMBHs at the centers of
galaxies. Observations up to 10keV have established that the cosmic
X-ray background (CXB) is mostly due to Seyferts with a peak in their
redshift distribution at z$\sim$0.7 \citep{Hasinger05}. Furthermore,
theoretical modeling of the observed number counts suggests that CXB
at 30keV is also dominated by obscured Seyferts at z$\sim$0.7
\citep{Gilli07, Worsley05}. Given the strong ionization field produced
by the accretion disk surrounding a SMBH, the dust present can be
heated to near sublimation temperatures, making an AGN appear very
luminous in the mid-infrared (mid-IR).

Mid-IR spectroscopy is a powerful tool to examine the nature of the
emission from AGNs, as well as the nuclear star-formation
activity. Since IR observations are much less affected by dust
extinction than those at shorter wavelengths, they have been
instrumental in the study of obscured emission from optically thick
regions in AGNs. This is crucial to understand the physical process of
galaxy evolution. With the advent of the {\em Infrared Space
  Observatory (ISO)}, local Seyferts have been studied by several
groups \citep[see][for a review]{Verma05}. Mid-IR diagnostic diagrams
to quantitatively disentangle the emission from AGNs, starbursts and
quiescent star-forming (SF) regions have been proposed, using both
spectroscopy and broad-band photometry
\citep[i.e.][]{Genzel98,Laurent00}. The recent launch of the {\em
  Spitzer} Space Telescope \citep{Werner04} has enabled the study of
AGN with substantially better sensitivity and spatial resolution. In
particular, using the Infrared Spectrograph (IRS\footnote{The IRS was
  a collaborative venture between Cornell University and Ball
  Aerospace Corporation funded by NASA through the Jet Propulsion
  Laboratory and the Ames Research Center.})  \citep{Houck04a} on
board {\em Spitzer}, \citet{Weedman05} demonstrated early into the
mission the variety in the morphology displayed by the mid-IR spectra
of eight classical AGNs. Since then, large samples of AGNs have been
studied in detail, in an effort to quantify their mid-IR properties
\citep{Buchanan06, Sturm06, Deo07, Gorjian07, Hao07, Tommasin08}. In
addition, new mid-IR diagnostics have been developed to probe the
physics of more complex sources, such as luminous and ultra luminous
infrared galaxies (LIRGs/ULIRGs), which may also harbor AGNs. These
were based on correlating the strength of the Polycyclic Aromatic
Hydrocarbons (PAHs), high excitation fine-structure lines, as well as
silicate features
\citep[i.e.][]{Armus07,Spoon07,Charmandaris08,Nardini08}.

The extended 12\,$\mu$m galaxy sample is a flux-limited (down to
0.22\,Jy at 12\,$\mu$m) sample of 893 galaxies selected from the IRAS
Faint Source Catalog 2 \citep{Rush93}. As discussed by
\citet{Spinoglio89}, all galaxies emit a nearly constant fraction
($\sim$7\%) of their bolometric luminosity at 12\,$\mu$m.  As a
result, selecting active galaxies based on their rest frame 12$\mu$m
fluxes is the best approach to reduce selection bias due to the
variations in their intrinsic spectral energy distributions (SED). A
total of 116 objects from this sample have been optically classified
as Seyfert galaxies (53 Sy 1s and 63 Sy 2s), providing one of the
largest IR selected unbiased AGN sample.  This sample also has
ancillary data in virtually all wavelengths, thus making it the most
complete data set for systematically studying the fundamental issues
of AGNs in the infrared.  Low-resolution 5.5-35\,$\mu$m Spitzer/IRS
spectra of 51 Seyferts from the 12\,$\mu$m sample have been published
by \citet{Buchanan06}, who focused on the study of the Seyfert types
and the shape of mid-IR SED using principal component analysis. Based
on this analysis and comparing with radio data, where available, they
estimate the starburst contribution to the observed spectrum and find
it to appear stronger in Sy 2s, in contrast to the unified
model. However, high resolution {\em Spitzer} spectroscopy on 29
objects by \citet{Tommasin08} does not find a clear indication of
stronger star formation in Sy 2s than Sy 1s.

In this paper, we extend earlier work and study the mid-IR properties
and nature of the dust enshrouded emission from 103 Seyferts of the
12\,$\mu$m Seyferts, nearly 90\% of the whole sample. We use low
resolution Spitzer/IRS spectra, focusing mainly on their broad
emission and absorption features and provide for the first time our
measurements of the PAH emission and strength of Si absorption
features. Our observations and data reduction are presented in \S 2. In
\S 3, we show our analysis on the mid-IR continuum shape, PAH emission
and silicate strength of Seyfert galaxies.  A new method to separate
the star-formation and AGN contribution in the 12\,$\mu$m continuum is
proposed in \S 4.  Finally, we summarize our conclusions in \S 5.

\section{Observations and Data Reduction}

\begin{figure}
  \epsscale{1.5}
  \plottwo{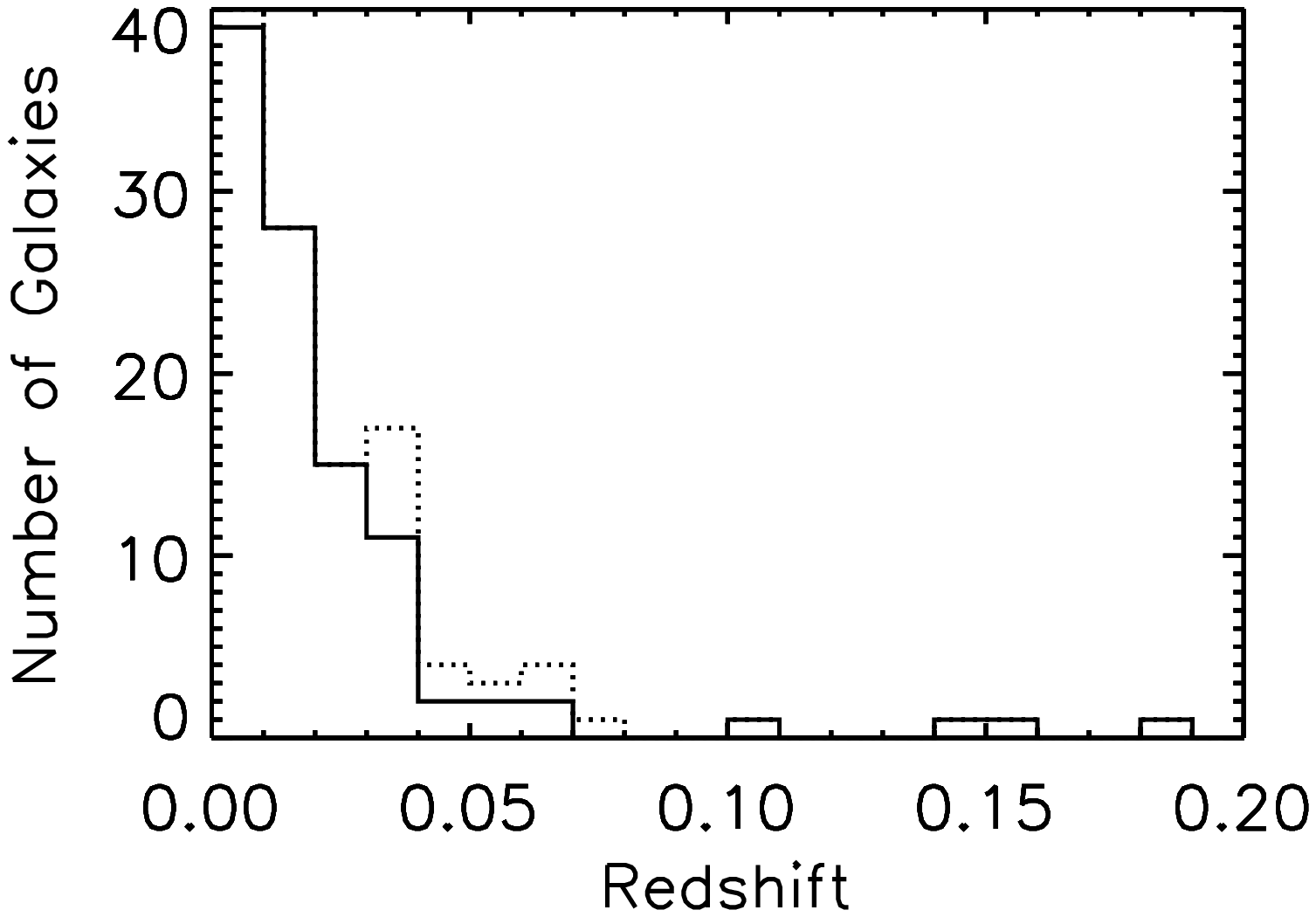}{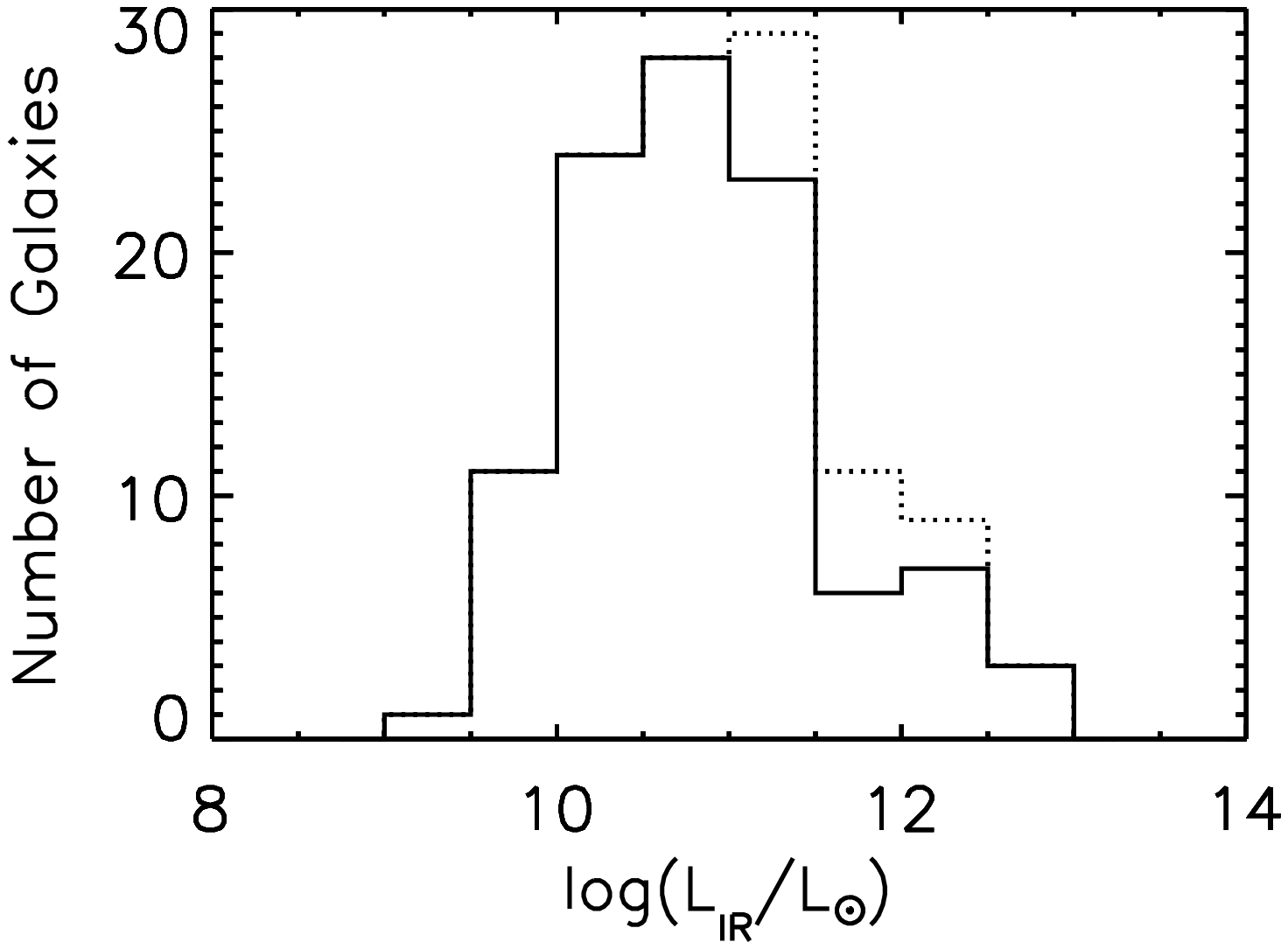}
  \caption{a) The redshift distribution of the 12\,$\mu$m Seyfert
    sample (dotted line) and those with available IRS data studied in
    this paper (solid line).  b) The luminosity distribution of the
    12\,$\mu$m Seyfert sample (dotted line) and those with available
    IRS data studied in this paper (solid line).}
  \label{fig:z_L}
\end{figure}

Since the launch of {\em Spitzer} in August 2003, a large fraction of
the 12\,$\mu$m Seyfert galaxies have been observed by various programs
using the low-resolution (R$\sim$64-128) and high-resolution
(R$\sim$600) modules of Spitzer/IRS. These observations are publicly
available in the {\em Spitzer} archive. A total of 84 galaxies have
been observed with the low-resolution spectral mapping mode of the IRS
by the GO program ``Infrared SEDs of Seyfert Galaxies: Starbursts and
the Nature of the Obscuring Medium'' (PID: 3269), and mid-IR spectra
extracted from the central regions of the maps for 51 galaxies were
published by \citet{Buchanan06}. Spectra for the remaining sources
have been taken as part of a number of Guaranteed, Open Time, as well
as Legacy programs with program identifications (PID) 14, 61, 86, 96,
105, 159, 3237, 3624, 30291, and 30572. For the purpose of this work,
which is mainly to use the PAH emission features to diagnose starburst
and AGN contribution, we are focusing on the low-resolution IRS
spectra (Short Low, SL: 5-15\,$\mu$m; Long Low, LL: 15-37\,$\mu$m). We
performed a complete search of the Spitzer Science Center (SSC) data
archive and retrieved a total of 103 sources with at least Short Low
observations. Among these objects, 47 are optically classified as Sy
1s and 56 as Sy 2s\footnote{The 12$\mu$m Seyferts not included in this
  study due to lack of Spitzer/IRS SL spectra are 6 Sy 1s (Mrk1034,
  M-3-7-11, Mrk618, F05563-3820, F15091-2107, E141-G55) and 7 Sy 2s
  (F00198-7926, F00521-7054, E541-IG12, NGC1068, F03362-1642, E253-G3,
  F22017+0319). }. We adopt the spectral classification of
\citet{Rush93} for the Seyfert types. Even though in a few cases, the
classification may be ambiguous or may have changed, we do not expect
our results, which are of statistical nature, to be affected. The
complete list of the objects analyzed in this paper including their
coordinates, IRAS fluxes, IR luminosity, redshift, Seyfert type and
{\em Spitzer} program identification number are presented in Table
\ref{tab1}.  The redshift and luminosity distribution of the
12\,$\mu$m Seyfert sample and the galaxies with IRS data studied in
this paper is displayed in Figure \ref{fig:z_L}.

With the exception of the data from the SINGS Legacy program (PID
159), of which we directly used spectra available at the
SSC\footnote{The SINGS data products are available at:
  http://data.spitzer.caltech.edu/popular/sings/. The nuclear spectra
  were extracted over a 50$\arcsec$$\times$33$\arcsec$ region on the
  nucleus.}, all other raw datasets are retrieved from the {\em
  Spitzer} archive and reduced in the following manner.

All except three\footnote{The three exceptions are NGC1097, NGC1566,
NGC5033, and they were also observed by SINGS.} of the 12\,$\mu$m
Seyferts with cz$<$10,000km\,s$^{-1}$ have been observed with the IRS
spectral mapping mode (PID 3269). This enabled us to also study the
circumnuclear activity and examine the contribution of the host galaxy
emission to the nuclear spectrum.  Using datasets from this program,
mid-IR spectra obtained from the central slit placement of each map
with point-source extraction were published by \citet{Buchanan06}.
However, as these authors have noted, since the observations had been
designed as a spectral map, blind telescope pointing was used. As a
result, no effort was made to accurately acquire each target and to
ensure that the central slit of each map was indeed well centered on
the source. Moreover, for sources where the mid-IR emission is
extended, using a point-source extraction method would likely result
in an overestimate of the flux densities, due to the slit loss
correction function applied. This would consequently affect the SED of
the galaxy.

\begin{figure}
  \epsscale{1.}
  \plotone{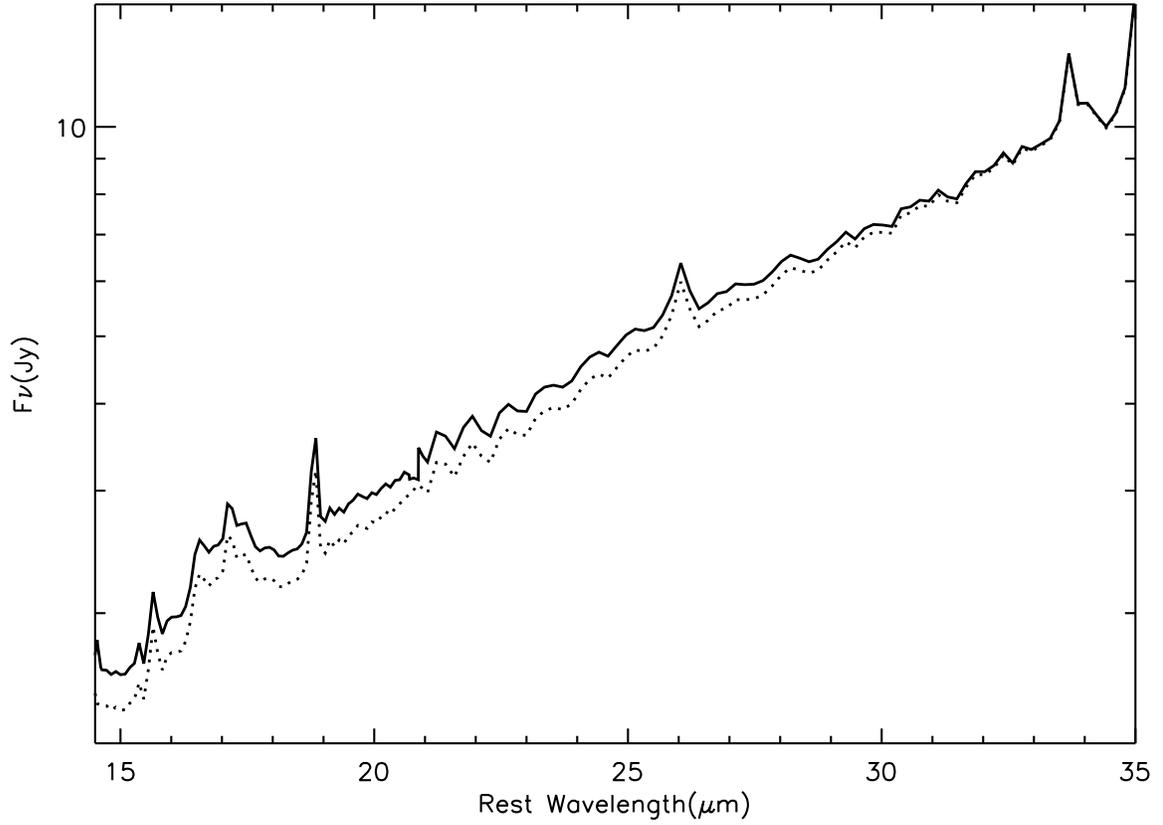}
  \caption{A comparison of the LL spectra of NGC1365. The solid and
    dotted line is before and after applying the image convolution
    method.}
  \label{fig:ngc1365}
\end{figure}

For most of the data reduction in this paper, we used The CUbe Builder
for IRS Spectra Maps \citep[CUBISM]{Smith07b}, in combination with an
image convolution method. This method is developed explicitly for
IRS mapping mode observations where the map size is small and involves
the following steps. Spectral cubes were built using CUBISM from the
IRS maps. Sky subtraction was performed by differencing the on and off
source observations of the same order in each module (SL and LL). The
observations from PID 3269 were designed to minimize redundancy and
maximize the number of sources that can be observed using a limited
amount of telescope time. As a result these maps consisted of only 13
steps in SL and 5 steps in LL with no repetition and for a source with
extended mid-IR emission, the small map may not encompass the entire
source, e.g. NGC1365. To obtain an accurate SED from the extracted
region of the galaxy, one needs to ensure that the same fraction of
source fluxes at all wavelengths is sampled. Since the point spread
function (PSF) changes from 5 to 35\,$\mu$m, we adopted an image
convolution method to account for the change in the full width half
maximum (FWHM) of the PSF: 2-dimensional images at each wavelength
were convolved to the resolution at the longest wavelength
(35\,$\mu$m). Then low-resolution spectra were extracted with matched
apertures, chosen to encompass the whole nuclear emission. Even
though the image convolution method dilutes the fluxes included in the
extraction aperture, it does ensure an accurate SED shape, especially
for small maps that cannot encompass the extended emission for the
source. A comparison of the LL spectra before and after image
convolution for NGC1365 can be found in Figure \ref{fig:ngc1365}.

\begin{figure}
  \epsscale{1}
  \plotone{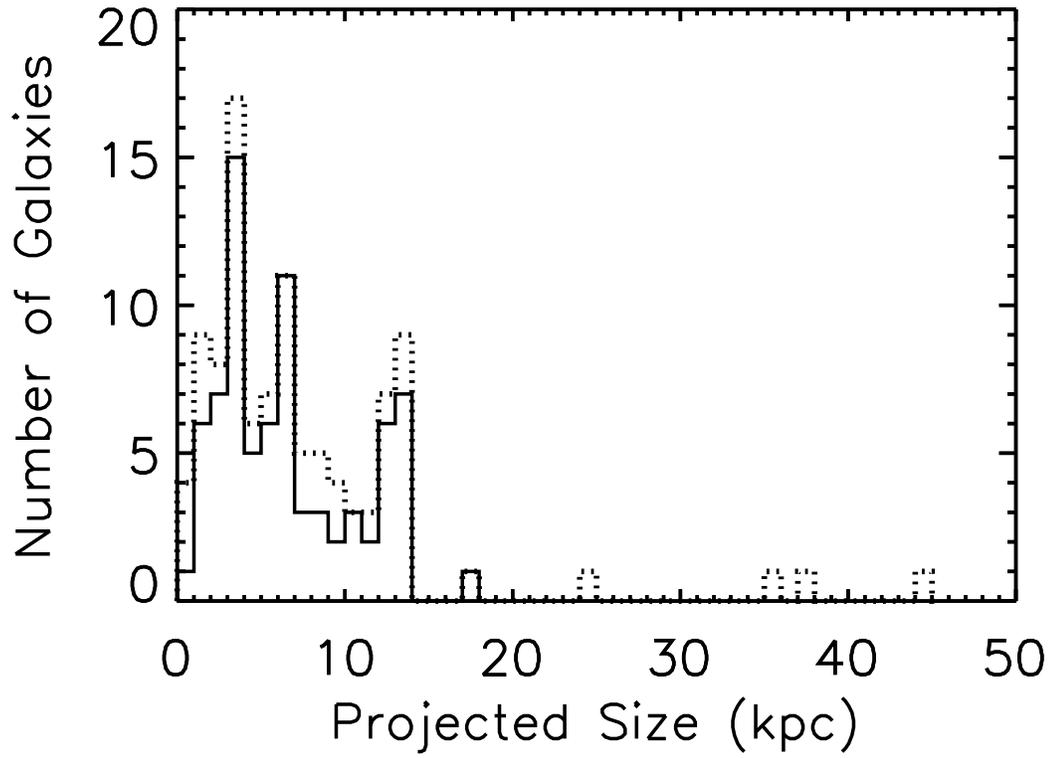}
  \caption{A histogram of the projected linear size of the IRS
    spectral extraction aperture at the distance of the corresponding
    target of the 12\,$\mu$m Seyfert sample.  The dotted line
    indicates the distribution of the whole sample, while the solid
    line indicates the distribution for those that have been observed
    using the spectral mapping mode.}
  \label{fig:size_hist}
\end{figure}

A number of tests with varying sizes for the spectral extraction
aperture were performed in order to select the optimum size.  An
aperture of 4$\times$3 pixels in LL was adopted, which corresponds to
an angular size of 20.4$\arcsec$$\times$15.3$\arcsec$. Then a complete
5--35$\mu$m spectrum of all galaxies centered on their nucleus was
extracted\footnote{For two Sy 2s NGC1143/4 and NGC4922, only SL data
  were available, thus only 5--15$\mu$m spectra were obtained.}. The
main motivation behind this choice was to ensure that we could produce
an accurate overall SED of the extracted region, even in cases where
the emission was extended. This was essential so that we have a
reliable measurement of the continuum emission, in order to calculate
the mid-IR slope, and to estimate the strength of the silicate
features. Further tests were performed by extracting just the SL
spectra of the sample using the smallest aperture possible (2$\times$2
pixels in SL or 3.6$\arcsec\times3.6\arcsec$).  It was found that, in
general, the measured fluxes and EWs of the PAH features did not
differ more than 20\% from the integrated spectra over the larger
apertures, which indicates that these objects are likely to be less
than 20\% more extended than the point-spread-function of point
sources. This suggests that most of the sources were not very extended
in the mid-IR and more than 80\% of their flux originates from a point
source unresolved to Spitzer. Scaling the spectra between different
orders and modules was not needed for most of the cases, and when an
order mismatch was detected, the SL spectra were scaled to match
LL. In Table \ref{tab2}, where we present our measurements, the sizes
of the extraction apertures as well as their corresponding projected
linear sizes on the sky are also listed. A histogram of the physical
size of extraction aperture for the whole sample is presented in
Figure \ref{fig:size_hist} with the dotted line, while in the same
figure, we also show the distribution for sources that are observed in
spectral mapping mode with the solid line. All the sources that are
extracted from a projected size of more than 20~kpc are observed using
the staring mode and reduced with point-source extraction method.

For data obtained with the IRS staring mode, the reduction was done in
the following manner. We started from intermediate pipeline products,
the ``droop'' files, which only lacked stray light removal and
flat-field correction.  Individual pointings to each nod position of
the slit were co-added using median averaging. Then on and off source
images were differenced to remove the contribution of the sky
emission. Spectra from the final 2-D images were extracted with the
Spectral Modeling, Analysis, and Reduction Tool
\citep[SMART,][]{Higdon04} in a point source extraction mode, which
scaled the extraction aperture with wavelength to recover the same
fraction of the diffraction limited instrumental PSF. Note that since
the width of both the SL and LL slits is 2 pixels (3.6$\arcsec$ and
10.2$\arcsec$ respectively), no information could be retrieved along
this direction from areas of the galaxy which were further away. The
spectra were flux calibrated using the IRS standard star $\alpha$ Lac,
for which accurate template was available \citep{Cohen03}.  Finally,
using the first order of LL (LL1, 20--36$\mu$m) spectrum to define the
absolute value of the continuum, the flux calibrated spectra of all
other low-resolution orders were scaled to it.

For nine sources in the 12\,$\mu$m Seyfert sample, both spectral
mapping and staring mode observations had been obtained and were
available in the SSC data archive. For these galaxies, the spectra
were extracted following the above mentioned recipes for all
observations. In order to ascertain again how extended the emission
from these source was, we compared the resulting spectra of the same
source. With two exceptions, NGC4151 and NGC7213, all other seven
spectra obtained in staring mode required a scaling factor larger than
1.15 between the SL and LL modules. This further suggests that the
nuclear emission in those sources is indeed extended, and point-source
extraction may not be appropriate. For galaxies with only staring mode
observations, we also checked the difference between the overlap
region ($\sim15\mu$m) in the SL and LL modules. None of them require a
scaling factor more than 1.15, suggesting that those objects are
point-like at least along the direction of the IRS slits.

\section{Results}                                        

\subsection{Global Mid-IR spectra of Seyfert Galaxies}

It has been well established that the mid-IR spectra of Seyfert
galaxies display a variety of features \citep[see][and references
therein]{Clavel00, Verma05, Weedman05, Buchanan06, Hao07}. This is
understood since, despite the optical classification of their nuclear
activity, emission from the circumnuclear region, as well as of the
host galaxy, also influences the integrated mid-IR spectrum of the
source. Our complete 12\,$\mu$m selected Seyfert sample provides an
unbiased framework to study the statistics on their mid-IR properties.
Earlier work by \citet{Buchanan06} on just 51 AGN from this sample
presented a grouping of them based on their continuum shapes and
spectral features. In this section, we explore the global shape of the
mid-IR spectra for the complete Spitzer/IRS set of data on the
12\,$\mu$m Seyferts.

\begin{figure}
  \epsscale{1.0}
  \plotone{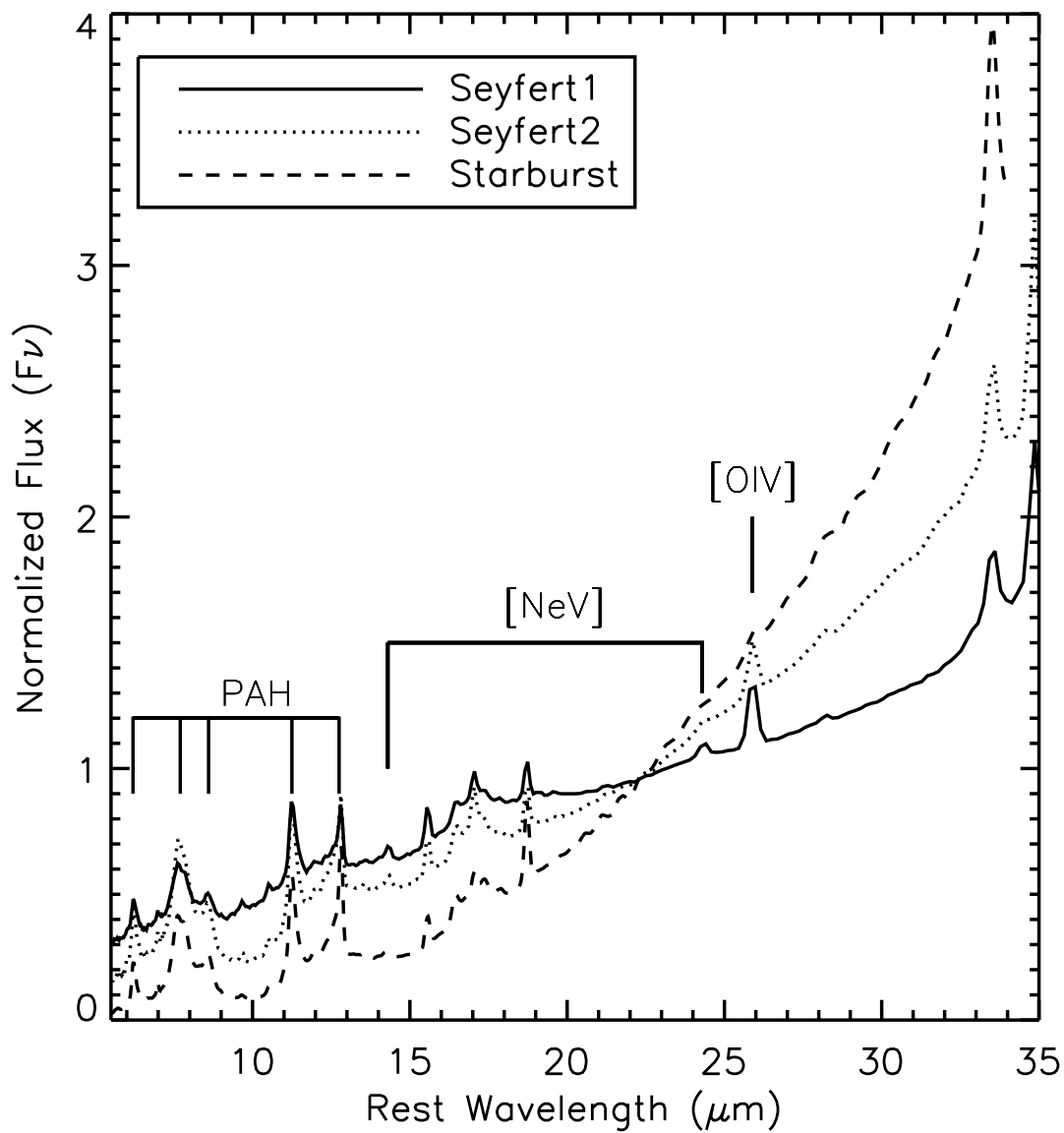}
  \caption{A comparison among the average mid-IR spectrum of Sy 1s
    (solid line) and Sy 2s (dotted line) of the 12\,$\mu$m sample, as
    well as the starbursts (dashed line) of \citet{Brandl06}. All
    spectra have been normalized at 22\,$\mu$m. Note that the
    high-ionization fine-structure lines of [OIV]\,25.89\,$\mu$m are
    present in all three spectra, while [NeV]\,14.3/24.3\,$\mu$m are
    only present in the average spectra of the two Seyfert types.}
  \label{fig:avespect}
\end{figure}

We examine if the average mid-IR spectrum of a Sy 1 galaxy is
systematically different from that of a Sy 2. The IRS spectra for 47
Sy 1s and 54 Sy 2s with full 5.5-35\,$\mu$m spectral coverage,
normalized at the wavelength of 22\,$\mu$m, are averaged and plotted
in Figure \ref{fig:avespect}. For comparison, we over-plot the average
starburst template from \citet{Brandl06}. It is clear that the mid-IR
continuum slope of the average Sy 1 spectrum is shallower than that of
Sy 2, while the starburst template has the steepest spectral slope,
indicating a different mixture of hot/cold dust component in these
galaxies \citep[also see][]{Hao07}. This would be consistent with the
interpretation that our mid-IR spectra of Sy 2s display a strong
starburst contribution, possibly due to circumnuclear star formation
activity included in the aperture we used to extract the spectra from,
consistent with the findings of \citet{Buchanan06} that star formation
is extended and it is not a purely nuclear phenomenon. PAH emission,
which is a good tracer of star formation activity \citep{Forster04},
can be detected in the average spectra of both Seyfert types, while it
is most prominent in the average starburst spectrum. PAH emission
originates from photo-dissociation region (PDR) and can easily be
destroyed by the UV/X-ray photons in strong a radiation field produced
near massive stars and/or an accretion disk surrounding a SMBH
\citep{Voit92, Laurent00, Clavel00, Sturm02, Verma05, Weedman05}. In
the 12\,$\mu$m Seyfert sample, we detect PAH emission in 37 Sy 1s and
53 Sy 2s, that is 78\% and 93\% for each type respectively. This is
expected since the apertures we used to extract the mid-IR spectra for
the 12$\mu$m sample correspond in most cases to areas of more than
1~kpc in linear dimensions (see Figure \ref{fig:size_hist}). As a
result, emission from the PDRs associated with the extended
circumnuclear region and the disk of the host galaxy is also
encompassed within the observed spectrum.  High ionization
fine-structure lines, such as [NeV]14.32\,$\mu$m/24.32\,$\mu$m, are
clearly detected even in the low-resolution average spectrum of Sy
1. This signature is also visible, though rather weak, in the average
spectrum of Sy 2, while it is absent in the average starburst
template. Due to photons of excitation energy higher than 97eV and
typically originating from the accretion disk of an AGN, [NeV] serves
as an unambiguous indicator of an AGN. Even though the low-resolution
module of IRS was not designed for studying fine-structure lines, we
are still able to detect [NeV] emission in 29 Sy 1s and 32 Sy 2s,
roughly 60\% for both types.  Another high ionization line,
[OIV]\,25.89\,$\mu$m, with ionization potential of 54eV, also appears
in both Seyfert types (42 Sy 1s and 41 Sy 2s), and is stronger in the
average spectrum of Sy 1. The [OIV] emission line can be powered by
shocks in intense star forming regions or AGNs \citep[see][]{Lutz98,
  Schaerer99, Bernard-Salas09, Hao09}. In our sample it is probably
powered by both, given the large aperture we adopted for spectral
extraction.  More details and a complete analysis of mid-IR
fine-structure lines for 29 galaxies from the 12\,$\mu$m Seyfert
sample are presented in \citet{Tommasin08}, while the work for the
entire sample is in progress (Tommasin et al. 2009).

An atlas with mid-IR low-resolution spectra of the 12\,$\mu$m Seyfert
sample is included at the end of this paper\footnote{All spectra are
  also available in electronic format.}. We find that a fraction of
our sources show a flattening or a local maximum in the mid-IR
continuum at $\sim$20\,$\mu$m, which had also been noted as a ``broken
power-law'' in some Seyfert galaxies by \citet{Buchanan06}. Another
more extreme such case was the metal-poor blue compact dwarf galaxy
SBS0335-052E \citep{Houck04b}, and it was interpreted with the
possible absence of larger cooler dust grains in the galaxy. Since the
change of continuum slope appears at $\sim$20\,$\mu$m, we use the flux
ratio at 20 and 30\,$\mu$m, F$_{20}$/F$_{30}$, to identify these
objects. After further examination of the spectra, we find 15 Sy 1s
and 4 Sy 2s, which have a F$_{20}$/F$_{30}$$\ge$0.95. We call these
objects ``20\,$\mu$m peakers''.

\begin{figure}
  \epsscale{1.0}
  \plotone{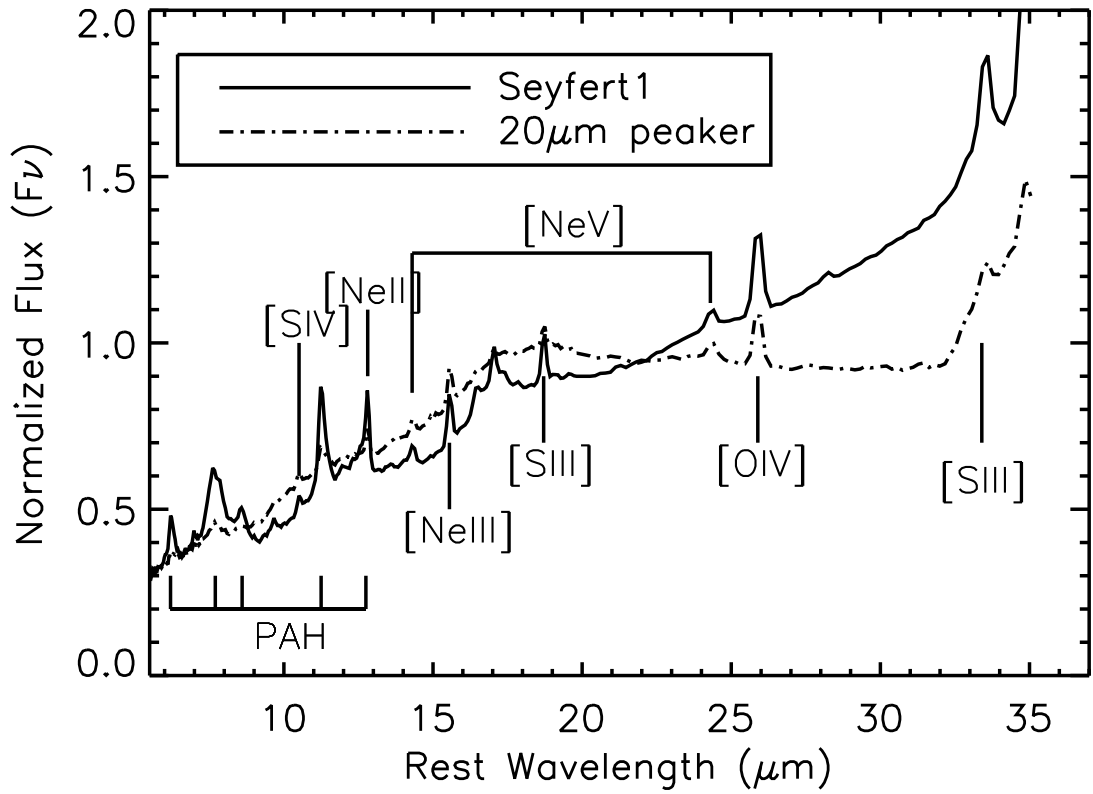}
  \caption{A comparison between the average mid-IR spectrum of
    ``20\,$\mu$m peakers'' (dash-dotted line) and Sy 1s (solid line) of
    our sample. All spectra have been normalized at 22\,$\mu$m.}
  \label{fig:ave_peaker}
\end{figure}

\begin{figure}
  \epsscale{1.0}
  \plotone{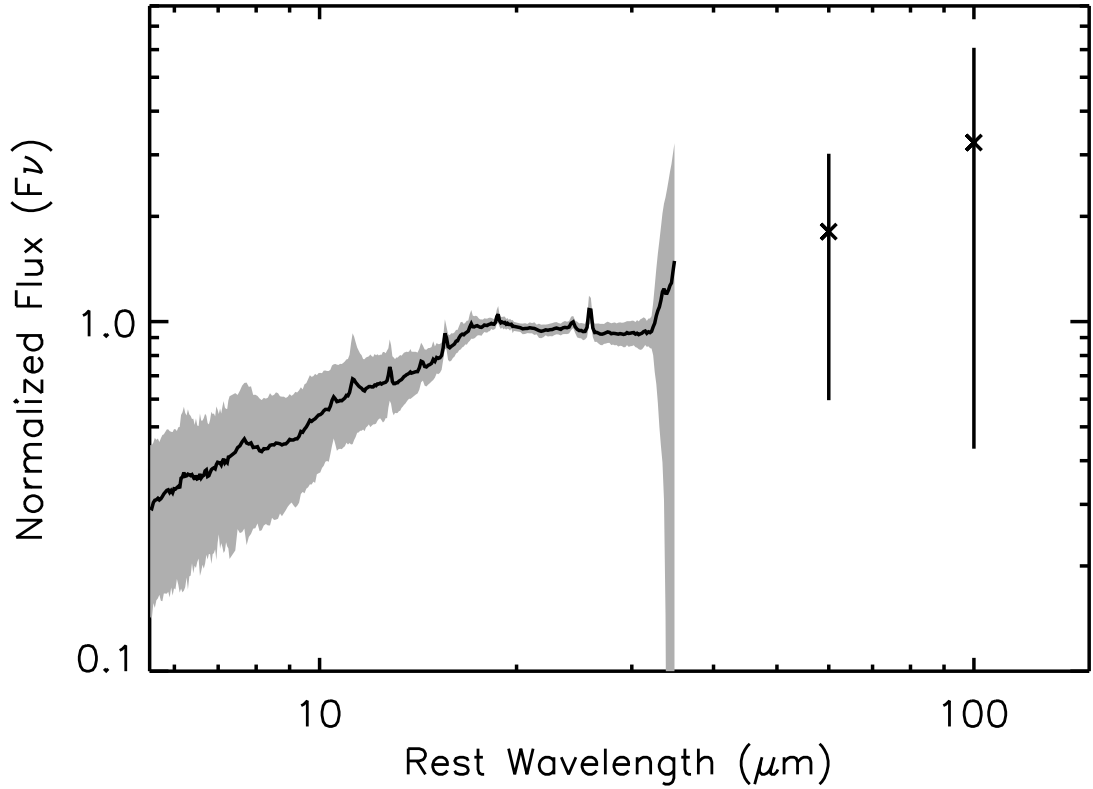}
  \caption{The global IR spectral energy distribution of the
    ``20\,$\mu$m peakers'' after normalizing at 22\,$\mu$m. The two
    data points at 60 and 100\,$\mu$m have been obtained by averaging
    the IRAS 60 and 100\,$\mu$m fluxes after normalization. The grey
    zone in the spectrum and the error bars indicate the 1-$\sigma$
    scatter of the averaged values.}
  \label{fig:peaker_sed}
\end{figure}

\begin{figure}
  \epsscale{1.0}
  \plotone{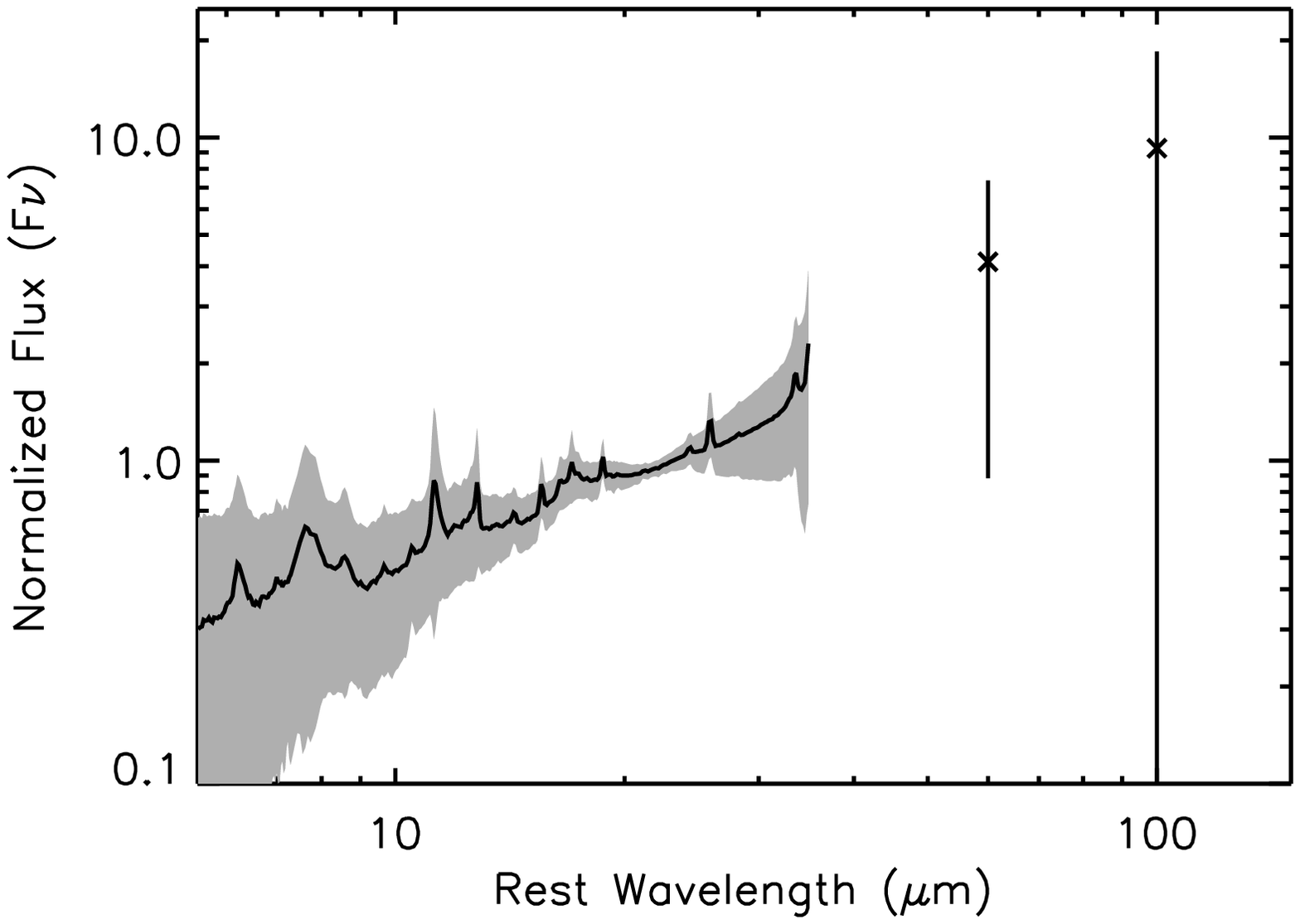}
  \caption{The average IR spectral energy distribution of the Seyfert
    1 galaxies after normalizing at 22\,$\mu$m. The two data points at
    60 and 100\,$\mu$m have been obtained by averaging the IRAS 60 and
    100\,$\mu$m fluxes after normalization. The grey zone in the
    spectrum and the error bars indicate the 1-$\sigma$ scatter of the
    averaged values.}
  \label{fig:s1_sed}
\end{figure}

\begin{figure}
  \epsscale{1.0}
  \plotone{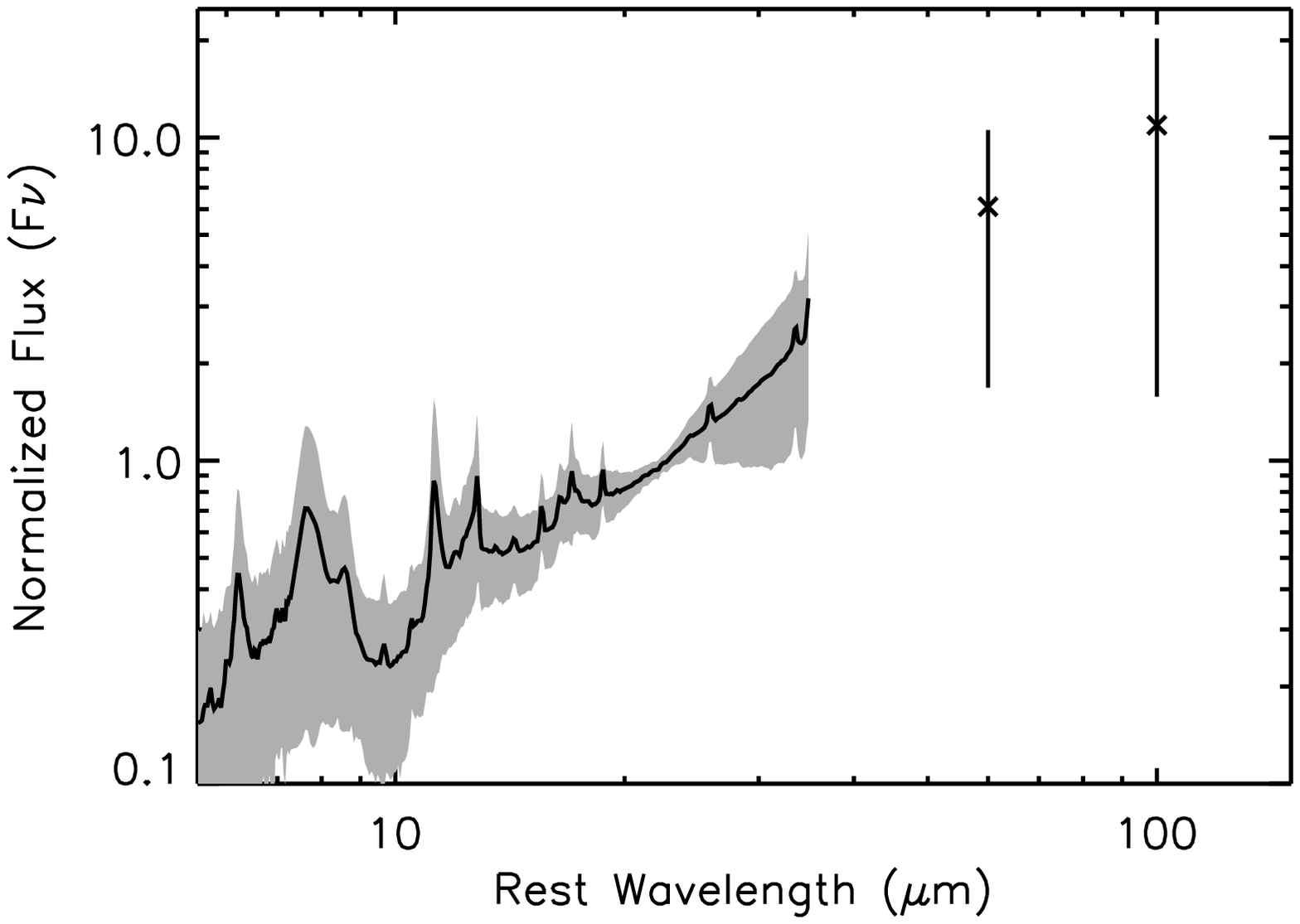}
  \caption{The average IR spectral energy distribution of the Seyfert
    2 galaxies after normalizing at 22\,$\mu$m. The two data points at
    60 and 100\,$\mu$m have been obtained by averaging the IRAS 60 and
    100\,$\mu$m fluxes after normalization. The grey zone in the
    spectrum and the error bars indicate the 1-$\sigma$ scatter of the
    averaged values. }
  \label{fig:s2_sed}
\end{figure}

To analyse the properties of ``20\,$\mu$m peakers'', we plot the
average IRS spectra of the 19 sources in Figure
\ref{fig:ave_peaker}. As most of these sources are type 1 Seyferts (15
out of 19), we also overplot the average Sy 1 spectrum for comparison.
In addition to their characteristic continuum shape, a number of other
differences between the ``20\,$\mu$m peakers'' and Sy 1s are also
evident. PAH emission, which is clearly detected in the average Sy 1
spectrum, appears to be rather weak in the average ``20\,$\mu$m
peaker'' spectrum. The high-ionization lines of [NeV] and [OIV] are
seen in both spectra with similar strength, while low-ionization
lines, especially [NeII] and [SIII], are much weaker in the average
spectrum of ``20\,$\mu$m peakers''. If we calculate the infrared color
of a galaxy using the ratio of F$_{25}$/F$_{60}$ (see section 3.3 for
more detailed discussion on this issue), we find an average value of
0.75 for the ``20\,$\mu$m peakers'', while it is 0.30 for the other
``non-20\,$\mu$m peaker'' Sy 1s in the 12\,$\mu$m sample.  Finally,
the average IR luminosities of the ``20\,$\mu$m peakers'' and Sy 1s do
not show significant difference, with log(L$_{\rm IR}/L_\odot$)=10.96
for the former and log(L$_{\rm IR}/L_\odot$)=10.86 for the
latter. These results are consistent with the ``20\,$\mu$m peakers''
being AGNs with a dominant hot dust emission from a small grain
population heated to effective temperatures of $\sim$ 150K and a
possible contribution due to the distinct emissivity of astronomical
silicates at 18$\mu$m.  Their radiation field must also be stronger
than a typical Sy 1, since it destroys the PAH molecules around the
nuclear region more efficiently. We should stress that unlike
SBS0335-052E, whose global IR SED peaks at $\sim$30\,$\mu$m, and
likely has limited quantities of cold dust, the mid-IR peak we observe
in these objects at $\sim$20\,$\mu$m is likely a local
one\footnote{However, some objects, such as Mrk335, Mrk704 and 3C234,
  do have IRAS ratios of F$_{\rm 25}$/F$_{\rm 60}>$1.}. This becomes
more evident in Figure \ref{fig:peaker_sed}, where we also include the
scaled average 60 and 100\,$\mu$m flux densities for the ``20\,$\mu$m
peakers''. It is clear that their far-IR emission increases with
wavelength, thus confirming the presence of ample quantities cold dust
in these objects.  To contrast the global SED of these objects with
the whole Seyfert sample we include in Figures \ref{fig:s1_sed} and
\ref{fig:s2_sed} the average SED of the Sy 1s and Sy 2s of the
sample. All SEDs have been normalized as in Figure \ref{fig:avespect}
at 22$\mu$m.  Unlike the ``20\,$\mu$m peakers'' one can easily observe
the regular increase of the flux from $\sim$15 to 60$\mu$m in the
average Sy 1 and Sy 2 SEDs.  A more detailed analysis of the dust
properties of the ``20\,$\mu$m peakers'' in comparison with typical 
active galaxies will be presented in a future paper.

\subsection{The PAH emission in the 12$\mu$m Seyferts}

In this section, we explore some of the properties of the PAH emission
in our sample and contrast our findings to the previous work. To
quantify the strength of PAH emission, we follow the usual approach
and measure the fluxes and equivalent widths (EWs) of the 6.2 and
11.2\,$\mu$m PAH features from their mid-IR spectra. Even though the
7.7$\mu$m PAH is the strongest of all PAHs \citep{Smith07a}, we prefer
not to include it in our analysis. This is due to the fact that its
measurement is affected by absorption and emission features next to it
and depends strongly both on the assumptions of the various
measurement methods (spline or Drude profile fitting) as well as the
underlying continuum. Furthermore, it often spans between the two SL
orders, which could also affect its flux estimate. The 6.2 and
11.2\,$\mu$m PAH EWs are derived by integrating the flux of the
features above an adopted continuum and then divide by the continuum
fluxes in the integration range. The baseline is determined by fitting
a spline function to the selected points (5.95-6.55\,$\mu$m for the
6.2\,$\mu$m PAH and 10.80-11.80\,$\mu$m for the 11.2\,$\mu$m PAH). The
PAH EWs as well as the integrated fluxes are listed in Table
\ref{tab2}. The errors including both flux calibration and measurement
are estimated to be less than $\sim$15\% on average.

\begin{figure}
  \epsscale{1.0}
  \plotone{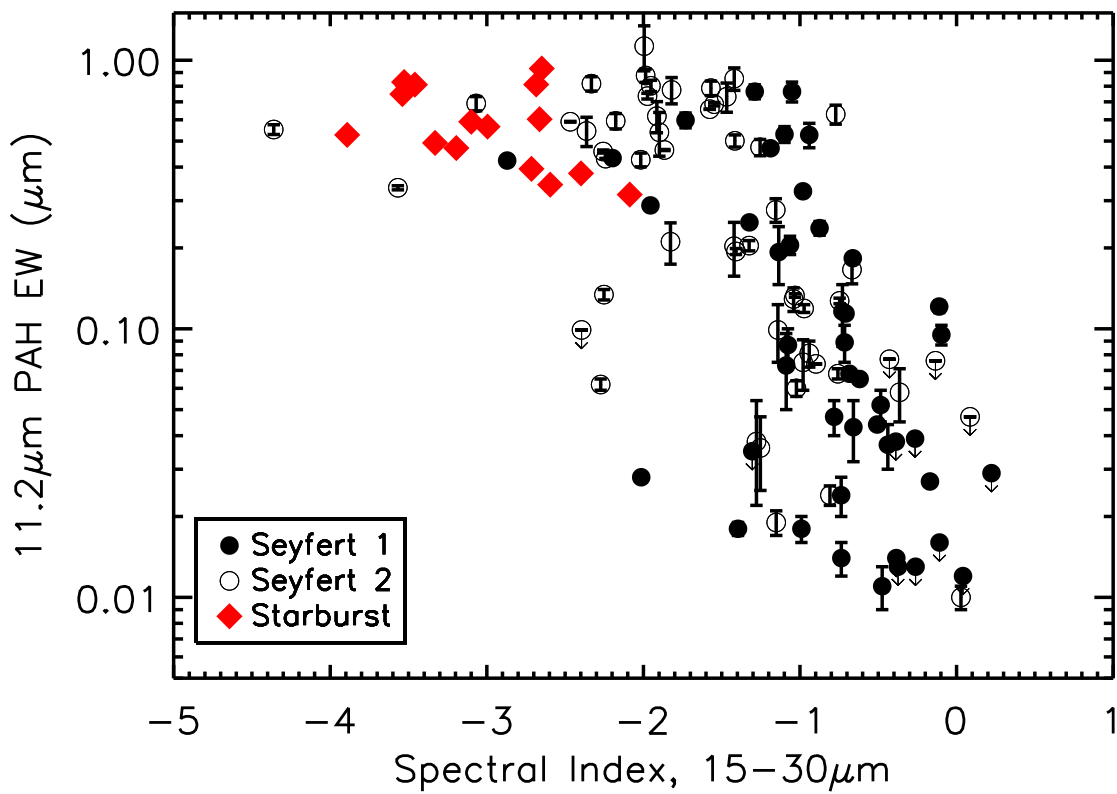}
  \caption{The 15-30\,$\mu$m spectral index vs 11.2\,$\mu$m PAH EW for
    the 12\,$\mu$m Seyfert sample.The filled circles are Seyfert 1s,
    the open circles are Seyfert 2s, while the diamonds denote the
    starburst galaxies from \citet{Brandl06}.Note that the PAH EWs of
    Seyferts are progressively suppressed as their 15 to 30\,$\mu$m
    continuum slopes flatten.}
  \label{fig:pah_index}
\end{figure}

The first study on the PAH properties of a large sample of Seyferts
was presented by \citet{Clavel00}, using ISO/PHOT-S 2.5--11$\mu$m
spectra and ISO/CAM broad band mid-IR imaging. The authors suggested
that there was a statistical difference in the strength of the PAH
emission and the underlying hot continuum ($\sim7\mu$m) emission
between type 1 and type 2 objects.  They also found that Sy 2s had
stronger PAHs than Sy 1s, while Sy 1s had higher hot continuum
associated with emission from the AGN torus. This was consistent with
an orientation depended depression of the continuum in Sy 2s. The
interpretation of these results was challenged by \citet{Lutz04}, who
attributed it to the large (24$\arcsec\times24\arcsec$) aperture of
ISO/PHOT-S, and the possible contamination from the host galaxy. More
recently, \citet{Deo07} have found a relation between the 6.2\,$\mu$m
PAH EWs and the 20-30\,$\mu$m spectral index\footnote{The spectral
  index $\alpha$ is defined as log(F1/F2)/log($\nu1$/$\nu2$).} , with
a steeper spectral slope seen in galaxies with a stronger starburst
contribution. This is understood since galaxies hosting an AGN are
``warmer'' and have an IR SED peaking at shorter wavelengths thus
appearing flatter in the mid-IR (see also next section). Given the
global correlations between star formation activity and PAH strength
\citep[i.e.]{Forster04, Peeters04, Wu05, Calzetti05, Calzetti07},
star-forming galaxies are expected to also have stronger PAH
features. In Figure \ref{fig:pah_index}, we plot the
15-30\,$\mu$m\footnote{We choose to use $\alpha_{15-30}$ so that we
  can directly make use of the spectral index measurement for the
  starburst galaxies in the \citet{Brandl06} sample.} spectral index
for the 12\,$\mu$m Seyfert sample as a function of their 11.2\,$\mu$m
PAH EWs. The diamonds indicate the starburst galaxies from
\citet{Brandl06}\footnote{We have excluded the 6 galaxies that have
  AGN signatures from the \citet{Brandl06} starburst sample.}. A
general trend of the PAH EWs decreasing as a function of 15-30\,$\mu$m
spectral index is observed in Figure \ref{fig:pah_index}, even though
it is much weaker than the anti-correlation presented by
\citet{Deo07}.  Starburst galaxies are located on the upper left
corner of the plot, having very steep spectral slopes, with
$<\alpha_{15-30}>$=-3.02$\pm$0.50, and large PAH EWs, nearly
0.7$\mu$m. Seyfert galaxies spread over a considerably larger range in
spectral slopes as well as PAH EWs. Sy 1s and Sy 2s are mixed on the
plot. On average, the 15-30\,$\mu$m spectral index is
$<\alpha_{15-30}>$=-0.85$\pm$0.61 for Sy 1s and
$<\alpha_{15-30}>$=-1.53$\pm$0.84 for Sy 2s. Note that although the
mean spectral slope is slightly steeper for Sy 2s, there is
substantial scatter, as is evident by the uncertainties of the mean
for each types (see also Figure 7 and 8).

\begin{figure}
  \epsscale{1.0}
  \plotone{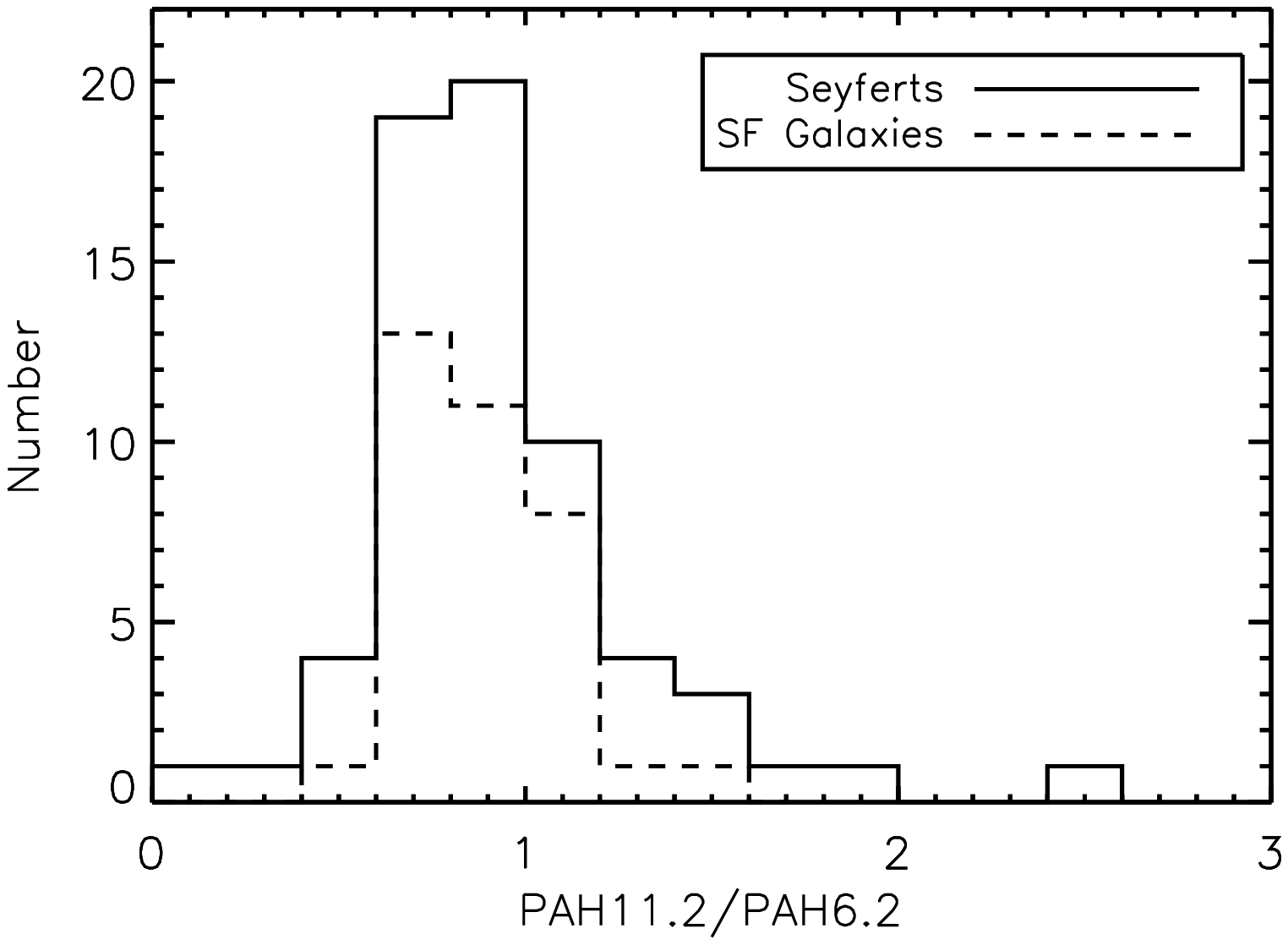}
  \caption{A histogram of the flux ratio of the 11.2\,$\mu$m PAH to
    the 6.2\,$\mu$m PAH feature. The solid line indicates the values
    of the 12\,$\mu$m Seyfert sample while the dashed line indicates
    those of the SF galaxies from the \citet{Brandl06} and
    \citet{Smith07a}. Galaxies with only upper limits measured for the
    aromatic features are excluded from this plot. Both the SF
    galaxies and the Seyferts appear to have similar distribution of
    the 11.2\,$\mu$m/6.2\,$\mu$m PAH flux ratios, indicating that
    globally the chemical structure of the aromatic features observed
    in Seyfert nuclei are likely very similar to those seen in SF
    galaxies.}
  \label{fig:pah_hist}
\end{figure}

It is well known that the flux ratio of different PAH emission bands
is a strong function of PAH size and their ionization state
\citep{Draine01}. The 6.2\,$\mu$m PAH emission is due to C-C
stretching mode and the 11.2\,$\mu$m feature is produced by C-H
out-of-plane bending mode \citep{Draine03}. In Figure
\ref{fig:pah_hist}, we display a histogram of the 11.2\,$\mu$m to
6.2\,$\mu$m PAH flux ratios for the 12\,$\mu$m Seyferts. Given the
relatively small number of starburst galaxies in the \citet{Brandl06}
sample (16 sources), we also included 20 HII galaxies from the SINGS
sample of \citet{Smith07a}, thus increasing the number of SF galaxies
to 36 sources and making its comparison with the Seyferts more
statistically meaningful. However, since \citet{Smith07a} adopted
multiple Drude profile fitting for the measurement of PAH features,
for reasons of consistency, we re-measured the PAH fluxes of the 20
SINGS galaxies with our spline fitting method. From Figure
\ref{fig:pah_hist}, we can see that both the Seyferts and SF galaxies,
indicated by the solid and dashed line respectively, appear to have
very similar distribution of PAH 11.2$\mu$m/6.2$\mu$m band ratios. The
average PAH flux ratios for the Seyfert sample is 0.94$\pm$0.37 while
for the SF galaxies is 0.87$\pm$0.24, which agree within
1-$\sigma$. This is also consistent with the findings of
\citet{Shi07}, who reported similar 11.2$\mu$m/7.7$\mu$m flux ratios
between a sample of higher redshift AGNs (PG, 2MASS and 3CR objects)
and the SINGS HII galaxies. This implies that even though the harsh
radiation field in AGNs may destroy a substantial amount of the
circumnuclear PAH molecules, and does so preferentially -- the smaller
PAHs being destroyed first \citep{Draine01,Smith07a} -- it likely does
not do so over a large volume. Enough molecules in the circumnuclear
regions do remain intact and as a result, the aromatic features that
we observe from Seyferts are essentially identical to those in SF
galaxies.

The relative strength of PAH emission can also be used to examine the
validity of the unified AGN model. As mentioned earlier, this model
attributes the variation in AGN types as the result of dust
obscuration and relative orientation of the line of sight to the
nucleus \citep{Antonucci93}. Sy 1s and Sy 2s are intrinsically the
same but appear to be different in the optical, mainly because of the
much larger extinction towards the nuclear continuum of Sy 2s when
viewed edge on.  The latest analysis of the IRS high-resolution
spectra of 87 galaxies from the 12\,$\mu$m Seyfert sample shows that
the average 11.2\,$\mu$m PAH EW is 0.29$\pm$0.38\,$\mu$m for Sy 1s and
0.37$\pm$0.35\,$\mu$m for Sy 2s (Tommasin et al. 2009, in
preparation).  As we show in Table \ref{tab3}, the 11.2\,$\mu$m PAH
EWs of the whole 12$\mu$m Seyfert sample (90 objects excluding upper
limits) is 0.21$\pm$0.22\,$\mu$m for the Sy 1s and 0.38$\pm$0.30$\mu$m
for the Sy 2s.  The difference we observe in the PAH EWs between the
two Seyfert types is somewhat larger than the one reported by Tommasin
et al. (2009, in preparation), though still consistent within
1$\sigma$. This further suggests that there is little discernible
difference between Sy 1s and Sy 2s at this wavelength.

If the observed AGN emission in the infrared does not depend on the
line of sight of the observer, one can compare the circumnuclear PAH
emission of Sy 1s and Sy 2s of similar bolometric luminosities to test
the unified model. If we bin the sources according to their IR
luminosity, we find an average PAH EW of 0.22$\pm$0.23\,$\mu$m for Sy
1s and 0.40\,$\pm$0.30\,$\mu$m for Sy 2s with L$_{\rm
  IR}<$10$^{11}$L$_\odot$; while the average PAH EWs are
0.19$\pm$0.19\,$\mu$m for Sy 1s and 0.37$\pm$0.30\,$\mu$m for Sy 2s
with L$_{\rm IR}\ge$10$^{11}$L$_\odot$ (see Table 3 for a summary of
these results). The difference between the two Seyfert types is still
less than 1$\sigma$. This result is not in agreement with the findings
of Buchanan et al. (2006) who based on a principle component analysis
find that in their subset of 51 galaxies the Sy 2s show a stronger
starburst eigenvector/template contribution than the Sy 1s. As the
authors suggest this might be due to a bias of selection effects. Our
result can be interpreted as indicating that there is some, but not
substantial obscuration in the mid-IR. As a consequence we are able to
probe deep into the nuclear region, sampling most of the volume
responsible for the mid-IR emission. This result is consistent with a
similar finding of \citet{Buchanan06} who compared the mid-IR to radio
ratio for their sample. They concluded that the observed factor of
$\sim$2 difference between the two types would imply either a smooth
torus which is optically thin in the mid-IR or a clumpy one containing
a steep radial distribution of optically thick dense clumps
\citep[][]{Nenkova08}.

\subsection{Cold/Warm AGN diagnostics}

\begin{figure}
  \epsscale{1.0}
  \plotone{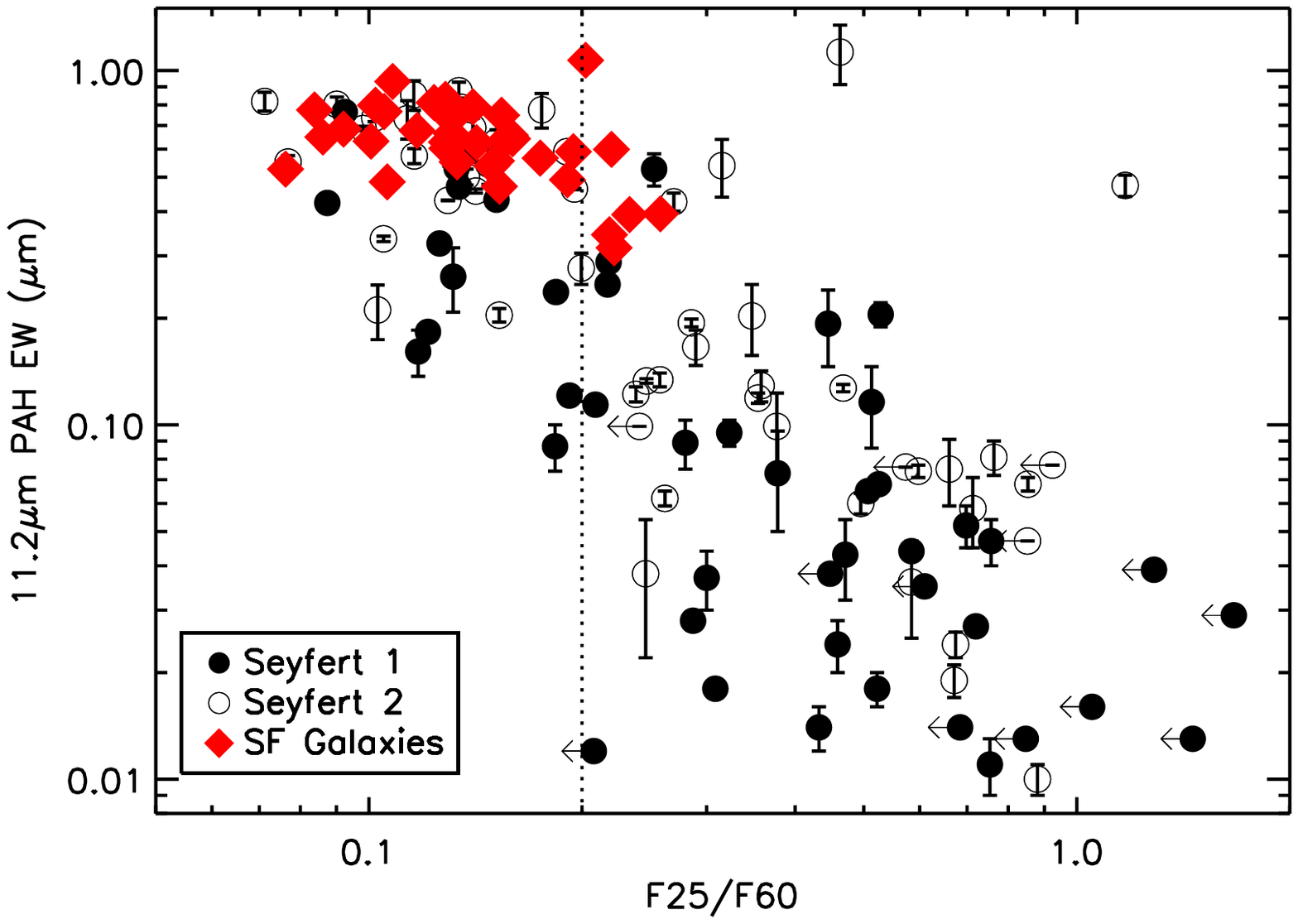}
  \caption{The IRAS 25 to 60$\mu$m flux ratio (F$_{25}$/F$_{60}$)as a
    function of the 11.2\,$\mu$m PAH EW for the 12\,$\mu$m Seyfert
    sample. The filled circles are Sy 1s, and the open circles are Sy
    2s. The diamonds represent the SF galaxies from \citet{Brandl06}
    and \citet{Smith07a}. The dotted line separates the warm and cold
    sources based on their IRAS colors. Note that the 11.2\,$\mu$m PAH
    EWs appear to anti-correlate with the dust temperature, as
    indicated by the ratio of F$_{25}$/F$_{60}$.}
  \label{fig:warm_cold}
\end{figure}

The {\em IRAS} 25 and 60\,$\mu$m flux ratio has long been used to
define the infrared color (``warm'' or ``cold'') of a galaxy, with
``warm'' galaxies having a ratio of F$_{25}$/F$_{60}>$0.2
\citep{Sanders88}. In Figure \ref{fig:warm_cold}, we plot the
11.2\,$\mu$m PAH EW as a function of the flux ratio between F$_{25}$
and F$_{60}$ for the 12\,$\mu$m Seyfert sample. The {\em IRAS} 25 and
60\,$\mu$m fluxes were compiled from \citet{Rush93} and
\citet{Sanders03} and are listed in Table \ref{tab1}. The aperture of
the {\em IRAS} broad band filters is on the order of a few arcminutes
on the sky\footnote{According to the IRAS Explanatory Supplement
  Document for unenhanced coadded IRAS images the resolution is
  approximately 1'$\times$ 5', 1'$\times$ 5', 2'$\times$ 5' and
  4'$\times$ 5' at 12, 25, 60 and 100\,$\mu$m, respectively ( see http://lambda.gsfc.nasa.gov/product/iras/docs/exp.sup ).}
and typically encompass the whole galaxy, while the PAH EW is measured
from a spectrum of a smaller region centered on the nucleus of the
galaxy (see Table \ref{tab2}). Nevertheless, we observe a clear trend
of the 11.2\,$\mu$m PAH EW decreasing with F$_{25}$/F$_{60}$ ratio in
Figure \ref{fig:warm_cold}. On this plot, we also include the SF
galaxies from \citet{Brandl06} and \citet{Smith07a}. All SF galaxies
appear to be clustered on the top left corner of the plot, having high
PAH EWs and low F$_{25}$/F$_{60}$ values, suggesting strong star
formation and cooler dust temperatures. The observed suppression of
PAH emission seen in the warm AGNs implies that the soft X-ray and UV
radiation of the accretion disk, which destroys the PAH molecules, is
also reprocessed by the dust and dominates the mid- and far-IR
colors. More specifically, warm Sy 1s have an average 11.2$\mu$m PAH
EW of 0.10$\pm$0.12$\mu$m, while for Sy 2s the value is
0.18$\pm$0.24$\mu$m. Similarly for the cold sources, the average PAH
EW is 0.40$\pm$0.23$\mu$m for Sy 1s and 0.59$\pm$0.19$\mu$m for Sy 2s.
We observe a $\sim$3$\sigma$ difference in the PAH EWs between the
cold and warm sources, independent of their Seyfert type. This
indicates that as the emission from the accretion disk surrounding the
SMBH of the active nucleus contributes progressively more to the IR
luminosity, its radiation field also destroys more of the PAH
molecules and thus diminishes their mid-IR emission.

The trend of PAH EWs decreasing with F$_{25}$/F$_{60}$ has also been
detected in a large sample of ULIRGs studied by \citet{Desai07}.  The
luminosities of the 12\,$\mu$m Seyfert sample are more comparable with
LIRGs, thus our work extends the correlation of \citet{Desai07} to a
lower luminosity range. This is rather interesting since deep
photometric surveys with {\em Spitzer} can now probe normal galaxies
as well as LIRGs at z$\sim$1 \citep{Lefloch05}, a fraction of
which are known to host an AGN based on optical spectra and mid-IR
colors \citep{Fadda02, Stern05, Brand06}.

We have also investigated the dependence of the 6.2\,$\mu$m PAH EW on
the F$_{25}$/F$_{60}$ ratio for our sample. A similar trend of the
6.2\,$\mu$m EWs decreasing with F$_{25}$/F$_{60}$ is observed as well,
though there is larger scatter than the one seen with the 11.2\,$\mu$m
feature. This is probably due to the fact that the 6.2\,$\mu$m PAH EW
is intrinsically fainter and only upper limits could be measured
for a number of source (see Table \ref{tab2}) .  Despite the scatter,
this trend is still a rather important finding, because for high
redshift galaxies (z$>$2.5), the 6.2 and 7.7$\mu$m PAHs might be the
only features available in the wavelength range covered by the IRS,
thus measuring them can reveal essential information on the
star-formation luminosity and dust composition of high-redshift
galaxies \citep[see][]{Houck05,Yan05,Teplitz07,Weedman08}.

\subsection{The Silicate Strength of the 12\,$\mu$m Seyferts}

In the mid-IR regime, one can examine not only the structure of
complex organic molecules and determine their aromatic or aliphatic
nature, but can also probe the chemistry of dust grains
\citep[see][]{vanDishoeck04}. One of the most prominent continuum
features in the 5-35$\mu$m range is the one associated with the
presence of astronomical silicates in the grains, which are
characterized by two peaks in the emissivity at 9.7 and 18$\mu$m
\citep[see][]{Dudley97}.  The silicate features had been detected in
absorption in SF galaxies, protostars and AGN for over 30 years
\citep[ie][]{Gillett75, Rieke75}, but it was the advent of space
observatories such as {\em ISO} and {\em Spitzer}, which allowed for
the first time of study over a wide range of astronomical objects and
physical conditions. According to the unified model, an edge-on view
through the cool dust in the torus will cause the silicate feature to
be seen in absorption, while with a face-on view, the hot dust at the
inner surface of the torus will cause an emission feature for the
silicate \citep{Efstathiou95}. Even though silicate in emission at
9.7$\mu$m had already been observed in the SF region N66
\citep{Contursi00}, emission at both 9.7\,$\mu$m and 18\,$\mu$m in
AGNs and Quasars was detected with {\em Spitzer}
\citep{Siebenmorgen05,Hao05,Sturm05}, providing strong support for the
unified model \citep{Antonucci93}.  Using Spitzer/IRS data,
\citet{Hao07} compiled a large, though inhomogeneous sample of AGNs
and ULIRGs, and uniformly studied the silicate features in these
galaxies. Using the same sample, \citet{Spoon07} proposed a new
diagnostic of mid-IR galaxy classification based on the strength of
silicate and PAH features. To put in the same context the properties
of the silicate feature in the 12\,$\mu$m Seyfert sample, we also
measured the strength of the silicate at 9.7\,$\mu$m, using the
definition and approach of \citet{Spoon07} as:

\begin{equation}
  S_{\mathrm{sil}}={\mathrm{ln}} \frac{f_{\mathrm{obs}}(9.7\mu m)}{f_{\mathrm{cont}}(9.7\mu m)}
\end{equation}
\\

where f$_{\mathrm{cont}}$(9.7\,$\mu$m) is the flux density of a local
mid-IR continuum, defined from the 5-35\,$\mu$m IRS spectrum. Sources
with silicate in emission have positive strength and those in
absorption negative. \citet{Buchanan06} did identify two AGN with
silicate emission and two more with broad silicate absorption out of
the 51 sources they studied. In this paper, we provide for the first
time measurements on the silicate features for this complete
sample. We follow the prescription of \citet{Spoon07} and
\citet{Sirocky08} for the continuum definition and identify the
sources to be PAH-dominated, continuum-dominated and
absorption-dominated. The values of S$_{\rm sil}$ measured from the
12\,$\mu$m Seyferts can be found in Table \ref{tab2}.

\begin{figure}
  \epsscale{1.8}
  \plottwo{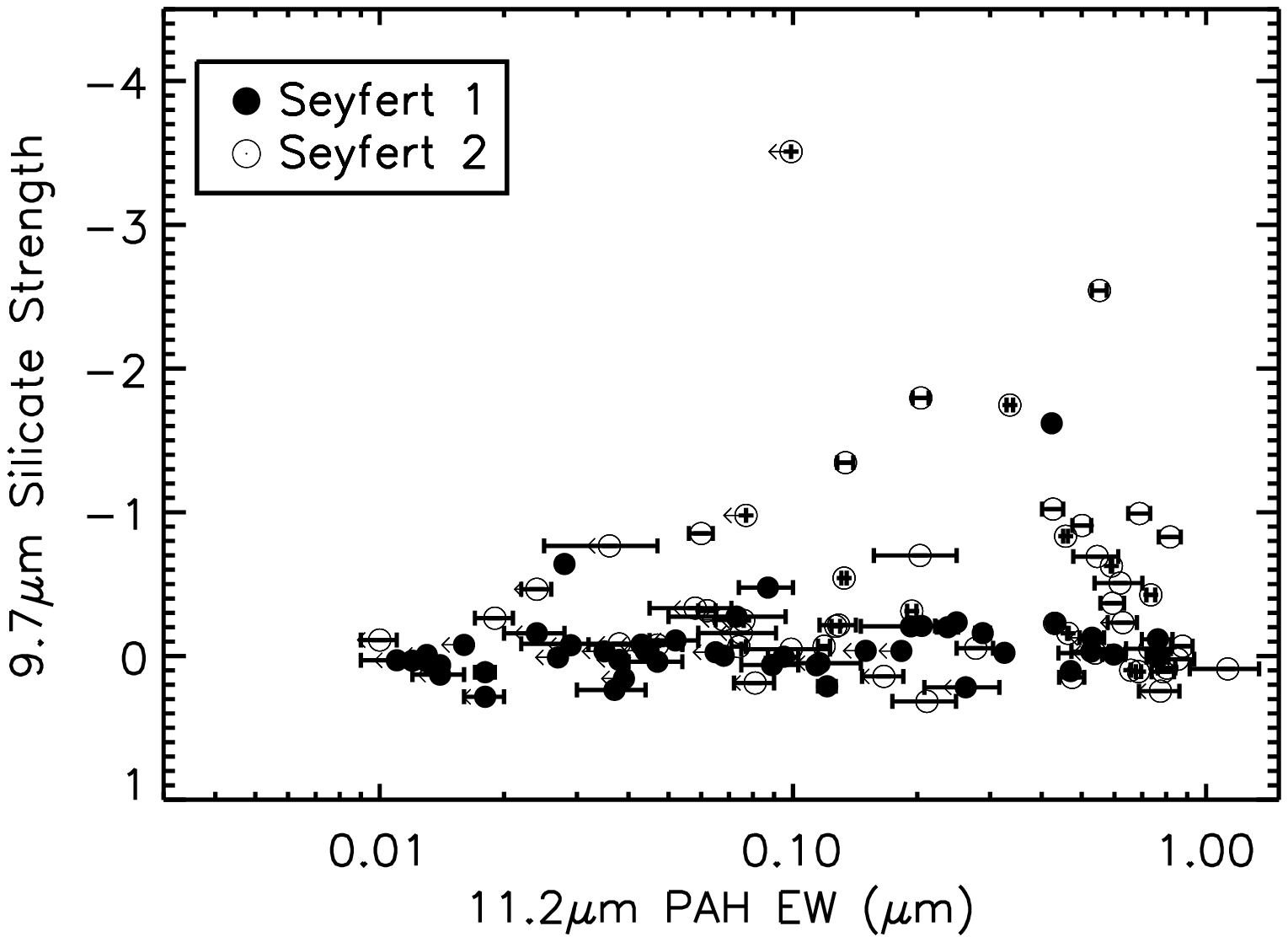}{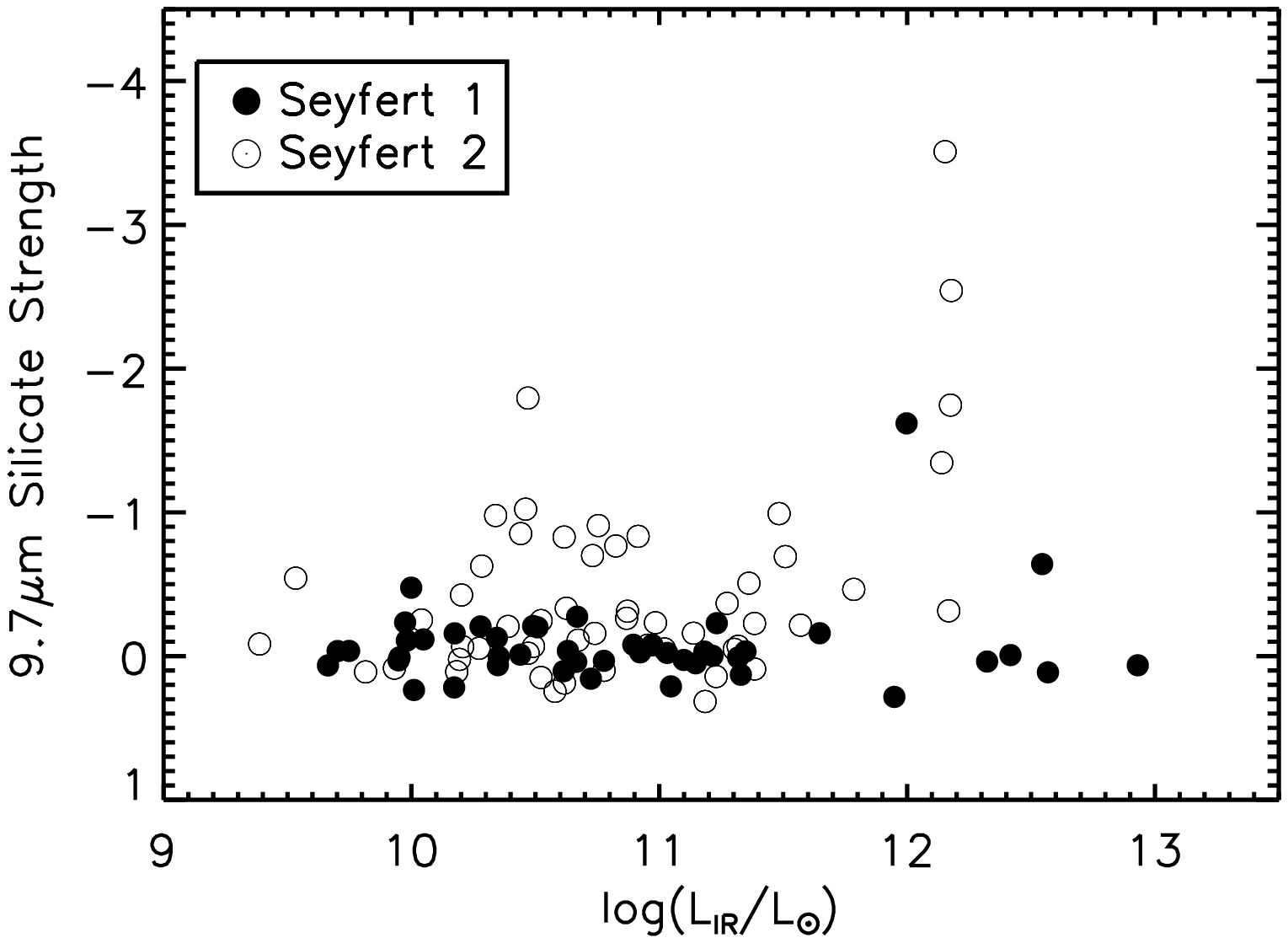}
  \caption{a) Top panel: Plot of the silicate strength at 9.7\,$\mu$m
    as a function of the PAH 11.2\,$\mu$m EW for the 12$\mu$m Seyfert
    sample. Upper limits are indicated with arrows. b) Bottom panel:
    The IR luminosity vs the 9.7\,$\mu$m silicate strength.  The
    symbols are defined in the same way as Figure 11. Note that Sy 2s
    display a larger range in possible values of Silicate strength
    than Sy 1s.}
  \label{fig:sil}
\end{figure}

In Figure \ref{fig:sil}a, we plot the 11.2\,$\mu$m PAH EWs as a
function of the 9.7\,$\mu$m silicate strength\footnote{A similar plot
  using the 6.2\,$\mu$m PAH EWs was proposed by \citet{Spoon07} as a
  mid-IR galaxy classification method}. We observe that most Sy 1s are
located close to S$_{\rm sil}$=0 and the range of their silicate
strength is rather narrow, with the exception of one galaxy, UGC5101,
which is also one of the ULIRGs of the Bright Galaxy Sample
\citep{Armus04,Armus07}. This is in agreement with the results of
\citet{Hao07}, who found that Sy 1s are equally likely to display the
9.7\,$\mu$m silicates feature in emission and in absorption. The Sy 2s
have a larger scatter in the value of the silicate strength, with most
of them showing the feature in absorption. The average silicate
strength of the 12\,$\mu$m Seyfert sample is -0.07$\pm$0.29 for Sy 1s
and -0.46$\pm$0.73 for Sy 2s, while the median values are -0.02 for Sy
1s and -0.23 for Sy 2s. Overall, though selected at 12\,$\mu$m, most
objects have a rather weak silicate strength, with only 18 Sy 2s and 2
Sy1s displaying values of S$_{\rm sil}<$-0.5. We also examined the
dependence of the silicate strength with the IR luminosity of the
objects and plot it in Figure \ref{fig:sil}b. Except for one galaxy
(NGC7172), all the sources with deep silicate absorption features have
IR luminosities larger than 10$^{12}$L$_{\odot}$, and thus are also
classified as ULIRGs. In Sy 1 galaxies, even when their luminosities
are larger than 10$^{12}$L$_\odot$, the 9.7\,$\mu$m silicate strength
is still near zero, while high-luminosity Sy 2s are more likely to
have deep silicate absorption features.

\begin{figure}
  \epsscale{1.8}
  \plottwo{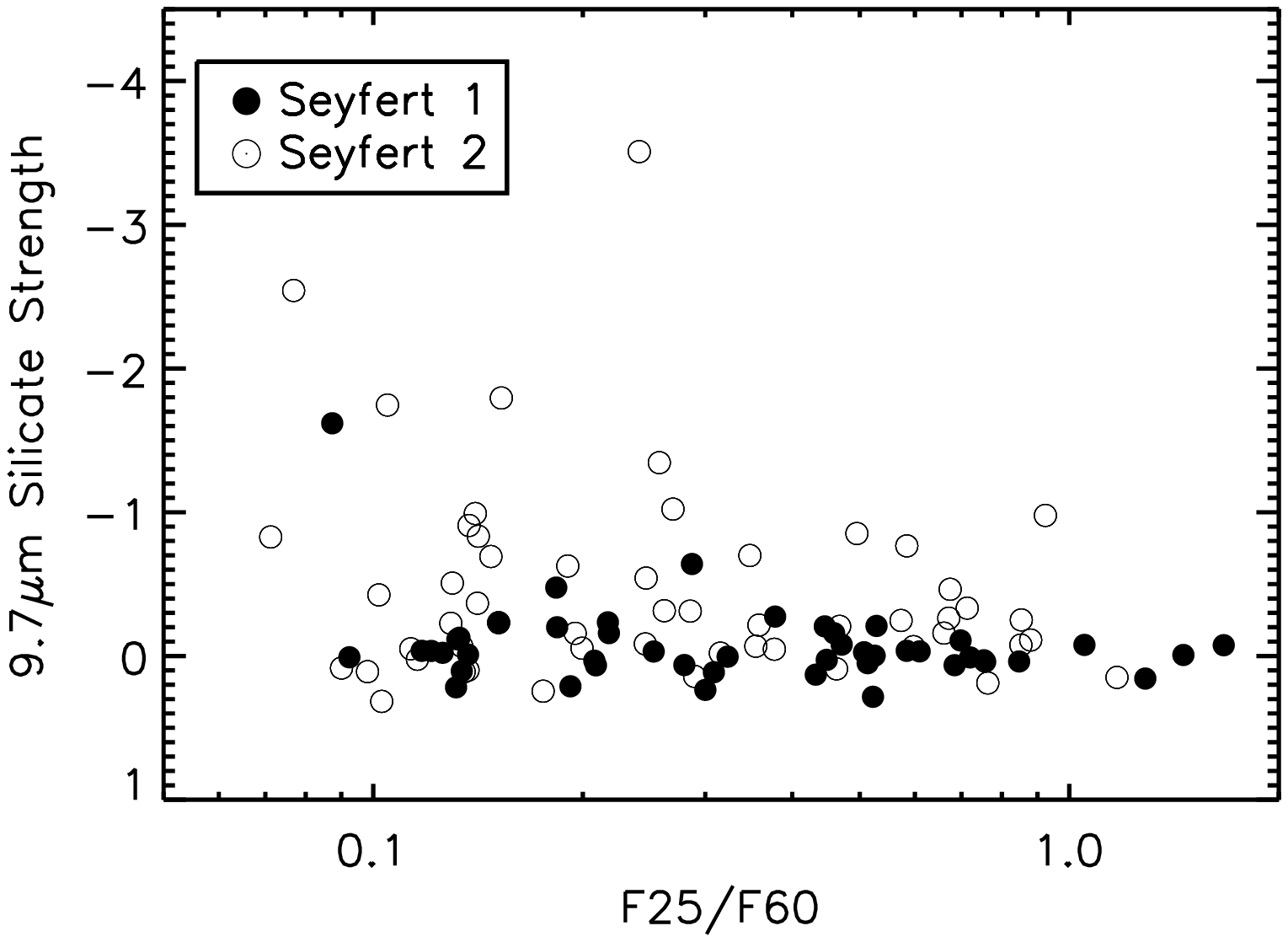}{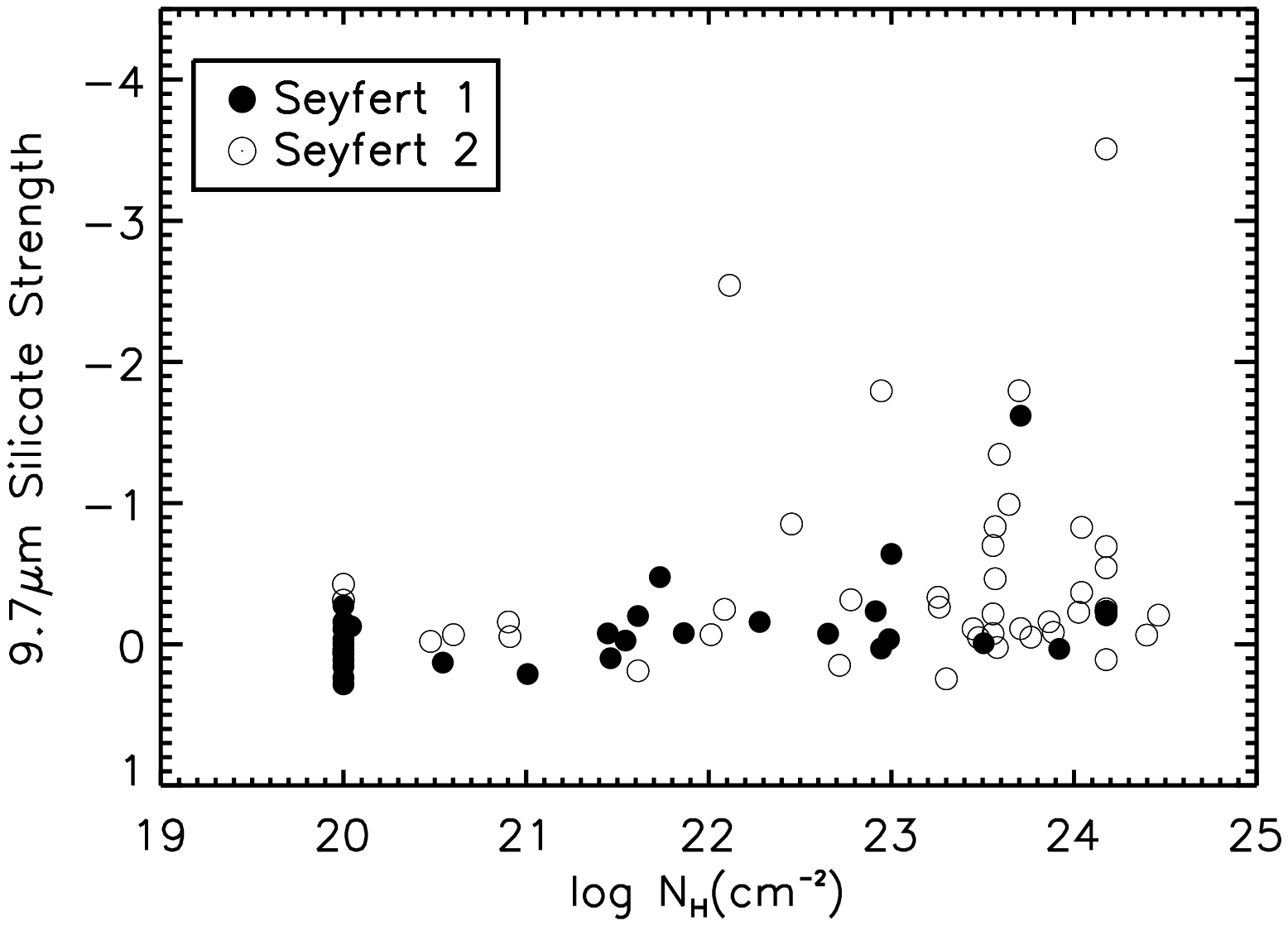}
  \caption{a) Top panel: The silicate strength at 9.7\,$\mu$m as a
    function of the galaxy color as indicated by the ratio of F25/F60.
    b) Bottom panel: The 9.7\,$\mu$m silicate strength versus hydrogen
    column density, as measured from the X-rays. The symbols are
    defined in the same way as Figure 11.}
  \label{fig:S_color_nh}
\end{figure}

We also compare the silicate strength with galaxy color, as defined in
\S 3.3. In Figure \ref{fig:S_color_nh}a, we plot the silicate strength
as a function of the IRAS flux ratio of F25/F60. Since the majority of the
galaxies do not have strong silicate features, no clear correlation
between the two parameters is seen. We notice that galaxies with
S$_{\rm sil}<-1$ also appear to have colder colors. This indicates
that more dust absorption is present in sources with the colder IR
SEDs, even though many sources with small F25/F60 ratios do not display
any silicate absorption features.

Finally, we investigate the relation between the mid-IR silicate
strength and the hydrogen column density\footnote{The values of the
  hydrogen column density have been taken from \citet{Markwardt05,
    Bassani06, Sazonov07, Shu07}.}, as measured from the X-rays.  The
latter can measure directly the absorption in active galaxies: the
power law spectrum in the 2-10keV range may be affected by a cutoff
due to photo-electric absorption, from which column densities of
$10^{22} - 10^{24} cm^{-2}$ are derived \citep[eg.][]{Maiolino98}. One
should note though that due to the substantially smaller physical size
of the nuclear region emitting in (hard) X-rays this measurement is
more affected by the clumpiness of the intervening absorbing
medium. The well established observational fact that Sy 2s are
preferentially more obscured than Sy 1s, as has been shown both in
the optical and in the X-ray spectra, is also apparent in the mid-IR
spectra by our results given in Figure \ref{fig:S_color_nh}b.  We find
that sources with weak silicate absorption or emission features span
over all values of the column densities. However, most of the sources
with a strong silicate absorption (S$_{\rm sil}<$ -0.5) have a $N_H >
10^{23}$ (cm$^{-2}$) (11 of 15 sources), and Sy 2s dominate this group
(13 of 15 sources).

\section{What powers the 12\,$\mu$m luminosity in the 12$\mu$m Seyferts?}

\begin{figure}
  \epsscale{1.0}
  \plotone{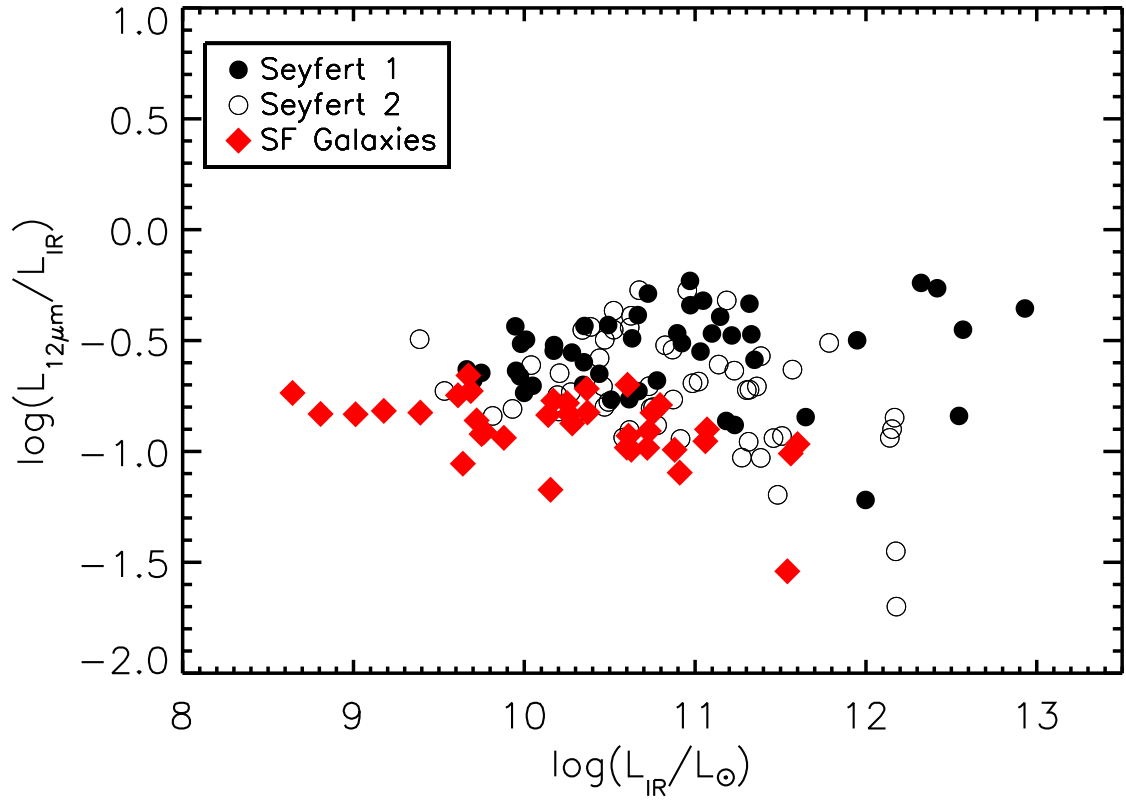}
  \caption{The ratio of L$_{\rm 12\mu m}$/L$_{\rm IR}$ versus the
    total infrared luminosity of the 12\,$\mu$m Seyfert sample and SF
    galaxies. For SF galaxies, their L$_{\rm 12\mu m}$ appear to
    account for a nearly constant fraction of the total infrared
    luminosity. For Seyfert galaxies, their L$_{\rm 12\mu m}$/L$_{\rm
      IR}$ ratios are higher than SF galaxies, and the scatter is also
    larger. The symbols are defined in the same way as Figure 11.}
  \label{fig:L12}
\end{figure}

The use of the global infrared dust emission as a tracer of the
absorbed starlight and associated star formation rate has been known
since the first results of IRAS (see Kennicutt 1998 and references
there in).  At 12\,$\mu$m, the flux obtained from the {\em IRAS}
broadband filter is dominated by the continuum emission, though it
could also be affected by several other discrete factors, including
the silicate features, the PAH emission, fine-structure lines, etc.
In Figure \ref{fig:L12}, we present the usual plot of L$_{\rm 12\mu
  m}$/L$_{\rm IR}$ versus the total IR luminosity\footnote{Calculated
  from the {\em IRAS} flux densities following the prescription of
  \citet{Sanders96}: L$_{\rm
    IR}$=5.6$\times$10$^5$D$_{Mpc}^2$($13.48S_{12}+5.16S_{25}+2.58S_{60}+S_{100}$).}
for the Seyfert galaxies and SF galaxies. A clear correlation between
these two parameters, originally presented for the 12$\mu$m sample by
\citet{Spinoglio95}, is seen. The two Seyfert types do not show
significant differences. SF galaxies appear to have a lower fractional
12\,$\mu$m luminosity when compared to Seyferts of similar total IR
luminosity. This can be explained by the presence of hot dust emission
(T$>$300K) originating from regions near the active nucleus,
contributing more strongly at the shortest wavelengths (5 to
15\,$\mu$m) of the IR SED. This is consistent with the results of
\citet{Spinoglio95}, who have shown that the 12\,$\mu$m luminosity is
$\sim$15\% of the bolometric luminosity\footnote{According to
  \citet{Spinoglio95}, the bolometric luminosity is derived by
  combining the blue photometry, the near-IR and FIR luminosities, as
  well as an estimate of the flux contribution from cold dust longward
  of 100\,$\mu$m.} in AGNs \citep{Spinoglio89}, while only $\sim$7\%
in starburst and normal galaxies.

Following these early IRAS results, a number of studies have explored
the issue of distinguishing AGN from the star formation signatures in
the mid-IR and to determine the fractional contribution of each
component to the IR luminosity for local \citep[i.e.][]{Genzel98,
  Laurent00, Peeters04, Buchanan06, Farrah07, Nardini08} and high
redshift sources \citep[i.e.][]{Brand06, Weedman06,Sajina07}.  More
recently, using the [OIV]25.89$\mu$m line emission as an extinction
free tracer of the AGN power, \citet{Melendez08b} were able to
decompose the stellar and AGN contribution to the [NeII]12.81$\mu$m
line. These authors have compiled a sample from existing {\em Spitzer}
observations by \citet{Deo07,Tommasin08, Sturm02, Weedman05}, as well
as X-ray selected sources from \citet{Melendez08a}. They found that Sy
1 and Sy 2 galaxies are different in terms of the relative
AGN/starburst contribution to the infrared emission, with star
formation being responsible for $\sim$25\% of the mid- and far-IR
continuum in Sy 1s, and nearly half of what was estimated for Sy 2s.

\begin{figure}
  \epsscale{1.6}
  \plottwo{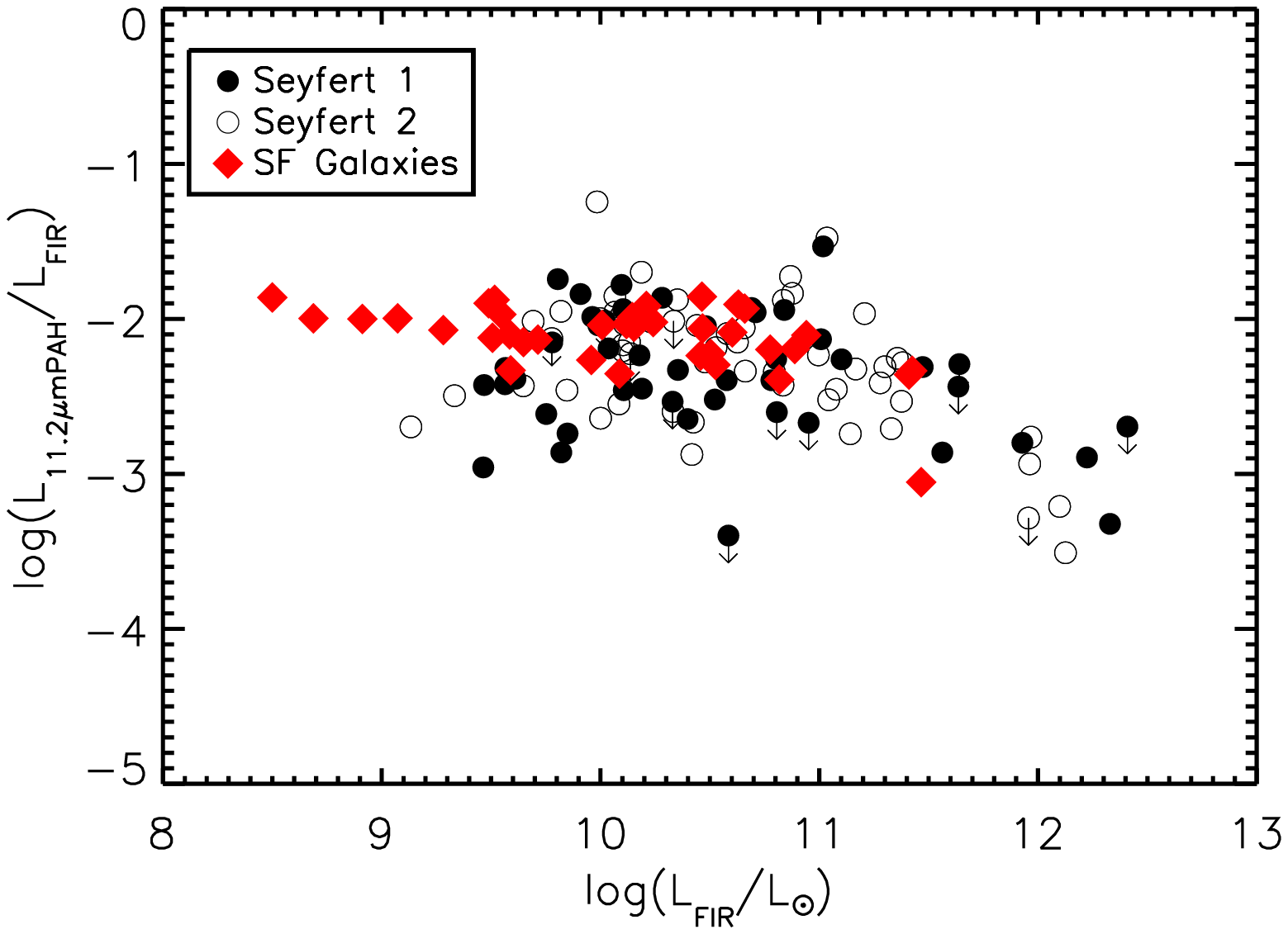}{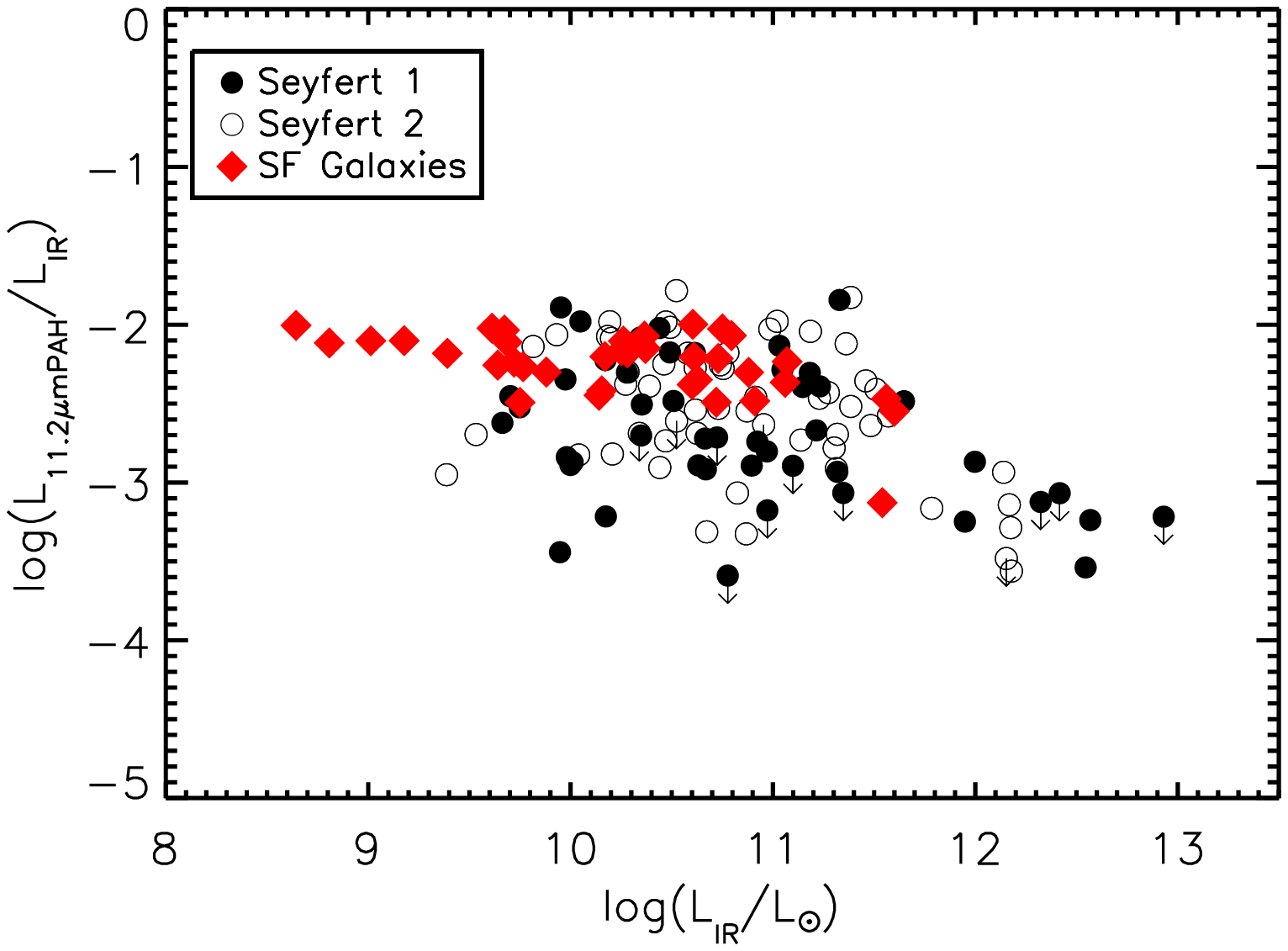}
  \caption{a) The L$_{\rm 11.2\mu m PAH}$/L$_{\rm FIR}$ versus the FIR
    luminosity of the 12\,$\mu$m Seyfert sample. b) Same as in a), but
    for the L$_{\rm 11.2\mu m PAH}$/L$_{\rm IR}$ versus the total IR
    luminosity. Both the SF galaxies and the Seyferts appear to have a
    constant ratio of L$_{\rm 11.2\mu m PAH}$/L$_{\rm FIR}$ and
    L$_{\rm 11.2\mu m PAH}$/L$_{\rm IR}$, though the Seyferts appear
    to have a lower fraction in L$_{\rm 11.2\mu m PAH}$/L$_{\rm
      IR}$. The symbols are defined in the same way as Figure 11.}
  \label{fig:L_pah}
\end{figure}

In Figure \ref{fig:L_pah}a and \ref{fig:L_pah}b, we plot L$_{\rm
  11.2\mu m PAH}$/L$_{\rm FIR}$ and L$_{\rm 11.2\mu m PAH}$/L$_{\rm
  IR}$ as a function of the far-infrared (FIR)
luminosity\footnote{Calculated from the {\em IRAS} flux densities
  following the prescription of \citet{Sanders96}: L$_{\rm
    FIR}$=5.6$\times$10$^5$D$_{Mpc}^2$($2.58S_{60}+S_{100}$)} and
the total IR luminosity for the Seyferts and SF galaxies. For both the SF
galaxies and the Seyferts, their PAH luminosity appear to have a nearly
constant fraction of their FIR luminosity, which would be expected
since both quantities have been used as indicators of the
star-formation rate. However, for a given PAH luminosity, Seyfert
galaxies display an excess in the total IR luminosity compared to
starburst systems. This is also understood as the total IR luminosity
is the sum of mid-IR and FIR luminosity and consequently it is
affected by the AGN emission in the mid-IR. We propose a simple method
to the mid-IR excess and estimate the AGN contribution to the
12\,$\mu$m luminosity.

\begin{figure}
  \epsscale{1.0}
  \plotone{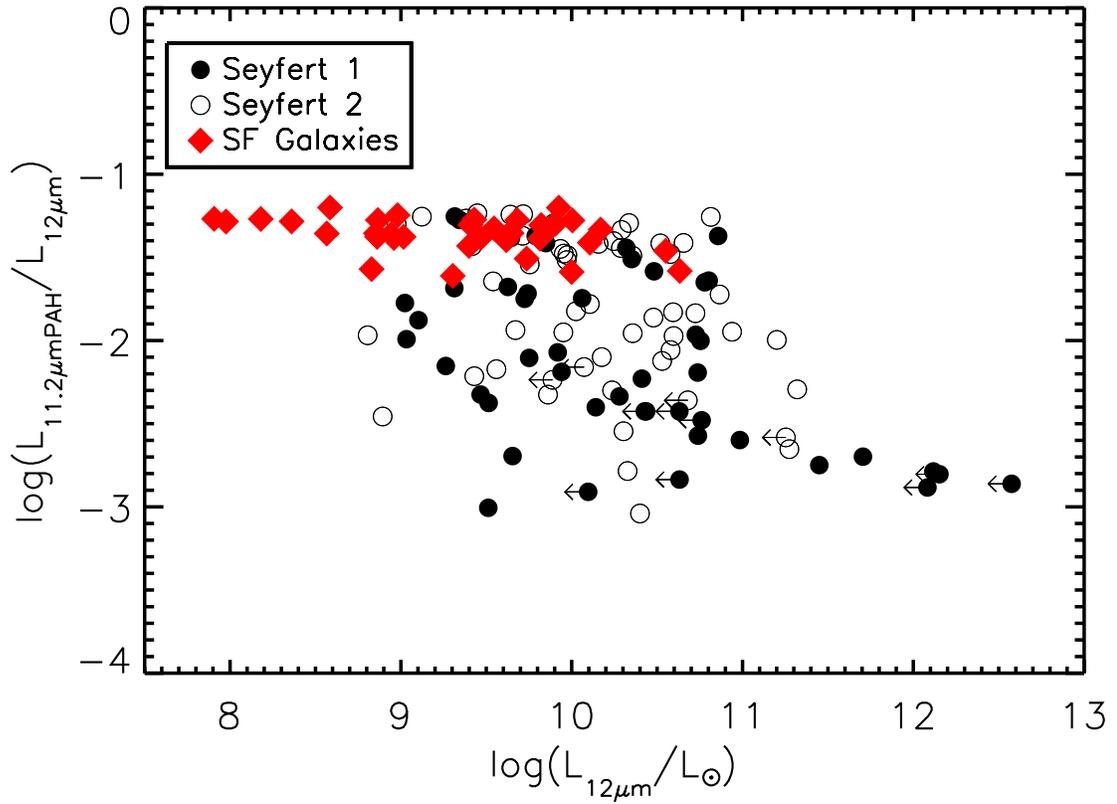}
  \caption{The L$_{\rm 11.2\mu m PAH}$/L$_{\rm 12\mu m}$ versus the
    IRAS 12\,$\mu$m luminosity of the 12\,$\mu$m Seyfert sample.  The
    SF galaxies have a nearly constant ratio of L$_{\rm 11.2\mu m
      PAH}$/L$_{\rm 12\mu m}$, while the Seyferts have more scatter.
    The symbols are defined in the same way as Figure 11.}
  \label{fig:L12_pah}
\end{figure}

In Figure \ref{fig:L12_pah}, we plot the L$_{\rm 11.2\mu m
  PAH}$/L$_{\rm 12\mu m}$ versus the 12\,$\mu$m luminosity for the
Seyfert and SF galaxies. There is a clear correlation for SF galaxies
with an average L(11.2\,$\mu$m PAH)/L(12\,$\mu$m) ratio of
0.044$\pm$0.010. Since there is no AGN contamination in the 12\,$\mu$m
luminosity for these galaxies, we can attribute all mid-IR continuum
emission to star formation. Seyfert galaxies display a larger scatter
on this plot, and we decompose their 12\,$\mu$m luminosity into two
parts: one contributed by the star formation activity, which is
proportional to their PAH luminosity, and one due to dust heated by
the AGN. If we assume that the star formation component in the
12\,$\mu$m luminosity of Seyferts is associated with the 11.2\,$\mu$m
PAH luminosity in the same manner as in SF galaxies, then we can
estimate the star formation contribution to the integrated 12\,$\mu$m
luminosity of the Seyfert sample. Subtracting this SF contribution
from the total 12\,$\mu$m luminosity, we can obtain, in a statistical
sense, the corresponding AGN contribution.

\begin{figure}
  \epsscale{1.0}
  \plotone{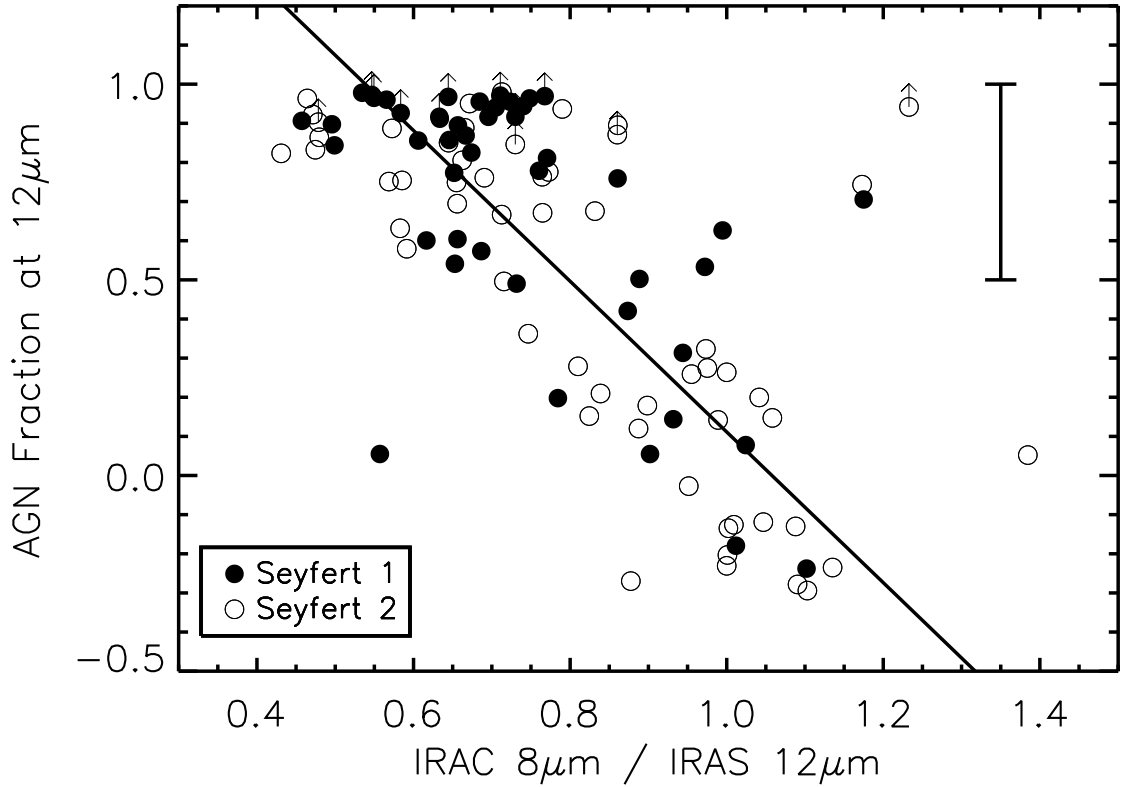}
  \caption{The AGN fraction (as defined in the text) at 12\,$\mu$m as
    a function of the IRAC 8\,$\mu$m to {\em IRAS} 12\,$\mu$m flux
    ratio. The Seyfert 1s are represented with filled circles and the
    Seyfert 2s are indicated with open circles. The solid line is a
    fit to the Seyfert galaxies excluding those with lower limits for
    the AGN fraction. The uncertainty on the ``AGN fraction'' is
    $\sim$25\%, as indicated at the top right corner of the plot.}
  \label{fig:agnfraction}
\end{figure}

To check the validity of this method, we plot in Figure
\ref{fig:agnfraction} the ``AGN fraction'' as a function of the IRAC
8\,$\mu$m to {\em IRAS} 12\,$\mu$m flux ratios. We define ``AGN
fraction'' as the AGN luminosity estimated using the above method
divided by the total 12\,$\mu$m luminosity: AGN fraction (12\,$\mu$m)
= (L$_{\rm 12\mu m}$-L$_{\rm SF}$)/L$_{\rm 12\mu m}$. The IRAC
8\,$\mu$m flux will be dominated by PAH emission when PAHs are
present, thus normalizing by the 12\,$\mu$m flux provides an estimate
of the PAH EW\footnote{A similar approach using Spitzer broad band
  filters was used successfully by \citet{Engelbracht08} to estimate
  the PAH contribution in starburst galaxies.}. As one would expect,
examining the Seyferts for which an AGN fraction was not a lower
limit, an anti-correlation between the two parameters is visible. This
suggests that our method of decomposing the 12\,$\mu$m luminosity is
reasonable. Since the scatter in the linear fit for SF galaxies in
Figure~\ref{fig:L12_pah} is $\sim$25\%, this translates directly to
the ``AGN fraction'' we have obtained. We estimate the uncertainty of
our calculated ``AGN fraction'' to be no better than $\sim$25\%.

\section{Conclusions}

We have analyzed Spitzer/IRS data for a complete unbiased sample of
Seyfert galaxies selected from the {\em IRAS} Faint Source Catalog
based on their 12\,$\mu$m fluxes. We extended earlier work on the same
sample by \citet{Buchanan06} who have published spectra for 51 objects
and explored the continuum shapes and the differences between Seyfert
types. In our study, we present 5--35\,$\mu$m low-resolution spectra
for 103 objects, nearly 90\% of the whole 12\,$\mu$m Seyfert
sample. The main results of our study are:

1. The 12\,$\mu$m Seyferts display a variety of mid-IR spectral
shapes. The mid-IR continuum slopes of Sy 1s and Sy 2s are on average
$<\alpha_{15-30}>$=-0.85$\pm$0.61 and -1.53$\pm$0.84 respectively,
though there is substantial scatter for both types. We identify a
group of objects with a local maximum in their mid-IR continuum at
$\sim$20\,$\mu$m, which is likely due to the presence of a warm
$\sim$150 K dust component and 18\,$\mu$m emission from astronomical
silicates. Emission lines, such as the
[NeV]\,14.3\,$\mu$m/24.3\,$\mu$m and [OIV]\,25.9\,$\mu$m lines, known
to be a signature of an AGN are stronger in the average spectra of Sy
1s than those of Sy 2s.

2. PAH emission is detected in both Sy 1s and Sy 2s, with no
statistical difference in the relative strength of PAHs between the
two types. This suggests that the volume responsible for the bulk of
their emission is likely optically thin at $\sim$12\,$\mu$m.

3. The 11.2\,$\mu$m PAH EW of the 12\,$\mu$m Seyfert sample correlates
well with the IRAS color of the galaxies as indicated by the flux
ratio of F$_{25}$/F$_{60}$. PAH emission is more suppressed in warmer
galaxies, in which the strong AGN activity may destroy the PAH
molecules.

4. The 9.7\,$\mu$m silicate feature is rather weak in Sy 1s (S$_{\rm
  sil}$=-0.07$\pm$0.29) while Sy 2s mostly display silicate in
absorption (S$_{\rm sil}$=-0.46$\pm$0.73). Deep silicate absorption is
observed in high luminosity Sy 2s which are classified as ULIRGs, and
those with high hydrogen column density estimated from their X-ray
emission.

5. The FIR luminosities of the 12\,$\mu$m Seyferts are dominated by
star-formation. Their mid-IR luminosity increases by the additional
AGN contribution. A method to estimate the AGN contribution to the
12\,$\mu$m luminosity, in a statistical sense, has been proposed and
applied to the sample.

\acknowledgments 

We would like to acknowledge L. Armus, M. Malkan, Y. Shi and M. Elvis
for helpful science discussions. We thank J.D. Smith with help on the
use of CUBISM to extract spectral mapping data. We would also like to
thank H. Spoon and L. Hao for help in measuring the silicate features.
We thank an anonymous referee whose comments help to improve this
manuscript.  V.C. acknowledges partial support from the EU ToK grant
39965. L.S. and S.T. acknowledge support from the Italian Space Agency
(ASI).

\clearpage

\begin{figure}
  \epsscale{1.0}
  \plotone{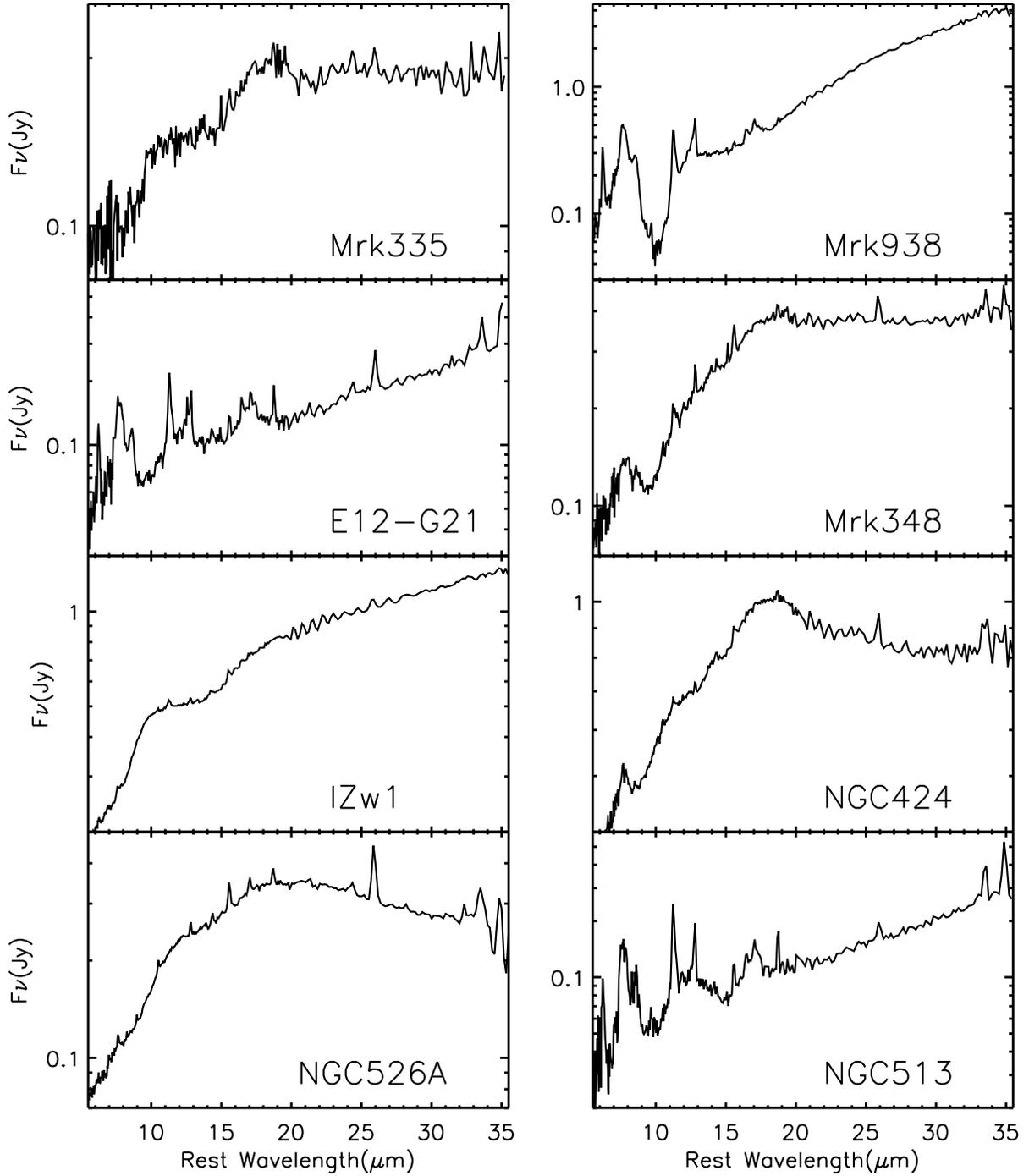}
  \caption{An Atlas of the Spitzer/IRS low-resolution 5-35$\mu$m
  spectra of the 12\,$\mu$m Seyfert sample.}
\end{figure}
\begin{figure}
  \setcounter{figure}{17}
  \epsscale{1.0}
  \plotone{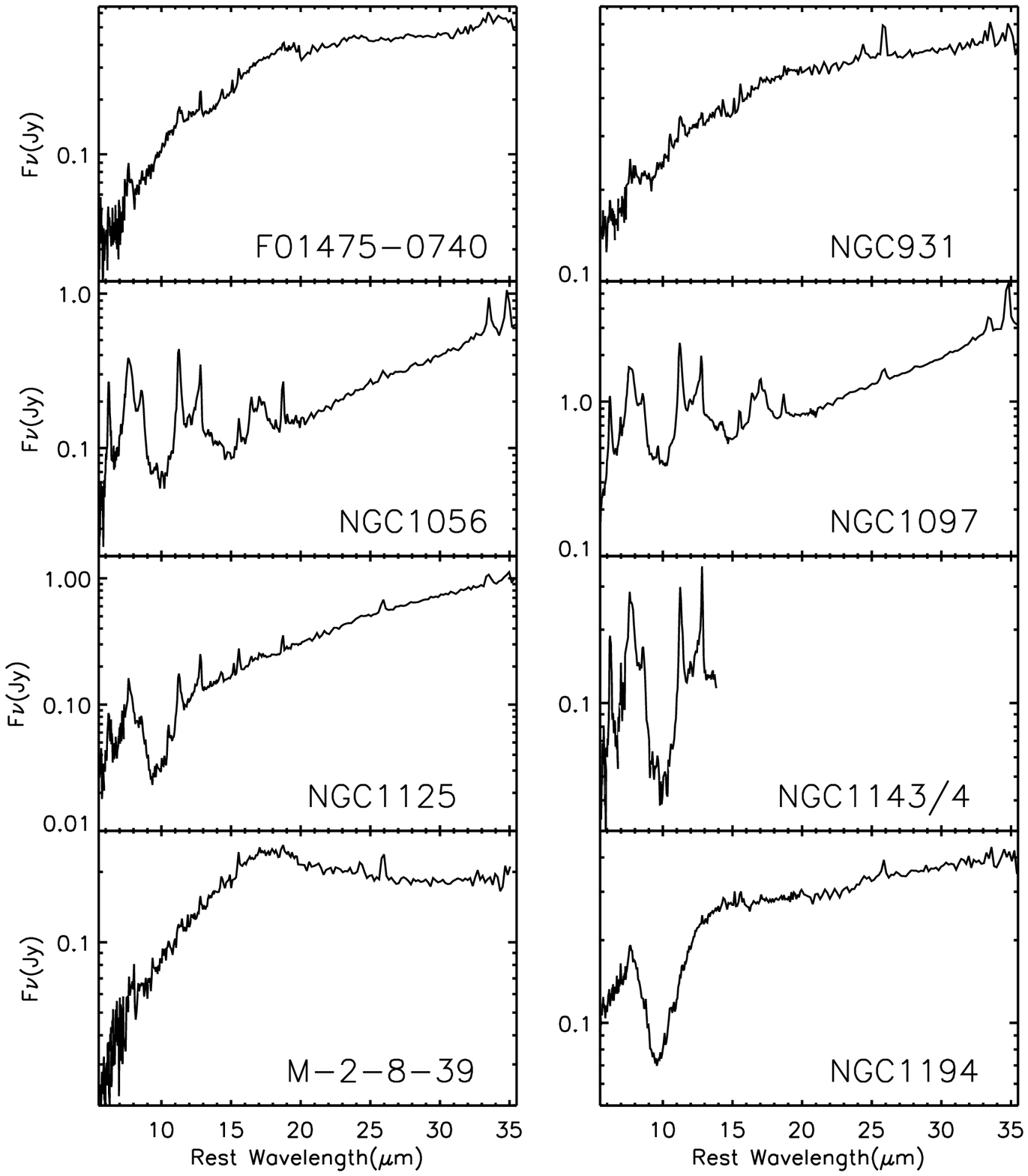}
  \caption{Continued.}
\end{figure}
\begin{figure}
  \setcounter{figure}{17}
  \epsscale{1.0}
  \plotone{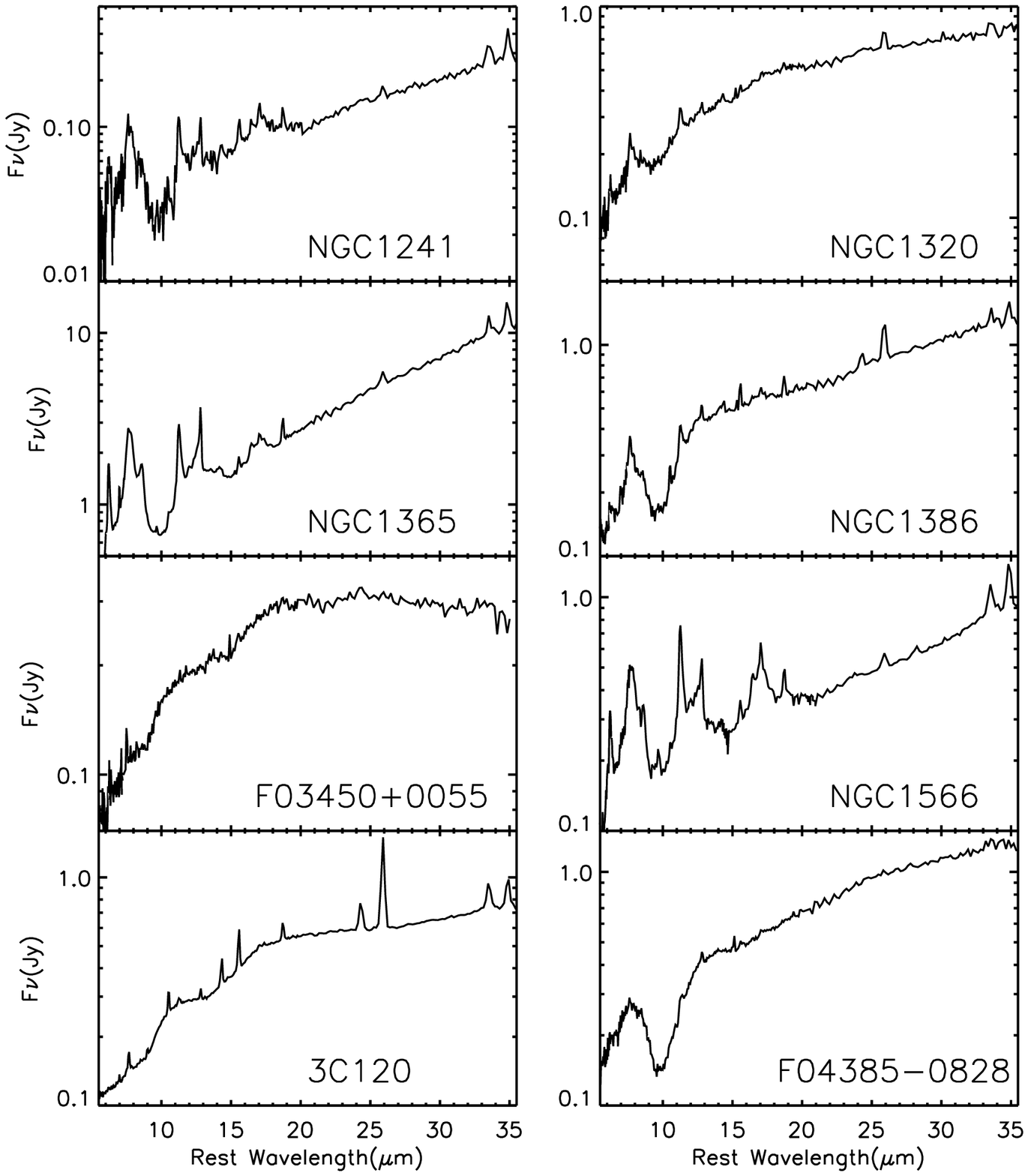}
  \caption{Continued.}
\end{figure}
\begin{figure}
  \setcounter{figure}{17}
  \epsscale{1.0}
  \plotone{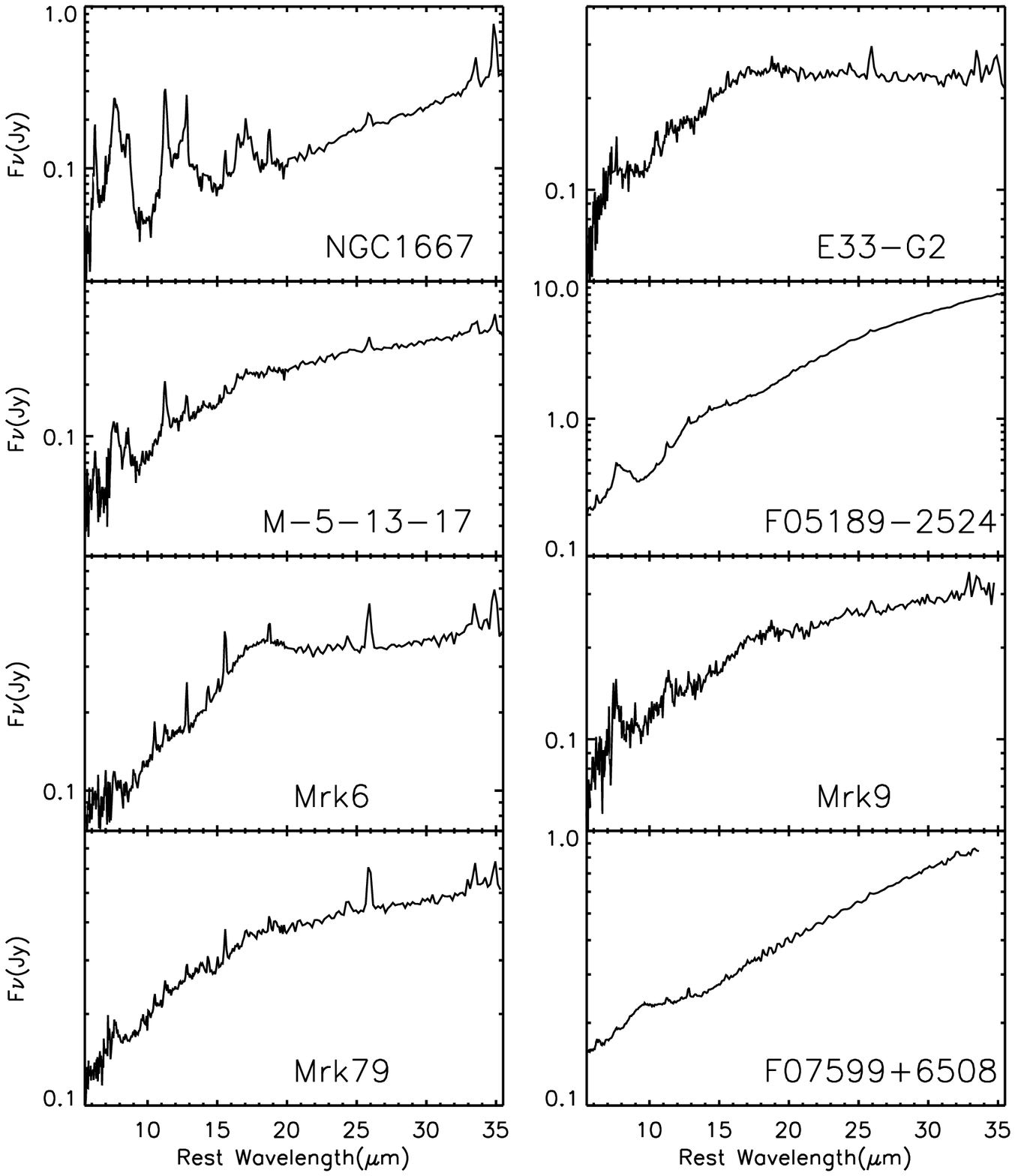}
  \caption{Continued.}
\end{figure}
\begin{figure}
  \setcounter{figure}{17}
  \epsscale{1.0}
  \plotone{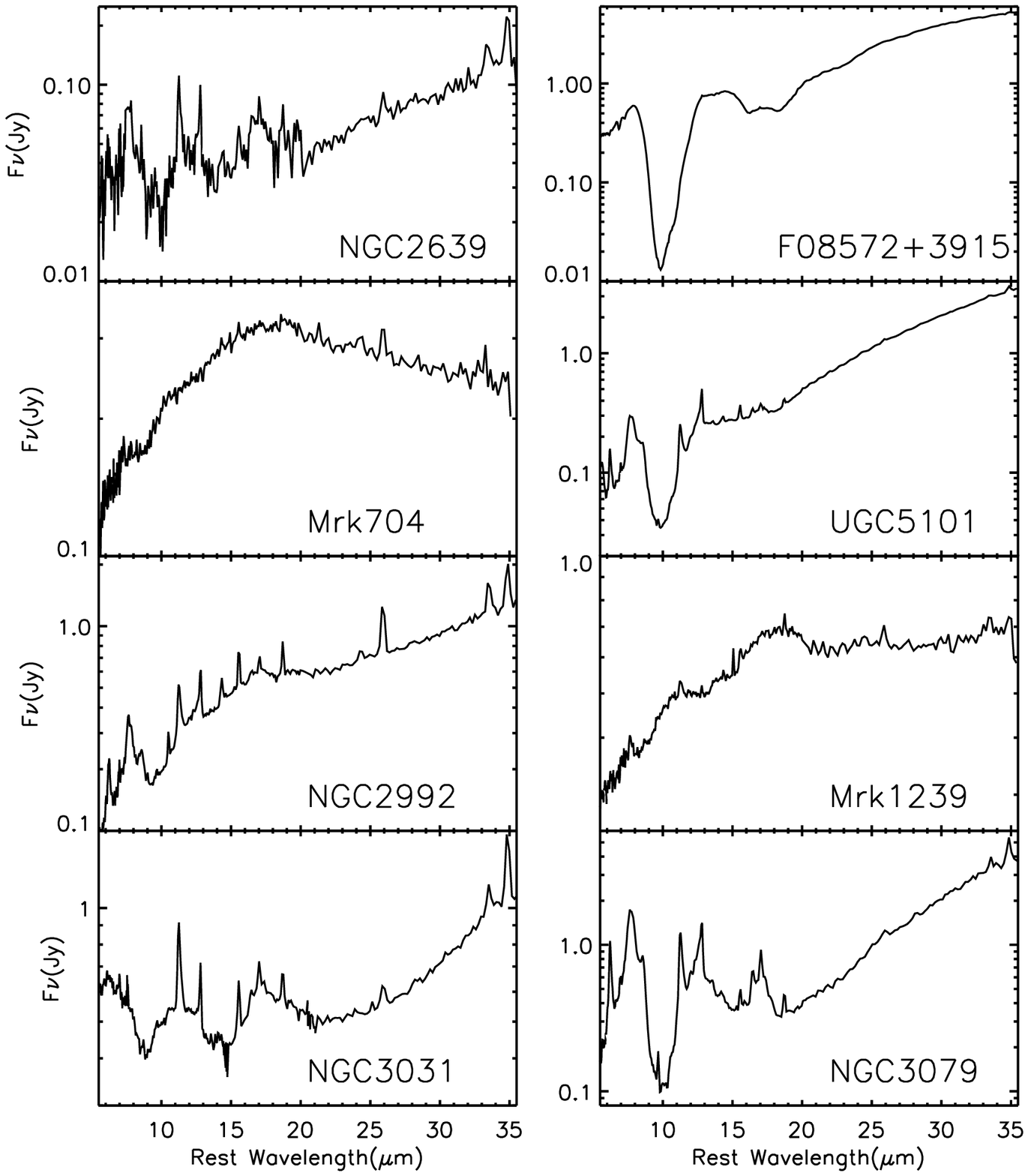}
  \caption{Continued.}
\end{figure}
\begin{figure}
  \setcounter{figure}{17}
  \epsscale{1.0}
  \plotone{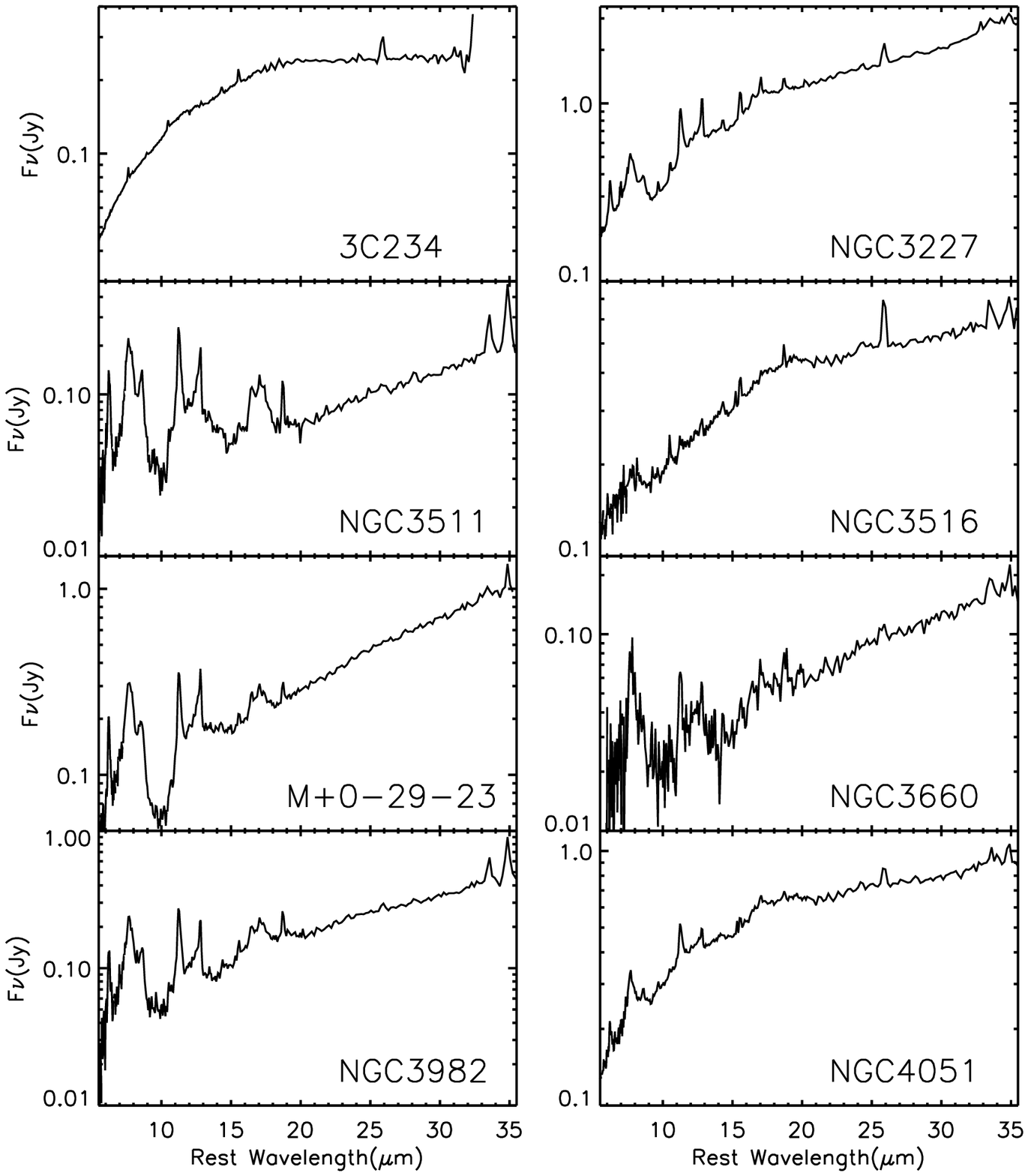}
  \caption{Continued.}
\end{figure}
\begin{figure}
  \setcounter{figure}{17}
  \epsscale{1.0}
  \plotone{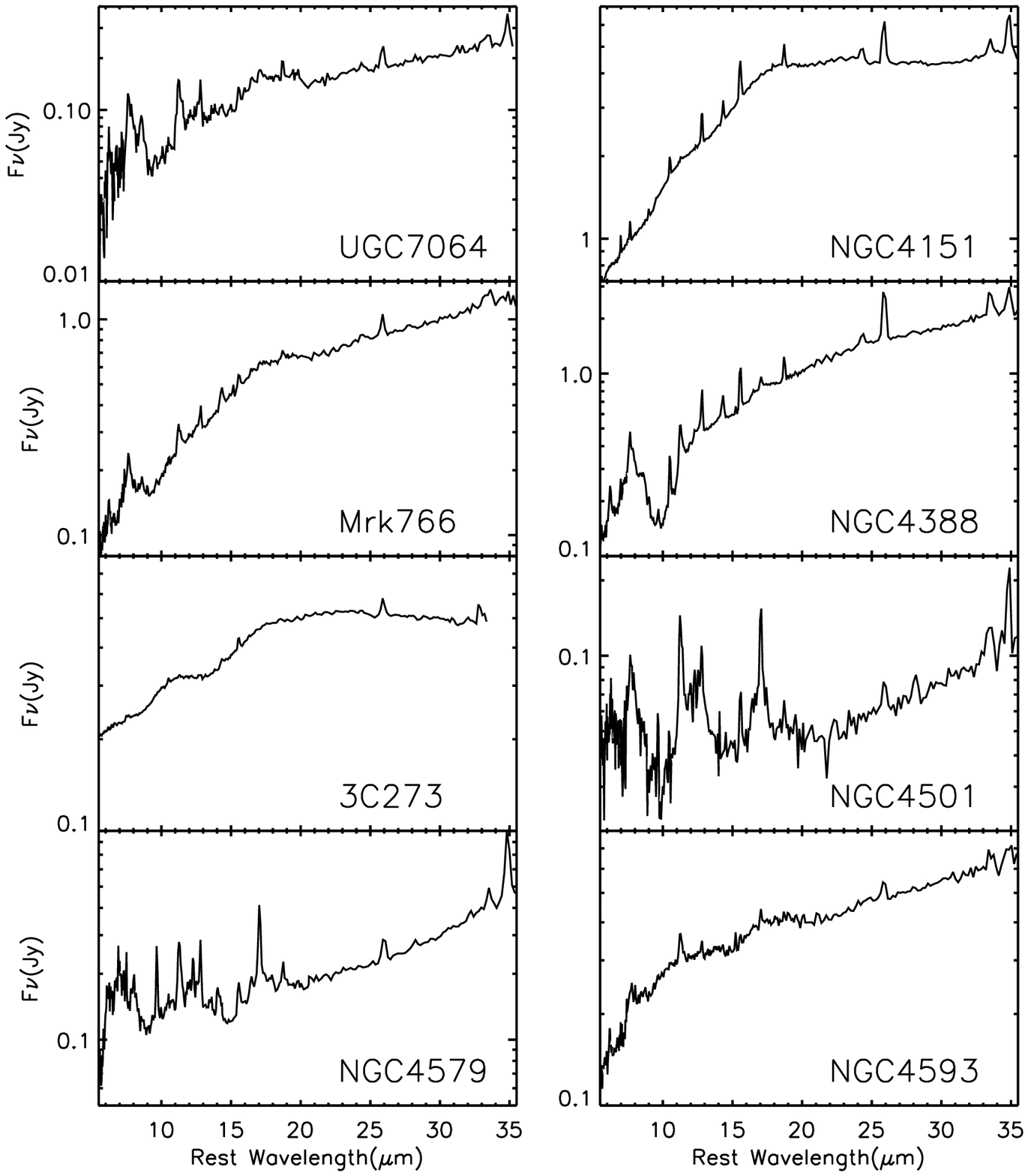}
  \caption{Continued.}
\end{figure}
\begin{figure}
  \setcounter{figure}{17}
  \epsscale{1.0}
  \plotone{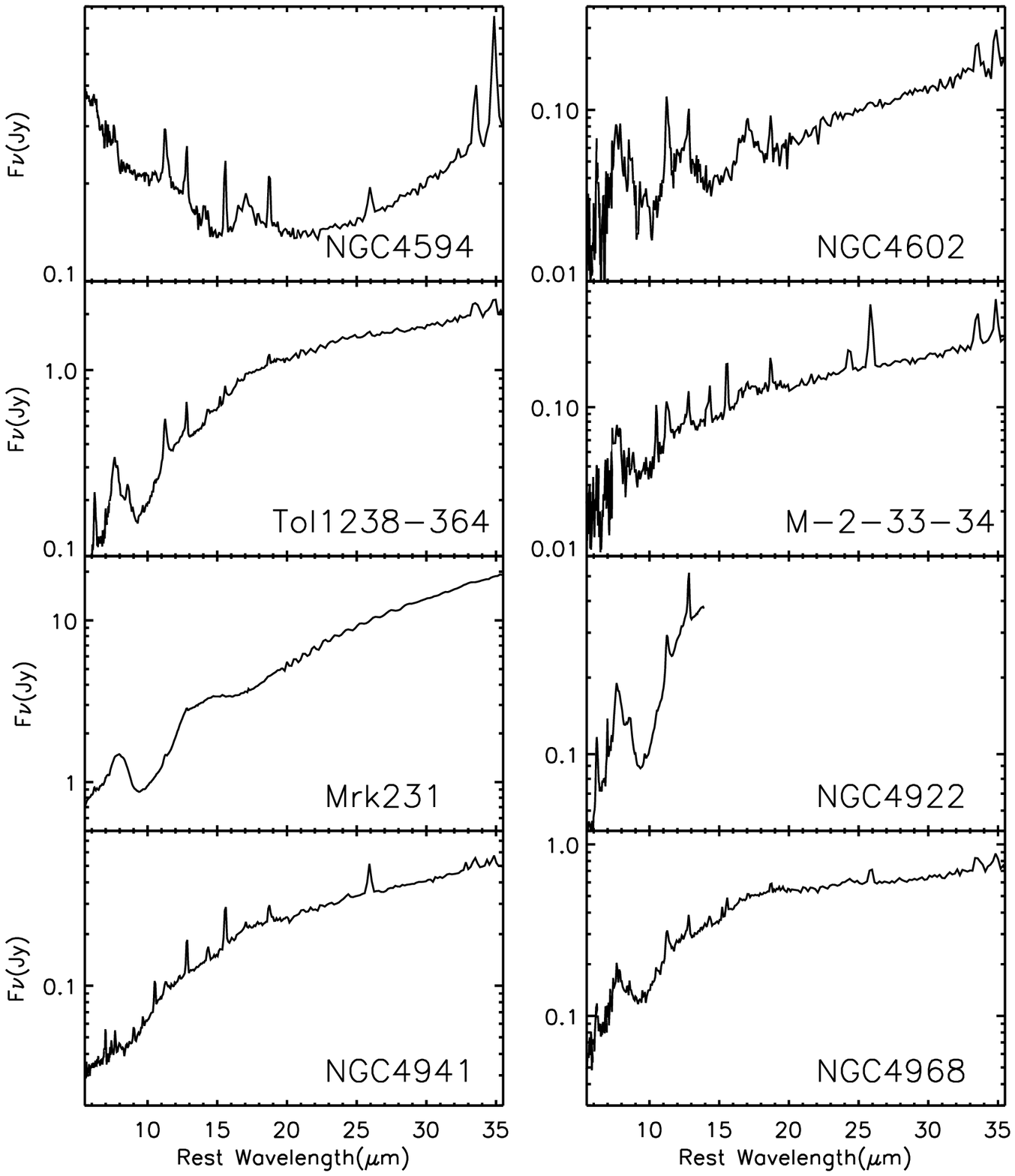}
  \caption{Continued.}
\end{figure}
\begin{figure}
  \setcounter{figure}{17}
  \epsscale{1.0}
  \plotone{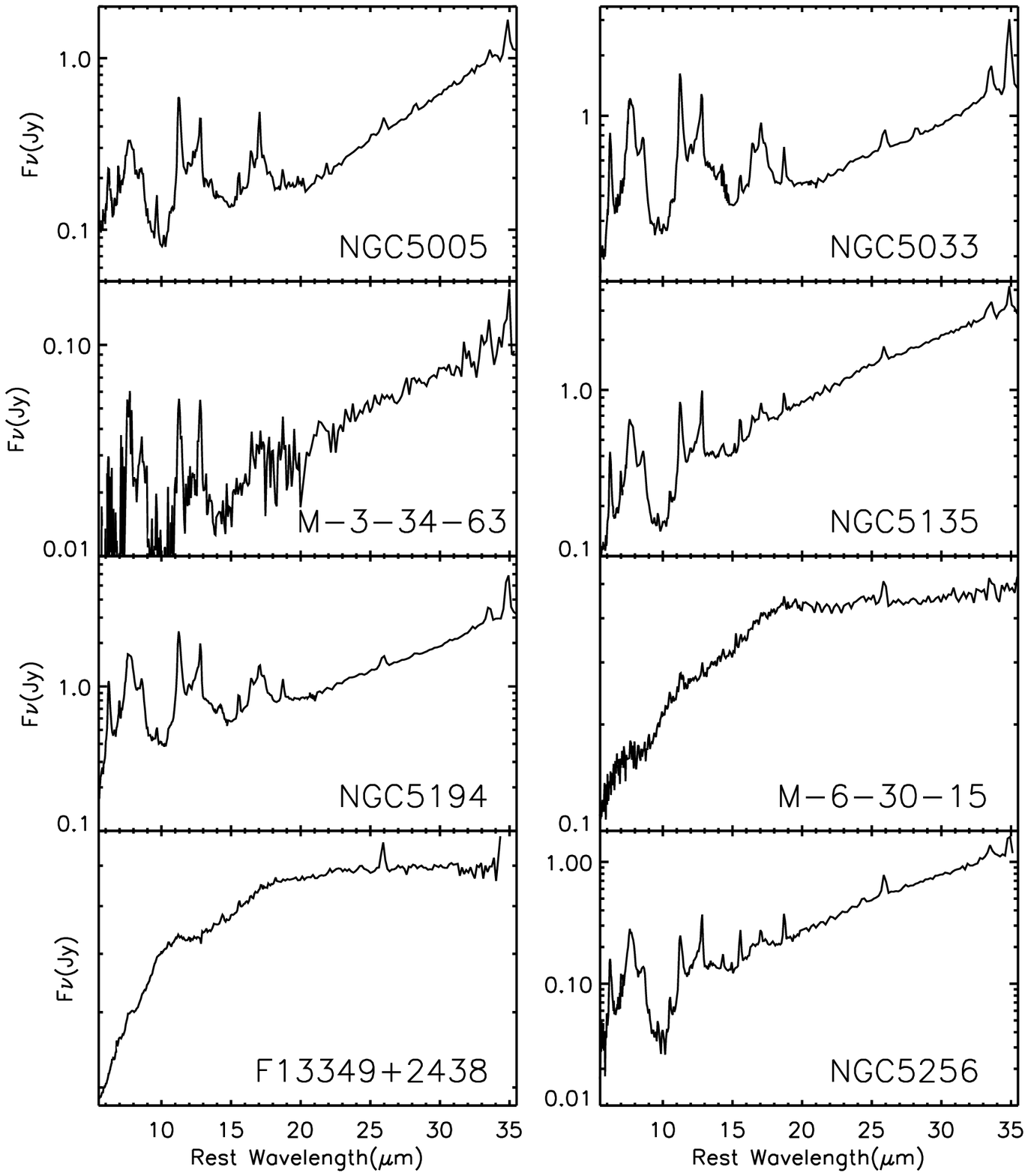}
  \caption{Continued.}
\end{figure}
\begin{figure}
  \setcounter{figure}{17}
  \epsscale{1.0}
  \plotone{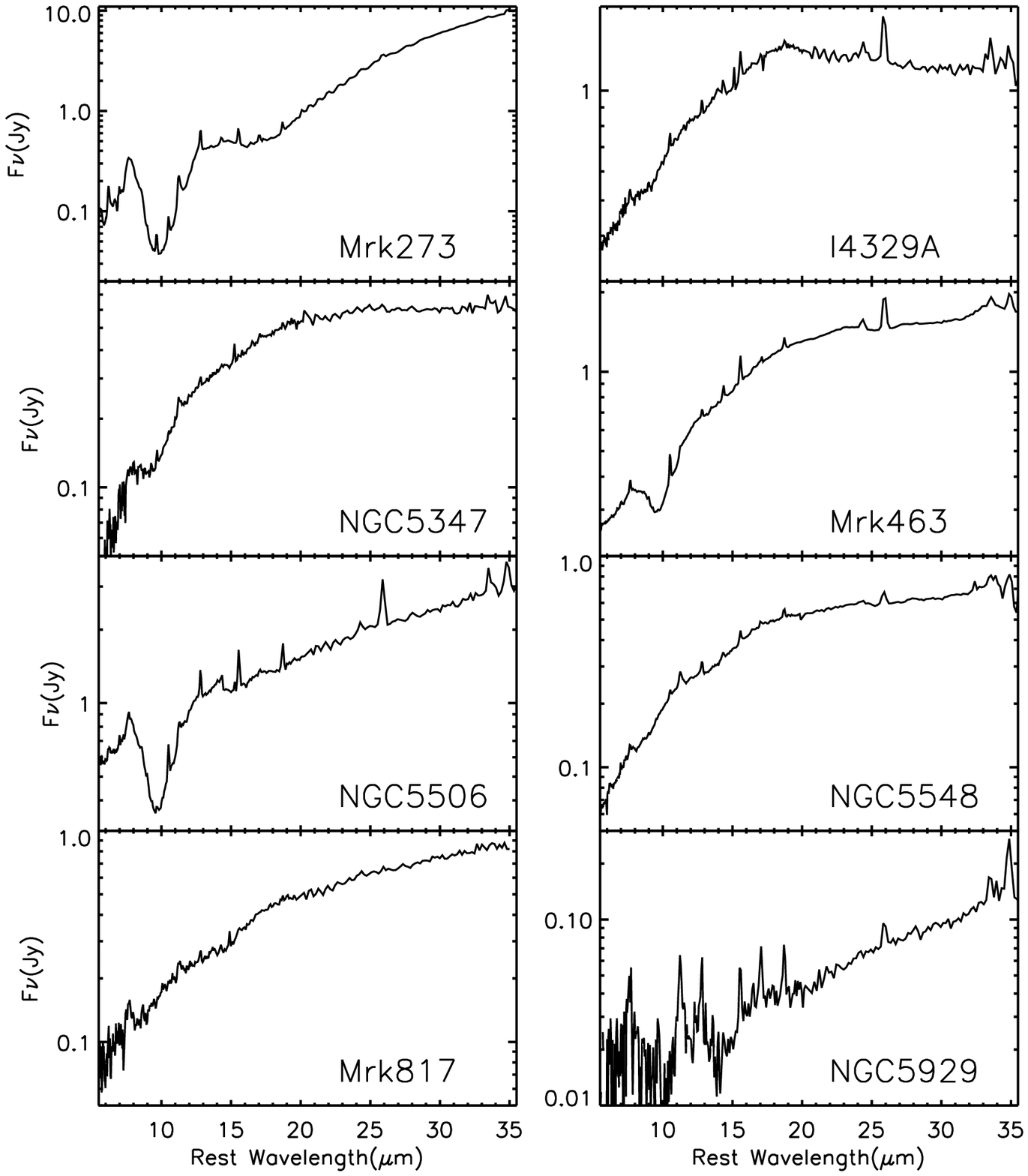}
  \caption{Continued.}
\end{figure}
\begin{figure}
  \setcounter{figure}{17}
  \epsscale{1.0}
  \plotone{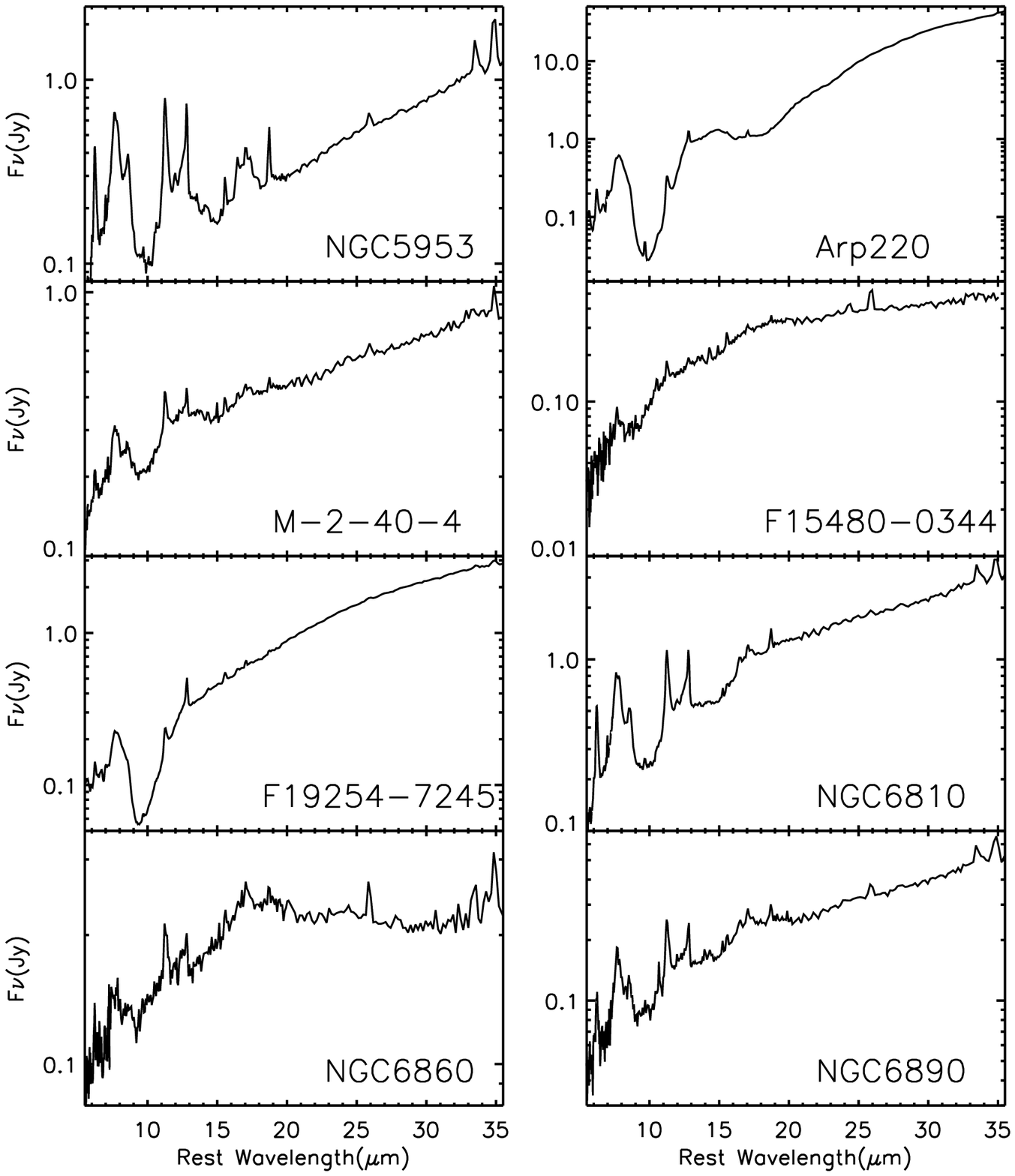}
  \caption{Continued.}
\end{figure}
\begin{figure}
  \setcounter{figure}{17}
  \epsscale{1.0}
  \plotone{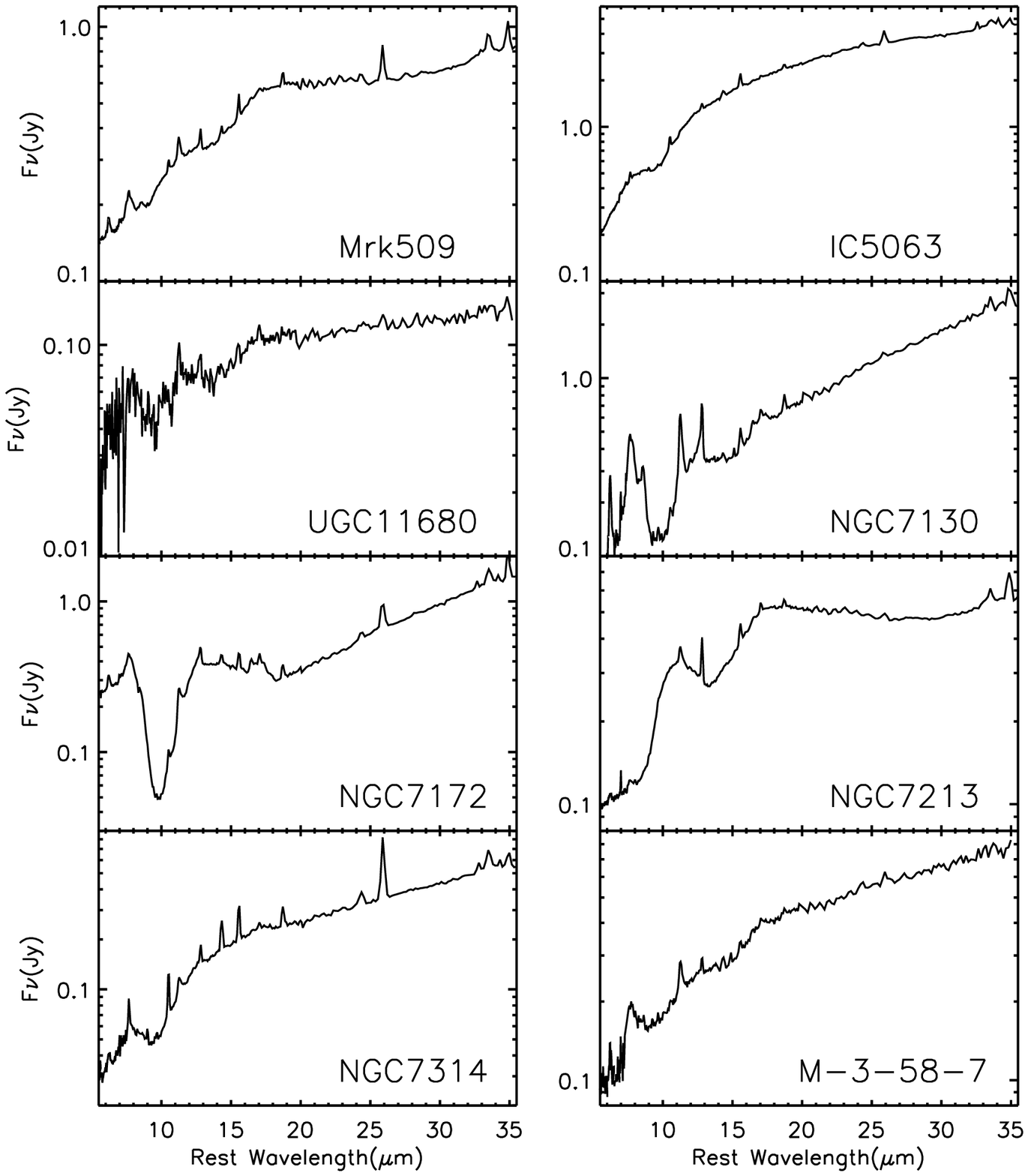}
  \caption{Continued.}
\end{figure}
\begin{figure}
  \setcounter{figure}{17}
  \epsscale{1.0}
  \plotone{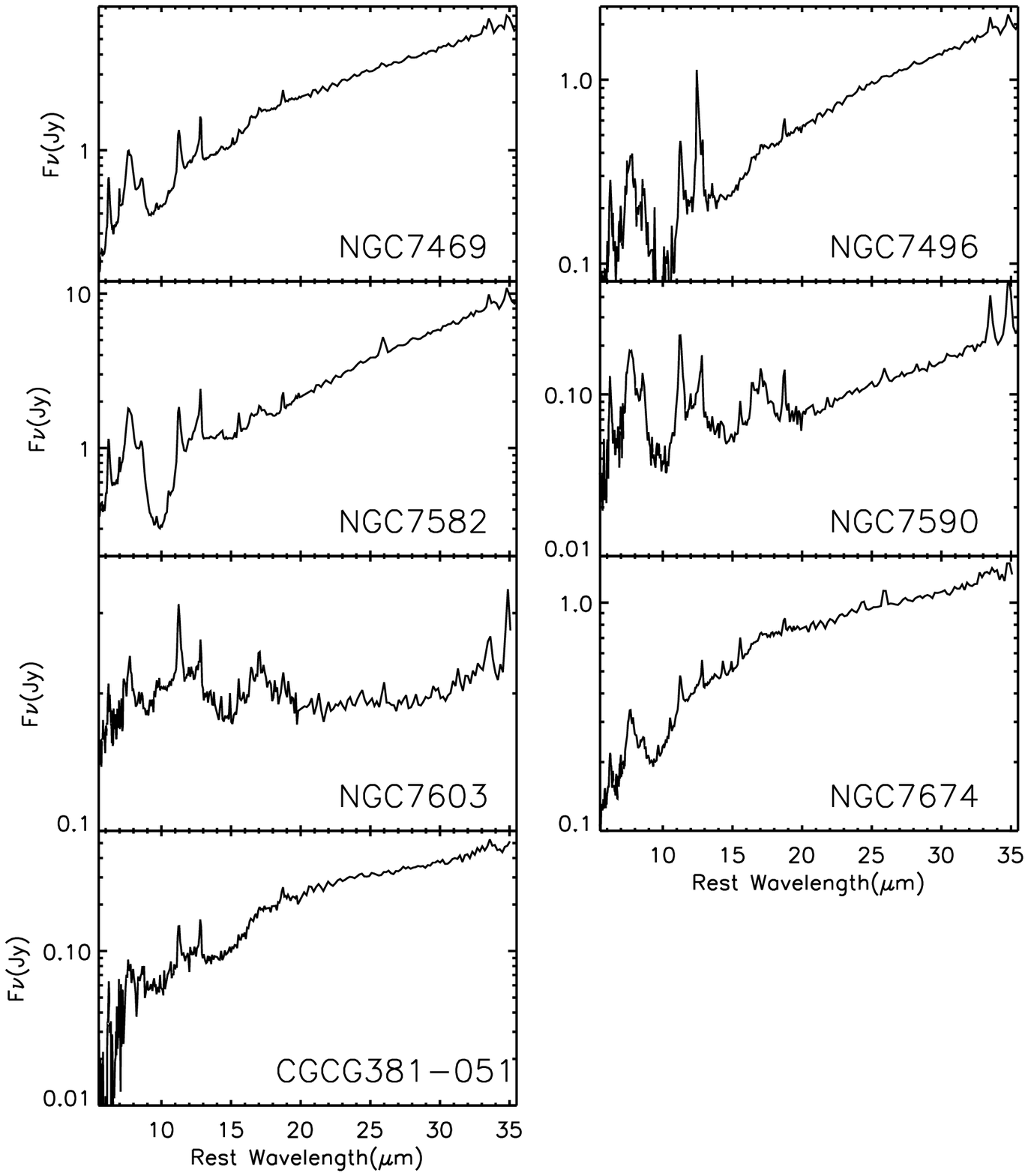}
  \caption{Continued.}
\end{figure}

\begin{deluxetable}{lrrrrrrrccc}
  \tabletypesize{\scriptsize}
  \setlength{\tabcolsep}{0.05in}
  \tablecaption{Properties of the Sample\label{tab1}}
  \tablewidth{0pc}
  \tablehead{
  \colhead{Object Name} & \colhead{RA} & \colhead{Dec} & \multicolumn{4}{c}{IRAS Flux (Jy)\tablenotemark{a}} &  \colhead{log(L$_{\rm IR}$)} & 
  \colhead{redshift} & \colhead{Type} & \colhead{PID}\\
  \colhead{} &  \colhead{J2000} & \colhead{J2000} & \colhead{12\,$\mu$m} & \colhead{25\,$\mu$m} &  \colhead{60\,$\mu$m} 
  & \colhead{100\,$\mu$m} & \colhead{(L$_\odot$)} & \colhead{} & \colhead{} & \colhead{} \\
  }
  \startdata
Mrk335      &  00h06m19.5s & +20d12m10s &   0.27   &   0.45  &  0.35  &   0.57   &  10.72  &  0.026 & Sy 1\tablenotemark{b} & 3269 \\
Mrk938      &  00h11m06.5s & -12d06m26s &   0.35   &   2.39  & 17.05  &   16.86  &  11.48  &  0.020 & Sy 2 & 3269 \\
E12-G21     &  00h40m46.1s & -79d14m24s &   0.22   &   0.19  &  1.51  &    3.22  &  11.03  &  0.030 & Sy 1 & 3269 \\
Mrk348	    &  00h48m47.1s & +31d57m25s &   0.49   &   1.02  &  1.43  &    1.43  &  10.62  &  0.015 & Sy 2\tablenotemark{b} & 3269 \\
IZw1        &  00h53m34.9s & +12d41m36s &   0.47   &   1.17  &  2.24  &    2.87  &  11.95  &  0.061 & Sy 1 & 14 \\
NGC424      &  01h11m27.6s & -38d05m00s &   1.22   &   1.76  &  2.00  &    1.74  &  10.67  &  0.012 & Sy 2\tablenotemark{b} & 3269 \\
NGC526A     &  01h23m54.4s & -35d03m56s	&   0.23   &   0.48  &  2.31  &    4.08  &  10.78  &  0.019 & Sy 1\tablenotemark{b} & 30572 \\
NGC513      &  01h24m26.8s & +33d47m58s &   0.25   &   0.48  &  0.41  &    1.32  &  10.52  &  0.020 & Sy 2 & 3269 \\
F01475-0740 &  01h50m02.7s & -07d25m48s &   0.32   &   0.84  &  1.10  &    1.05  &  10.62  &  0.018 & Sy 2 & 3269 \\
NGC931      &  02h28m14.5s & +31d18m42s &   0.62   &   1.42  &  2.80  &    5.66  &  10.92  &  0.017 & Sy 1 & 3269 \\
NGC1056     &  02h42m48.3s & +28d34m27s &   0.34   &   0.48  &  5.33  &   10.20  &   9.93  &  0.005 & Sy 2 & 3269 \\
NGC1097     &  02h46m19.0s & -30d16m30s &   2.96   &   7.30  & 53.35  &  104.79  &  10.78  &  0.004 & Sy 2 & 159 \\
NGC1125     &  02h51m40.3s & -16d39m04s &   0.32   &   1.00  &  3.71  &    4.04  &  10.46  &  0.011 & Sy 2 & 3269 \\
NGC1143/4   &  02h55m12.2s & -00d11m01s &   0.26   &   0.62  &  5.35  &   11.60  &  10.46  &  0.029 & Sy 2 & 3269\tablenotemark{c} \\
M-2-8-39    &  03h00m30.6s & -11d24m57s &   0.35   &   0.46  &  0.54  &    0.85  &  10.95  &  0.029 & Sy 2\tablenotemark{b} & 3269 \\
NGC1194     &  03h03m49.1s & -01d06m13s &   0.28   &   0.85  &  0.92  &    0.71  &  10.34  &  0.014 & Sy 2 & 3269 \\
NGC1241     &  03h11m14.6s & -08d55m20s &   0.33   &   0.60  &  4.37  &   10.74  &  10.75  &  0.014 & Sy 2 & 3269 \\
NGC1320     &  03h24m48.7s & -03d02m32s &   0.33   &   1.32  &  2.21  &    2.82  &  10.21  &  0.009 & Sy 2 & 3269 \\
NGC1365     &  03h33m36.4s & -36d08m25s &   5.12   &  14.28  & 94.31  &  165.67  &  11.23  &  0.005 & Sy 1 & 3269 \\
NGC1386     &  03h36m46.2s & -35d59m57s &   0.52   &   1.46  &  5.92  &    9.55  &   9.53  &  0.003 & Sy 2 & 3269 \\
F03450+0055 &  03h47m40.2s & +01d05m14s &   0.29   &   0.39  &  0.87  &    3.92  &  11.10  &  0.031 & Sy 1\tablenotemark{b} & 3269 \\
NGC1566     &  04h20m00.4s & -54d56m16s &   1.91   &   3.02  & 22.53  &   58.05  &  10.61  &  0.005 & Sy 1 & 159 \\
3C120       &  04h33m11.1s & +05d21m16s &   0.43   &   0.67  &  1.55  &    4.82  &  11.33  &  0.033 & Sy 1 & 86 \\
F04385-0828 &  04h40m54.9s & -08d22m22s &   0.59   &   1.70  &  2.91  &    3.55  &  10.82  &  0.015 & Sy 2 & 3269 \\
NGC1667     &  04h48m37.1s & -06d19m12s &   0.63   &   0.71  &  6.27  &   14.92  &  11.02  &  0.015 & Sy 2 & 3269 \\
E33-G2      &  04h55m58.9s & -75d32m28s &   0.24   &   0.47  &  0.82  &    1.84  &  10.52  &  0.018 & Sy 2\tablenotemark{b} & 3269 \\
M-5-13-17   &  05h19m35.8s & -32d39m28s &   0.23   &   0.57  &  1.28  &    2.34  &  10.28  &  0.012 & Sy 1 & 3269 \\
F05189-2524 &  05h21m01.5s & -25d21m45s &   0.74   &   3.47  & 13.25  &   11.84  &  12.17  &  0.043 & Sy 2 & 86 \\
Mrk6        &  06h52m12.2s & +74d25m37s &   0.26   &   0.73  &  1.25  &    0.90  &  10.63  &  0.019 & Sy 1\tablenotemark{b} & 3269 \\
Mrk9        &  07h36m57.0s & +58d46m13s &   0.23   &   0.39  &  0.76  &    0.98  &  11.15  &  0.040 & Sy 1 & 3269 \\
Mrk79       &  07h42m32.8s & +49d48m35s &   0.36   &   0.73  &  1.55  &    2.35  &  10.90  &  0.022 & Sy 1 & 3269 \\
F07599+6508 &  08h04m33.1s & +64d59m49s &   0.33   &   0.54  &  1.75  &    1.47  &  12.57  &  0.148 & Sy 1 & 105 \\
NGC2639     &  08h43m38.1s & +50d12m20s &   0.24   &   0.27  &  2.03  &    7.18  &  10.34  &  0.011 & Sy 1 & 3269 \\
F08572+3915 &  09h00m25.4s & +39d03m54s &   0.33   &   1.76  &  7.30  &    4.77  &  12.15  &  0.058 & Sy 2 & 105 \\
Mrk704      &  09h18m26.0s & +16d18m19s &   0.42   &   0.60  &  0.36  &    0.45  &  10.97  &  0.029 & Sy 1\tablenotemark{b} & 704 \\
UGC5101     &  09h35m51.6s & +61d21m11s &   0.25   &   1.02  & 11.68  &   19.91  &  12.00  &  0.039 & Sy 1 & 105 \\
NGC2992     &  09h45m42.0s & -14d19m35s &   0.63   &   1.38  &  7.51  &   17.22  &  10.51  &  0.008 & Sy 1 & 3269 \\
Mrk1239     &  03h10m53.7s & -02d33m11s &   0.76   &   1.21  &  1.68  &    2.42  &  11.32  &  0.029 & Sy 1\tablenotemark{b} & 3269 \\
NGC3031     &  09h55m33.2s & +69d03m55s &   5.86   &   5.42  & 44.73  &  174.02  &   9.70  &  0.001 & Sy 1 & 159 \\
3C234       &  10h01m49.5s & +28d47m09s &   0.22   &   0.35  &  0.24  &    0.34  &  12.42  &  0.185 & Sy 1\tablenotemark{b} & 3624 \\
NGC3079     &  10h01m57.8s & +55d40m47s &   2.54   &   3.61  & 50.67  &  104.69  &  10.62  &  0.004 & Sy 2 & 3269 \\
NGC3227     &  10h23m30.6s & +19d51m54s &   0.94   &   1.83  &  8.42  &   17.30  &   9.97  &  0.004 & Sy 1 & 96 \\
NGC3511     &  11h03m23.8s & -23d05m12s &   1.03   &   0.83  &  8.98  &   21.87  &   9.95  &  0.004 & Sy 1 & 3269 \\
NGC3516     &  11h06m47.5s & +72d34m07s &   0.39   &   0.96  &  2.09  &    2.73  &  10.17  &  0.009 & Sy 1 & 3269 \\
M+0-29-23   &  11h21m12.2s & -02d59m03s &   0.48   &   0.76  &  5.85  &    9.18  &  11.36  &  0.025 & Sy 2 & 3269 \\
NGC3660     &  11h23m32.3s & -08d39m31s &   0.42   &   0.64  &  2.03  &    4.47  &  10.47  &  0.012 & Sy 2 & 3269 \\
NGC3982     &  11h56m28.1s & +55d07m31s &   0.47   &   0.97  &  7.18  &   16.24  &   9.81  &  0.004 & Sy 2 & 3269 \\
NGC4051     &  12h03m09.6s & +44d31m53s &   1.35   &   2.20  & 10.53  &   24.93  &   9.66  &  0.002 & Sy 1 & 3269 \\
UGC7064     &  12h04m43.3s & +31d10m38s &   0.22   &   0.88  &  3.48  &    6.25  &  11.18  &  0.025 & Sy 1 & 3269 \\
NGC4151     &  12h10m32.6s & +39d24m21s &   2.01   &   4.87  &  6.46  &    8.88  &   9.95  &  0.003 & Sy 1\tablenotemark{b} & 14 \\
Mrk766      &  12h18m26.5s & +29d48m46s &   0.35   &   1.47  &  3.89  &    4.20  &  10.67  &  0.013 & Sy 1 & 3269 \\
NGC4388     &  12h25m46.7s & +12d39m44s &   1.01   &   3.57  & 10.27  &   14.22  &  10.73  &  0.008 & Sy 2 & 3269 \\
3C273       &  12h29m06.7s & +02d03m09s &   0.82   &   1.43  &  2.09  &    2.53  &  12.93  &  0.158 & Sy 1\tablenotemark{b} & 105 \\
NGC4501     &  12h31m59.2s & +14d25m14s &   2.29   &   2.98  & 19.68  &   62.97  &  10.98  &  0.008 & Sy 2 & 3269 \\
NGC4579     &  12h37m43.5s & +11d49m05s &   1.12   &   0.78  &  5.93  &   21.39  &  10.17  &  0.005 & Sy 1 & 159 \\
NGC4593     &  12h39m39.4s & -05d20m39s &   0.47   &   0.96  &  3.43  &    6.26  &  10.35  &  0.009 & Sy 1 & 3269 \\
NGC4594     &  12h39m59.4s & -11d37m23s &   0.74   &   0.50  &  4.26  &   22.86  &   9.75  &  0.003 & Sy 1 & 159 \\
NGC4602     &  12h40m36.8s & -05d07m59s &   0.58   &   0.65  &  4.75  &   13.30  &  10.44  &  0.008 & Sy 1 & 3269 \\
Tol1238-364 &  12h40m52.8s & -36d45m21s &   0.72   &   2.54  &  8.90  &   13.79  &  10.87  &  0.011 & Sy 2 & 3269 \\
M-2-33-34   &  12h52m12.4s & -13d24m53s &   0.36   &   0.65  &  1.23  &    2.36  &  10.49  &  0.015 & Sy 1 & 3269 \\
Mrk231      &  12h56m14.2s & +56d52m25s &   1.83   &   8.84  & 30.80  &   29.74  &  12.54  &  0.042 & Sy 1 & 105 \\
NGC4922     &  13h01m24.9s & +29d18m40s	&   0.27   &   1.48  &  6.21  &	   7.33	 &  11.31  &  0.024 & Sy 2 & 3237\tablenotemark{c}\\
NGC4941     &  13h04m13.1s & -05d33m06s &   0.39   &   0.46  &  1.87  &	   4.79  &   9.39  &  0.004 & Sy 2 & 30572 \\
NGC4968     &  13h07m06.0s & -23d40m37s &   0.62   &   1.16  &  2.48  &    3.39  &  10.39  &  0.010 & Sy 2 & 3269 \\
NGC5005     &  13h10m56.2s & +37d03m33s &   1.65   &   2.26  & 22.18  &   63.40  &  10.20  &  0.003 & Sy 2 & 3269 \\
NGC5033     &  13h13m27.5s & +36d35m38s &   1.77   &   2.14  & 16.20  &   50.23  &  10.05  &  0.003 & Sy 1 & 159 \\
M-3-34-63   &  13h22m19.0s & -16d42m30s &   0.95   &   2.88  &  6.22  &    6.37  &  11.38  &  0.021 & Sy 2 & 3269 \\
NGC5135     &  13h25m44.0s & -29d50m01s &   0.63   &   2.38  & 16.86  &   30.97  &  11.27  &  0.014 & Sy 2 & 3269 \\
NGC5194     &  13h29m52.7s & +47d11m43s &   7.21   &   9.56  & 97.42  &  221.21  &  10.18  &  0.002 & Sy 2 & 159 \\
M-6-30-15   &  13h35m53.8s & -34d17m44s &   0.33   &   0.97  &  1.39  &    2.26  &   9.98  &  0.008 & Sy 1\tablenotemark{b} & 3269 \\
F13349+2438 &  13h37m18.7s & +24d23m03s &   0.61   &   0.72  &  0.85  &    0.90  &  12.32  &  0.108 & Sy 1\tablenotemark{b} & 61 \\
NGC5256     &  13h38m17.5s & +48d16m37s &   0.32   &   1.07  &  7.25  &   10.11  &  11.51  &  0.028 & Sy 2 & 3269 \\
Mrk273      &  13h44m42.1s & +55d53m13s &   0.24   &   2.36  & 22.51  &   22.53  &  12.17  &  0.038 & Sy 2 & 105 \\
I4329A      &  13h49m19.2s & -30d18m34s &   1.11   &   2.26  &  2.15  &    2.31  &  10.97  &  0.016 & Sy 1\tablenotemark{b} & 3269 \\
NGC5347     &  13h53m17.8s & +33d29m27s &   0.30   &   1.22  &  1.43  &    3.33  &  10.04  &  0.008 & Sy 2 & 3269 \\
Mrk463      &  13h56m02.9s & +18d22m19s &   0.47   &   1.49  &  2.21  &    1.87  &  11.78  &  0.050 & Sy 2 & 105 \\
NGC5506     &  14h13m14.8s & -03d12m27s &   1.29   &   4.17  &  8.42  &    8.87  &  10.44  &  0.006 & Sy 2 & 3269 \\
NGC5548     &  14h17m59.5s & +25d08m12s &   0.43   &   0.81  &  1.07  &	   2.07  &  10.66  &  0.017 & Sy 1 & 30572 \\
Mrk817      &  14h36m22.1s & +58d47m39s &   0.38   &   1.42  &  2.33  &    2.35  &  11.35  &  0.031 & Sy 1 & 3269 \\
NGC5929     &  15h26m06.1s & +41d40m14s &   0.43   &   1.67  &  9.52  &   13.84  &  10.58  &  0.008 & Sy 2 & 3269 \\
NGC5953     &  15h34m32.4s & +15d11m38s &   0.82   &   1.58  & 11.79  &   19.89  &  10.49  &  0.007 & Sy 2 & 3269 \\
Arp220      &  15h34m57.1s & +23d30m11s &   0.61   &   8.00  &104.09  &  115.29  &  12.18  &  0.018 & Sy 2 & 105 \\
M-2-40-4    &  15h48m24.9s & -13d45m28s &   0.41   &   1.45  &  4.09  &    7.06  &  11.32  &  0.025 & Sy 2 & 3269 \\
F15480-0344 &  15h50m41.5s & -03d53m18s &   0.24   &   0.72  &  1.09  &    4.05  &  11.14  &  0.030 & Sy 2 & 3269 \\
F19254-7245 &  19h31m21.4s & -72d39m18s &   0.26   &   1.35  &  5.24  &    8.03  &  12.14  &  0.062 & Sy 2 & 105 \\
NGC6810     &  19h43m34.4s & -58d39m21s &   1.27   &   3.55  & 18.20  &   32.60  &  10.74  &  0.007 & Sy 2 & 3269 \\
NGC6860     &  20h08m46.9s & -61d06m01s &   0.25   &   0.31  &  0.96  &   2.19   &  10.35  &  0.015 & Sy 1\tablenotemark{b} & 3269 \\
NGC6890     &  20h18m18.1s & -44d48m25s &   0.36   &   0.80  &  4.01  &   8.26   &  10.27  &  0.008 & Sy 2 & 3269 \\
Mrk509      &  20h44m09.7s & -10d43m25s &   0.30   &   0.73  &  1.39  &   1.36   &  11.21  &  0.034 & Sy 1 & 86 \\
I5063       &  20h52m02.3s & -57d04m08s &   1.11   &   3.94  &  5.87  &   4.25	 &  10.87  &  0.011 & Sy 2 & 30572 \\
UGC11680    &  21h07m43.6s & +03d52m30s &   0.37   &   0.86  &  2.97  &   5.59   &  11.23  &  0.026 & Sy 2 & 3269 \\
NGC7130     &  21h48m19.5s & -34d57m05s &   0.58   &   2.16  & 16.71  &  25.89   &  11.38  &  0.016 & Sy 2 & 3269 \\
NGC7172     &  22h02m01.9s & -31d52m11s	&   0.42   &   0.88  &  5.76  &  12.42   &  10.47  &  0.009 & Sy 2 & 30572 \\
NGC7213     &  22h09m16.2s & -47d10m00s &   0.65   &   0.81  &  2.70  &   8.99   &  10.01  &  0.006 & Sy 1\tablenotemark{b} & 86 \\
NGC7314     &  22h35m46.2s & -26d03m01s &   0.55   &   0.96  &  5.24  &  16.57   &  10.00  &  0.005 & Sy 1 & 30572 \\
M-3-58-7    &  22h49m37.1s & -19d16m26s &   0.25   &   0.98  &  2.60  &   3.62   &  11.30  &  0.031 & Sy 2 & 3269 \\
NGC7469     &  23h03m15.6s & +08d52m26s &   1.59   &   5.96  & 27.33  &  35.16   &  11.65  &  0.016 & Sy 1 & 3269 \\
NGC7496     &  23h09m47.3s & -43d25m41s &   0.58   &   1.93  & 10.14  &  16.57   &  10.28  &  0.006 & Sy 2 & 3269 \\
NGC7582     &  23h18m23.5s & -42d22m14s &   2.30   &   7.39  & 52.20  &  82.86   &  10.91  &  0.005 & Sy 2 & 3269 \\
NGC7590     &  23h18m54.8s & -42d14m21s &   0.69   &   0.89  &  7.69  &  20.79   &  10.19  &  0.005 & Sy 2 & 3269 \\
NGC7603     &  23h18m56.6s & +00d14m38s &   0.40   &   0.24  &  1.25  &   2.00   &  11.05  &  0.030 & Sy 1\tablenotemark{b} & 3269 \\
NGC7674     &  23h27m56.7s & +08d46m45s &   0.68   &   1.92  &  5.36  &   8.33   &  11.57  &  0.029 & Sy 2 & 3269 \\
CGCG381-051 &  23h48m41.7s & +02d14m23s &   0.51   &   0.18  &  1.75  &   2.76   &  11.19  &  0.031 & Sy 2 & 3269 \\
\enddata
\tablenotetext{a}{The IRAS fluxes are adopted from \citet{Rush93} and
  \citet{Sanders03}. The uncertainties in the flux measurements range
  from a few percent to over 20\%, averaging $\sim$10\%.}
\tablenotetext{b}{We call these sources ``20\,$\mu$m peakers'', which
  have F20/F30$\ge$0.95.}  \tablenotetext{c}{SL only.}

\end{deluxetable}
\clearpage
\begin{deluxetable}{lccccccc}
  \tabletypesize{\scriptsize}
  \setlength{\tabcolsep}{0.02in}
  \tablecaption{PAH and Silicate Strength of the 12$\mu$m Seyfert Sample\label{tab2}}
  \tablewidth{0pc}
  \tablehead{
    \colhead{Object Name} & \colhead{EW(6.2\,$\mu$m)} & \colhead{F(6.2\,$\mu$m)}  & \colhead{EW(11.2\,$\mu$m)} & \colhead{F(11.2\,$\mu$m)} 
    & \colhead{Silicate Strength\tablenotemark{a}} & \colhead{Aperture Size} & \colhead{Projected Size}\\
    \colhead{} &  \colhead{($\mu$m)} & \colhead{($\times$10$^{-20}$\,W\,cm$^{-2}$)} &  \colhead{($\mu$m)} & \colhead{($\times$10$^{-20}$\,W\,cm$^{-2}$)} & \colhead{} & \colhead{($\arcsec$)} & \colhead{(kpc)} \\
    }
  \startdata
Mrk335      & $<$0.074           & $<$5.17           & $<$0.039           & $<$1.33            &  0.157 & 20.4$\times$15.3 & 11.3$\times$8.4 \\
Mrk938      &    0.440$\pm$0.018 &   33.5$\pm$0.9    &    0.690$\pm$0.042 &   22.1$\pm$0.7     & -0.991 & 20.4$\times$15.3 &  8.6$\times$6.5 \\
E12-G21     &    0.278$\pm$0.017 &   10.9$\pm$0.3    &    0.325$\pm$0.005 &    7.53$\pm$0.03   & -0.022 & 20.4$\times$15.3 & 13.0$\times$9.8 \\
Mrk348      & $<$0.083           & $<$5.78           &    0.058$\pm$0.013 &    2.32$\pm$0.50   & -0.333 & 20.4$\times$15.3 &  6.4$\times$4.8 \\
IZw1        & $<$0.018           & $<$2.90           &    0.018$\pm$0.002 &    2.10$\pm$0.27   &  0.284 & 10.2\tablenotemark{b} & 13.6 \\
NGC424      & $<$0.024           & $<$6.95           &    0.010$\pm$0.001 &    1.53$\pm$0.09   & -0.111 & 20.4$\times$15.3 &  5.1$\times$3.9 \\
NGC526A     & $<$0.240           & $<$1.60           & $<$0.012           & $<$0.62            &  0.033 & 10.2\tablenotemark{b} & 4.1 \\ 
NGC513      &    0.334$\pm$0.107 &    9.50$\pm$1.5   &    0.474$\pm$0.033 &    9.35$\pm$0.37   &  0.149 & 20.4$\times$15.3 &  8.6$\times$6.5 \\
F01475-0740 & $<$0.177           & $<$4.74           &    0.081$\pm$0.009 &    2.75$\pm$0.22   &  0.188 & 20.4$\times$15.3 &  7.7$\times$5.8 \\
NGC931      & $<$0.060           & $<$7.47           &    0.065$\pm$0.002 &    4.53$\pm$0.20   & -0.026 & 20.4$\times$15.3 &  7.3$\times$5.5 \\
NGC1056     &    0.486$\pm$0.125 &   26.9$\pm$3.6    &    0.803$\pm$0.039 &   22.4$\pm$0.5     &  0.084 & 20.4$\times$15.3 &  2.1$\times$1.6 \\
NGC1097     &    0.327$\pm$0.003 &  103$\pm$3        &    0.657$\pm$0.001 &  116$\pm$1         &  0.099 & 50$\times$33     &  4.2$\times$2.8 \\
NGC1125     &    0.258$\pm$0.049 &    7.75$\pm$1.17  &    0.426$\pm$0.025 &    7.16$\pm$0.15   & -1.022 & 20.4$\times$15.3 &  4.7$\times$3.5 \\
NGC1143/4   &    0.343$\pm$0.067 &   18.2$\pm$1.5    &    0.574$\pm$0.028 &   13.0$\pm$0.3     &\nodata\tablenotemark{c} & 19.8$\times$16.2 & 12.2$\times$10.0 \\
M-2-8-39    & $<$0.128           & $<$2.98           & $<$0.047           & $<$1.19            & -0.076 & 20.4$\times$15.3 & 12.6$\times$9.4 \\
NGC1194     & $<$0.059           & $<$5.05           & $<$0.077           & $<$2.47            & -0.978 & 20.4$\times$15.3 &  6.0$\times$4.5 \\
NGC1241     &    0.461$\pm$0.024 &    8.13$\pm$0.34  &    0.501$\pm$0.026 &    5.01$\pm$0.01   & -0.908 & 20.4$\times$15.3 &  6.0$\times$4.5 \\
NGC1320     &    0.082$\pm$0.019 &    6.86$\pm$1.24  &    0.074$\pm$0.003 &    4.60$\pm$0.11   & -0.065 & 20.4$\times$15.3 &  3.8$\times$2.9 \\
NGC1365     &    0.368$\pm$0.003 &  173$\pm$1        &    0.432$\pm$0.015 &  120$\pm$3         & -0.229 & 20.4$\times$15.3 &  2.1$\times$1.6 \\
NGC1386     &    0.053$\pm$0.022 &    5.89$\pm$1.81  &    0.133$\pm$0.002 &    9.13$\pm$0.07   & -0.542 & 20.4$\times$15.3 &  1.3$\times$1.0 \\
F03450+0055 & $<$0.103           & $<$6.58           & $<$0.038           & $<$1.62            &  0.027 & 20.4$\times$15.3 & 13.5$\times$10.1 \\
NGC1566     &    0.223$\pm$0.034 &   27.6$\pm$2.8    &    0.470$\pm$0.004 &   32.2$\pm$0.4     &  0.105 & 50$\times$33     &  5.2$\times$3.4 \\
3C120       & $<$0.015           & $<$1.35           &    0.014$\pm$0.002 &    0.904$\pm$0.142 &  0.130 & 10.2\tablenotemark{b} & 7.2 \\
F04385-0828 & $<$0.058           & $<$8.03           &    0.036$\pm$0.011 &    2.16$\pm$0.62   & -0.766 & 20.4$\times$15.3 &  6.4$\times$4.8 \\
NGC1667     &    0.391$\pm$0.073 &   17.3$\pm$1.5    &    0.731$\pm$0.091 &   15.6$\pm$1.0     & -0.050 & 20.4$\times$15.3 &  6.4$\times$4.8 \\
E33-G2      & $<$0.102           & $<$6.25           & $<$0.076           & $<$2.70            & -0.247 & 20.4$\times$15.3 &  7.7$\times$5.8 \\
M-5-13-17   &    0.200$\pm$0.004 &    6.65$\pm$0.06  &    0.193$\pm$0.047 &    5.22$\pm$0.92   & -0.206 & 20.4$\times$15.3 &  5.1$\times$3.9 \\
F05189-2524 &    0.037$\pm$0.001 &    6.54$\pm$0.16  &    0.062$\pm$0.003 &    8.02$\pm$0.35   & -0.315 & 10.2\tablenotemark{b} & 9.3 \\
Mrk6        & $<$0.097           & $<$6.32           &    0.044$\pm$0.001 &    1.59$\pm$0.03   & -0.036 & 20.4$\times$15.3 &  8.2$\times$6.1 \\
Mrk9        & $<$0.142           & $<$8.12           &    0.116$\pm$0.030 &    3.54$\pm$0.82   &  0.050 & 20.4$\times$15.3 & 17.5$\times$13.1 \\
Mrk79       & $<$0.039           & $<$4.05           &    0.043$\pm$0.011 &    2.22$\pm$0.54   & -0.079 & 20.4$\times$15.3 &  9.5$\times$7.1 \\
F07599+6508 &    0.027$\pm$0.001 &    3.36$\pm$0.14  &    0.018$\pm$0.001 &    0.97$\pm$0.06   &  0.113 & 10.2\tablenotemark{b} & 35.0 \\
NGC2639     & $<$0.207           & $<$5.40           &    0.530$\pm$0.036 &    4.70$\pm$0.19   & -0.127 & 20.4$\times$15.3 &  4.7$\times$3.5 \\
F08572+3915 & $<$0.021           & $<$5.25           & $<$0.099           & $<$2.36            & -3.509 & 10.2\tablenotemark{b} & 12.9 \\
Mrk704      & $<$0.071           & $<$7.38           & $<$0.029           & $<$1.58            & -0.075 & 20.4$\times$15.3 & 12.6$\times$9.4 \\
UGC5101     &    0.229$\pm$0.004 &   12.3$\pm$0.2    &    0.423$\pm$0.007 &   10.1$\pm$0.1     & -1.619 & 10.2\tablenotemark{b} & 8.6 \\
NGC2992     &    0.151$\pm$0.006 &   15.3$\pm$0.2    &    0.237$\pm$0.014 &   15.8$\pm$0.6     & -0.200 & 20.4$\times$15.3 &  3.4$\times$2.6 \\
Mrk1239     & $<$0.029           & $<$7.18           &    0.027$\pm$0.001 &    3.19$\pm$0.20   &  0.010 & 20.4$\times$15.3 & 12.6$\times$9.4 \\
NGC3031     & $<$0.034           &$<$16.8            &    0.183$\pm$0.002 &   23.2$\pm$0.2     & -0.035 & 50$\times$33     &  1.0$\times$0.7 \\
3C234       & $<$0.019           & $<$0.82           & $<$0.013           & $<$0.45            & -0.007 & 10.2\tablenotemark{b} & 44.6 \\
NGC3079     &    0.458$\pm$0.006 &  111$\pm$2        &    0.818$\pm$0.050 &   62.6$\pm$2.0     & -0.828 & 20.4$\times$15.3 &  1.7$\times$1.3 \\
NGC3227     &    0.138$\pm$0.003 &   24.0$\pm$0.5    &    0.249$\pm$0.007 &   29.0$\pm$0.6     & -0.234 & 10.2\tablenotemark{b} & 0.8 \\
NGC3511     &    0.638$\pm$0.168 &   18.0$\pm$1.5    &    0.764$\pm$0.046 &   12.7$\pm$0.4     &  0.009 & 20.4$\times$15.3 &  1.7$\times$1.3 \\
NGC3516     & $<$0.061           & $<$6.57           &    0.024$\pm$0.004 &    1.22$\pm$0.24   & -0.158 & 20.4$\times$15.3 &  3.8$\times$2.9 \\
M+0-29-23   &    0.437$\pm$0.099 &   21.0$\pm$2.5    &    0.619$\pm$0.082 &   16.7$\pm$1.1     & -0.507 & 20.4$\times$15.3 & 10.8$\times$8.1 \\
NGC3660     & $<$0.434           & $<$5.10           &    0.539$\pm$0.100 &    2.97$\pm$0.34   & -0.020 & 20.4$\times$15.3 &  5.1$\times$3.9 \\
NGC3982     &    0.467$\pm$0.058 &   14.4$\pm$1.1    &    0.787$\pm$0.049 &   14.0$\pm$0.3     &  0.110 & 20.4$\times$15.3 &  1.7$\times$1.3 \\
NGC4051     &    0.079$\pm$0.015 &    9.20$\pm$1.20  &    0.114$\pm$0.002 &    9.99$\pm$0.13   &  0.065 & 20.4$\times$15.3 &  0.8$\times$0.6 \\
UGC7064     &    0.364$\pm$0.015 &    7.90$\pm$0.21  &    0.527$\pm$0.055 &    7.87$\pm$0.43   & -0.031 & 20.4$\times$15.3 & 10.8$\times$8.1 \\
NGC4151     & $<$0.011           & $<$6.98           &    0.011$\pm$0.002 &    4.82$\pm$0.91   &  0.030 & 10.2\tablenotemark{b} & 0.7 \\
Mrk766      &    0.036$\pm$0.013 &    3.08$\pm$1.06  &    0.073$\pm$0.023 &    4.38$\pm$1.21   & -0.274 & 20.4$\times$15.3 &  5.6$\times$4.2 \\
NGC4388     &    0.128$\pm$0.013 &   15.0$\pm$1.2    &    0.203$\pm$0.046 &   14.3$\pm$2.5     & -0.699 & 20.4$\times$15.3 &  3.4$\times$2.6 \\
3C273       & $<$0.013           & $<$2.17           & $<$0.014           & $<$1.03            &  0.064 & 10.2\tablenotemark{b} & 37.6 \\
NGC4501     & $<$0.187           & $<$6.96           &    0.629$\pm$0.051 &    6.91$\pm$0.26   & -0.232 & 20.4$\times$15.3 &  3.4$\times$2.6 \\
NGC4579     &    0.152$\pm$0.039 &     14.4$\pm$2.6  &    0.262$\pm$0.054 &    8.86$\pm$1.43   &  0.218 & 50$\times$33     &  5.3$\times$3.5 \\
NGC4593     &    0.044$\pm$0.005 &    4.82$\pm$0.57  &    0.089$\pm$0.014 &    6.10$\pm$0.85   &  0.063 & 20.4$\times$15.3 &  3.8$\times$2.9 \\
NGC4594     &    0.052$\pm$0.002 &    13.1$\pm$0.5   &    0.161$\pm$0.024 &    7.56$\pm$0.97   & -0.036 & 50$\times$33     &  3.6$\times$2.3 \\
NGC4602     &    0.326$\pm$0.194 &    4.61$\pm$1.54  &    0.598$\pm$0.038 &    5.21$\pm$0.17   & -0.010 & 20.4$\times$15.3 &  3.4$\times$2.6 \\
Tol1238-364 &    0.185$\pm$0.003 &   15.1$\pm$0.2    &    0.194$\pm$0.005 &   14.6$\pm$0.3     & -0.312 & 20.4$\times$15.3 &  4.7$\times$3.5 \\
M-2-33-34   & $<$0.304           & $<$5.11           &    0.205$\pm$0.016 &    3.01$\pm$0.17   & -0.209 & 20.4$\times$15.3 &  6.4$\times$4.8 \\
Mrk231      &    0.011$\pm$0.001 &    7.50$\pm$0.39  &    0.028$\pm$0.001 &    8.77$\pm$0.35   & -0.640 & 10.2\tablenotemark{b} & 9.2 \\
NGC4922     &    0.152$\pm$0.008 &    7.22$\pm$0.35  &    0.122$\pm$0.006 &    5.82$\pm$0.29   & \nodata\tablenotemark{c} & 10.2 & 5.1 \\
NGC4941     & $<$0.041           & $<$1.06           &    0.038$\pm$0.016 &    0.82$\pm$0.34   & -0.084 & 10.2\tablenotemark{b} & 0.8 \\
NGC4968     &    0.172$\pm$0.010 &    9.11$\pm$0.44  &    0.127$\pm$0.003 &    6.56$\pm$0.13   & -0.206 & 20.4$\times$15.3 &  4.3$\times$3.2 \\
NGC5005     &    0.192$\pm$0.015 &   18.2$\pm$1.3    &    0.735$\pm$0.018 &   28.8$\pm$0.1     & -0.425 & 20.4$\times$15.3 &  1.3$\times$1.0 \\
NGC5033     &    0.319$\pm$0.060 &   76.8$\pm$8.8    &    0.764$\pm$0.065 &   84.4$\pm$3.3     & -0.116 & 50$\times$33     &  3.1$\times$2.1 \\
M-3-34-63   &    0.915$\pm$0.325 &    4.27$\pm$0.61  &    1.128$\pm$0.215 &    3.24$\pm$0.23   &  0.090 & 20.4$\times$15.3 &  9.1$\times$6.8 \\
NGC5135     &    0.384$\pm$0.032 &   41.6$\pm$1.9    &    0.594$\pm$0.039 &   39.2$\pm$1.3     & -0.367 & 20.4$\times$15.3 &  6.0$\times$4.5 \\
NGC5194     &    0.372$\pm$0.014 &  111$\pm$3        &    0.686$\pm$0.010 &  118$\pm$1         &  0.109 & 50$\times$33     &  2.1$\times$1.4 \\
M-6-30-15   & $<$0.063           & $<$6.33           &    0.052$\pm$0.007 &    3.01$\pm$0.40   & -0.108 & 20.4$\times$15.3 &  3.4$\times$2.6 \\
F13349+2438 & $<$0.008           & $<$2.06           & $<$0.013           & $<$1.69            &  0.038 & 10.2\tablenotemark{b} & 24.7 \\
NGC5256     &    0.608$\pm$0.045 &   19.9$\pm$0.9    &    0.545$\pm$0.068 &   10.8$\pm$0.9     & -0.692 & 20.4$\times$15.3 & 12.1$\times$9.1 \\
Mrk273      &    0.192$\pm$0.003 &   13.0$\pm$0.2    &    0.335$\pm$0.006 &    8.35$\pm$0.12   & -1.746 & 10.2\tablenotemark{b} & 8.2\\
I4329A      & $<$0.020           & $<$6.20           & $<$0.016           & $<$2.85            & -0.077 & 20.4$\times$15.3 &  6.9$\times$5.1 \\
NGC5347     & $<$0.112           & $<$4.37           &    0.068$\pm$0.003 &    3.29$\pm$0.15   & -0.251 & 20.4$\times$15.3 &  3.4$\times$2.6 \\
Mrk463      & $<$0.008           & $<$1.79           &    0.024$\pm$0.002 &    2.78$\pm$0.22   & -0.464 & 10.2\tablenotemark{b} & 11.1 \\
NGC5506     &    0.023$\pm$0.001 &   10.8$\pm$0.3    &    0.060$\pm$0.004 &    9.80$\pm$0.67   & -0.852 & 20.4$\times$15.3 &  2.6$\times$1.9 \\
NGC5548     &    0.018$\pm$0.005 &    1.08$\pm$0.32  &    0.047$\pm$0.007 &    2.71$\pm$0.38   &  0.040 & 10.2\tablenotemark{b} & 3.7 \\
Mrk817      & $<$0.109           & $<$7.60           & $<$0.035           & $<$1.72            & -0.032 & 20.4$\times$15.3 & 13.5$\times$10.1 \\
NGC5929     & $<$0.480           & $<$4.44           &    0.775$\pm$0.087 &    3.50$\pm$0.22   &  0.245 & 20.4$\times$15.3 &  3.4$\times$2.6 \\
NGC5953     &    0.684$\pm$0.039 &   52.9$\pm$1.4    &    0.877$\pm$0.050 &   40.7$\pm$0.1     & -0.068 & 20.4$\times$15.3 &  3.0$\times$2.2 \\
Arp220      &    0.344$\pm$0.006 &   22.2$\pm$0.3    &    0.552$\pm$0.023 &   15.4$\pm$0.6     & -2.543 & 10.2\tablenotemark{b} & 3.9 \\
M-2-40-4    &    0.066$\pm$0.002 &    8.25$\pm$0.17  &    0.119$\pm$0.004 &    8.29$\pm$0.21   & -0.068 & 20.4$\times$15.3 & 10.8$\times$8.1 \\
F15480-0344 & $<$0.190           & $<$5.29           &    0.075$\pm$0.016 &    2.45$\pm$0.48   & -0.159 & 20.4$\times$15.3 & 13.0$\times$9.8 \\
F19254-7245 &    0.064$\pm$0.002 &    5.09$\pm$0.18  &    0.134$\pm$0.006 &    5.06$\pm$0.22   & -1.345 & 10.2\tablenotemark{b} & 13.7 \\
NGC6810     &    0.419$\pm$0.020 &   56.1$\pm$1.4    &    0.463$\pm$0.002 &   46.4$\pm$0.3     & -0.158 & 20.4$\times$15.3 &  3.0$\times$2.2 \\
NGC6860     &    0.084$\pm$0.018 &    6.32$\pm$1.28  &    0.095$\pm$0.008 &    3.59$\pm$0.28   &  0.005 & 20.4$\times$15.3 &  6.4$\times$4.8 \\
NGC6890     &    0.237$\pm$0.020 &    9.66$\pm$0.57  &    0.277$\pm$0.028 &    8.33$\pm$0.58   & -0.054 & 20.4$\times$15.3 &  3.4$\times$2.6 \\
Mrk509      &    0.042$\pm$0.002 &    4.92$\pm$0.19  &    0.068$\pm$0.002 &    4.77$\pm$0.16   & -0.002 & 10.2\tablenotemark{b} & 7.5 \\
I5063       &    0.011$\pm$0.002 &    2.14$\pm$0.34  &    0.019$\pm$0.002 &    3.96$\pm$0.40   & -0.263 & 10.2\tablenotemark{b} & 2.4 \\
UGC11680    & $<$0.334           & $<$7.77           &    0.166$\pm$0.019 &    2.50$\pm$0.25   &  0.142 & 20.4$\times$15.3 & 11.3$\times$8.4 \\
NGC7130     &    0.493$\pm$0.044 &   32.7$\pm$1.9    &    0.430$\pm$0.001 &   25.2$\pm$0.1     & -0.227 & 20.4$\times$15.3 &  6.9$\times$5.2 \\
NGC7172     &    0.045$\pm$0.001 &    9.25$\pm$0.29  &    0.204$\pm$0.009 &    7.52$\pm$0.05   & -1.795 & 10.2\tablenotemark{b} & 1.9 \\
NGC7213     &    0.022$\pm$0.002 &    1.73$\pm$0.20  &    0.037$\pm$0.007 &    2.88$\pm$0.51   &  0.236 & 10.2\tablenotemark{b} & 1.2 \\
NGC7314     &    0.063$\pm$0.007 &    1.63$\pm$0.18  &    0.087$\pm$0.013 &    1.87$\pm$0.27   & -0.476 & 10.2\tablenotemark{b} & 1.0 \\
M-3-58-7    &    0.074$\pm$0.002 &    5.58$\pm$0.13  &    0.099$\pm$0.024 &    5.03$\pm$1.10   & -0.047 & 20.4$\times$15.3 & 13.5$\times$10.1 \\
NGC7469     &    0.293$\pm$0.003 &   60.8$\pm$0.5    &    0.288$\pm$0.012 &   45.4$\pm$1.2     & -0.159 & 20.4$\times$15.3 &  6.9$\times$5.2 \\
NGC7496     &    0.912$\pm$0.189 &   44.6$\pm$4.5    &    0.590$\pm$0.002 &   21.6$\pm$0.1     & -0.626 & 20.4$\times$15.3 &  2.6$\times$1.9 \\
NGC7582     &    0.274$\pm$0.023 &  101$\pm$5        &    0.457$\pm$0.006 &   77.1$\pm$0.3     & -0.833 & 20.4$\times$15.3 &  2.2$\times$1.6 \\
NGC7590     &    0.496$\pm$0.005 &   14.0$\pm$0.1    &    0.854$\pm$0.082 &   13.0$\pm$0.5     &  0.023 & 20.4$\times$15.3 &  2.2$\times$1.6 \\
NGC7603     &    0.056$\pm$0.004 &    7.08$\pm$0.53  &    0.121$\pm$0.006 &    6.22$\pm$0.24   &  0.211 & 20.4$\times$15.3 & 13.0$\times$9.8 \\
NGC7674     &    0.132$\pm$0.006 &   14.2$\pm$0.4    &    0.129$\pm$0.013 &   10.0$\pm$0.8     & -0.215 & 20.4$\times$15.3 & 12.6$\times$9.4 \\
CGCG381-051 &    0.542$\pm$0.031 &    5.07$\pm$0.16  &    0.211$\pm$0.037 &    4.04$\pm$0.47   &  0.316 & 20.4$\times$15.3 & 13.5$\times$10.1 \\
\enddata

\tablenotetext{a}{The error in the silicate measurement is determined
  by the S/N of the continuum flux, which has an absolute uncertainty
  of $\sim$5\%.}
\tablenotetext{b}{Spectrum extracted from IRS staring mode. The size
  of the LL slit is 2 pixel wide, where 1 pixel$\sim$5.1$\arcsec$.}
\tablenotetext{c}{To measure the silicate feature, a complete
  5-35\,$\mu$m spectrum is needed. This source only has an SL spectrum
  from 5-14.5\,$\mu$m.}
\end{deluxetable}
\clearpage

\begin{deluxetable}{lcccc}
  \tabletypesize{\scriptsize}
  \setlength{\tabcolsep}{0.2in}
  \tablecaption{11.2\,$\mu$m PAH EWs for the 12\,$\mu$m Seyfert sample\label{tab3}}
  \tablewidth{0pc}
  \tablehead{
  \colhead{} &  \multicolumn{2}{c}{Seyfert 1} &  \multicolumn{2}{c}{Seyfert 2} \\
  \colhead{} &  \colhead{EW($\mu$m)} & \colhead{Sources} & \colhead{EW($\mu$m)} & \colhead{Sources} 
  }
  \startdata
Whole sample\tablenotemark{a}                     & 0.154$\pm$0.216 & 47 & 0.352$\pm$0.312 & 56 \\
Whole sample\tablenotemark{b}                     & 0.205$\pm$0.215 & 37 & 0.384$\pm$0.299 & 52 \\
L$_{\rm IR}$$<$10$^{11}$L$_\odot$\tablenotemark{a}  & 0.171$\pm$0.235 & 29 & 0.356$\pm$0.318 & 35 \\
L$_{\rm IR}$$\ge$10$^{11}$L$_\odot$\tablenotemark{a}& 0.127$\pm$0.184 & 18 & 0.345$\pm$0.310 & 21 \\
L$_{\rm IR}$$<$10$^{11}$L$_\odot$\tablenotemark{b}  & 0.217$\pm$0.232 & 24 & 0.395$\pm$0.303 & 32 \\
L$_{\rm IR}$$\ge$10$^{11}$L$_\odot$\tablenotemark{b}& 0.185$\pm$0.188 & 13 & 0.367$\pm$0.300 & 20 \\
F25/F60$>$0.2\tablenotemark{a}                    & 0.068$\pm$0.116 & 33 & 0.143$\pm$0.237 & 30 \\
F25/F60$\le$0.2\tablenotemark{a}                  & 0.358$\pm$0.262 & 14 & 0.592$\pm$0.191 & 26 \\   
F25/F60$>$0.2\tablenotemark{b}                    & 0.102$\pm$0.119 & 24 & 0.177$\pm$0.237 & 26 \\
F25/F60$\le$0.2\tablenotemark{b}                  & 0.397$\pm$0.227 & 13 & 0.592$\pm$0.191 & 26 \\   
  \enddata
\tablenotetext{a}{Values calculated including the upper limits on PAH EWs.}
\tablenotetext{b}{Values calculated excluding the upper limits on PAH EWs.}
\end{deluxetable}

\end{document}